\documentclass[journal]{vgtc}                     

\onlineid{0}

\vgtccategory{Research}

\vgtcpapertype{please specify}

\title{Make the Fastest Faster: Importance Mask Synthesis for Interactive Volume Visualization using Reconstruction Neural Networks}

\author{%
  \authororcid{Jianxin Sun}{0000-0002-9627-9397},
  \authororcid{David Lenz}{0000-0002-2587-2783}, 
  \authororcid{Hongfeng Yu}{0000-0002-0596-8227}, and
  \authororcid{Tom Peterka}{0000-0002-0525-3205}
}

\authorfooter{
  \item
  	Jianxin Sun and Hongfeng Yu are with the University of Nebraska-Lincoln.
  	E-mail: jianxin.sun@huskers.unl.edu, yu@cse.unl.edu.
  \item
  	David Lenz and Tom Peterka are with Argonne National Laboratory.
  	E-mail: dlenz@anl.gov, tpeterka@mcs.anl.gov.

}

\abstract{
Visualizing a large-scale volumetric dataset with high resolution is challenging due to the substantial computational time and space complexity. Recent deep learning-based image inpainting methods significantly improve rendering latency by reconstructing a high-resolution image for visualization in constant time on GPU from a partially rendered image where only a portion of pixels go through the expensive rendering pipeline. However, existing solutions need to render every pixel of either a predefined regular sampling pattern or an irregular sample pattern predicted from a low-resolution image rendering. Both methods require a significant amount of expensive pixel-level rendering. In this work, we provide Importance Mask Learning (IML) and Synthesis (IMS) networks, which are the first attempts to directly synthesize important regions of the regular sampling pattern from the user's view parameters, to further minimize the number of pixels to render by jointly considering the dataset, user behavior, and the downstream reconstruction neural network. Our solution is a unified framework to handle various types of inpainting methods through the proposed differentiable compaction/decompaction layers. Experiments show our method can further improve the overall rendering latency of state-of-the-art volume visualization methods using reconstruction neural network for free when rendering scientific volumetric datasets. Our method can also directly optimize the off-the-shelf pre-trained reconstruction neural networks without elongated retraining.
}


\teaser{
  \centering
  \includegraphics[trim=0 0 0 0,clip,width=\linewidth]{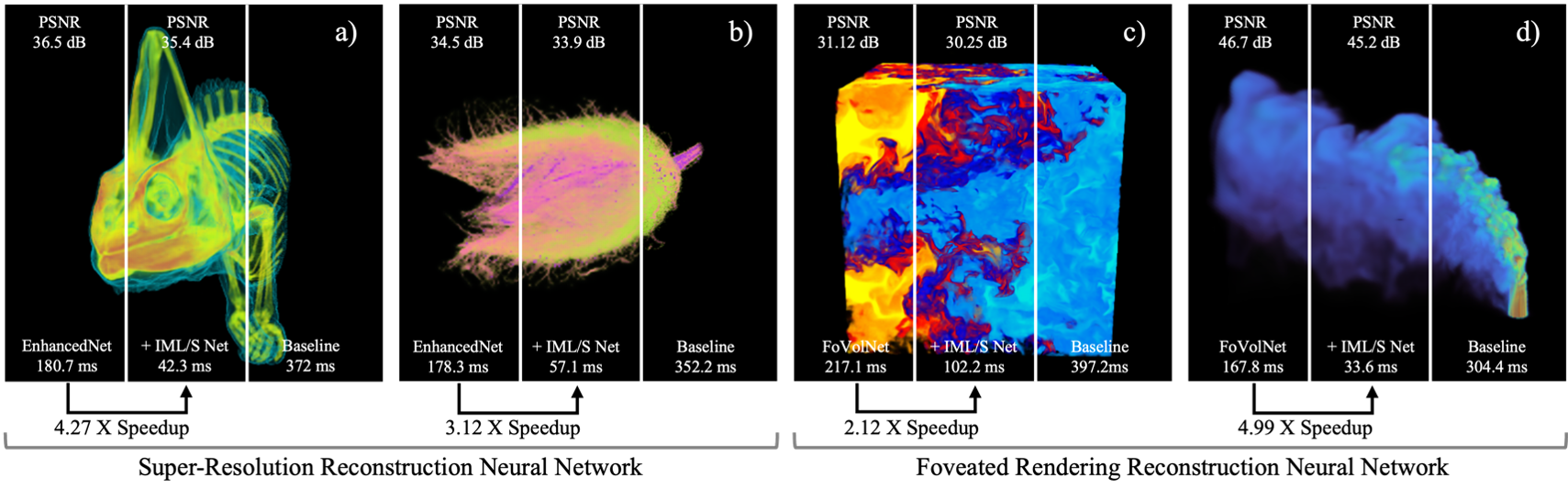}
  \caption{The proposed IML and IMS Networks can further reduce the rendering latency of already fast volume visualization methods using reconstruction neural networks, like the super-resolution EnhanceNet (a and b) and foveated rendering FoVolNet (c and d), with similar rendering quality. For each subfigure, the rightmost portion is the baseline using the traditional ray casting method, the leftmost portion is the interactive visualization method using reconstruction neural network, and the middle portion is the proposed method using our IML/S Net that enhances the performance of respective reconstruction neural network.
  }
  \label{fig:teaser}
}

\graphicspath{{figs/}{figures/}{pictures/}{images/}{./}} 

\usepackage{tabu}                      
\usepackage{booktabs}                  
\usepackage{lipsum}                    
\usepackage{mwe}                       

\usepackage{mathptmx}                  

\usepackage{microtype}                 
\PassOptionsToPackage{warn}{textcomp}  
\usepackage{textcomp}                  
\usepackage{mathptmx}                  
\usepackage{times}                     
\usepackage{cite}                      
\usepackage{tabu}                      
\usepackage{booktabs}                  
\usepackage{adjustbox}
\usepackage{tabularx} 
\usepackage{array}
\usepackage{caption}
\usepackage{subcaption}
\usepackage{gensymb}
\usepackage{amsmath}
\usepackage{amssymb}    
\usepackage{enumitem}
\usepackage{algorithm}
\usepackage{algpseudocode}
\usepackage{multirow}
\usepackage{array}
\usepackage{wrapfig}
\usepackage{graphicx}
\usepackage[normalem]{ulem}

\newcolumntype{P}[1]{>{\centering\arraybackslash}p{#1}}
\newcolumntype{M}[1]{>{\centering\arraybackslash}m{#1}}
\newcolumntype{C}{>{\centering\arraybackslash}X}

\usepackage{interval}

\usepackage{mathrsfs}

\begin{document}




\firstsection{Introduction}
\maketitle
Interactive volume visualization techniques are crucial for enabling researchers across various disciplines to efficiently discover insightful features within scientific datasets from fields like medical imaging, geophysics, meteorology, materials science, and physical simulations. A visualization system that quickly responds to user interactions, adapting to data and viewing adjustments, can greatly enhance the efficiency and effectiveness of exploring complex scientific datasets. However, interactive volume visualization is challenging because of the following factors: First, the size of volumetric data generated nowadays is growing exponentially, necessitating an efficient storage and I/O system to facilitate large-scale data streaming between different computing units for generating comprehensive visualization results. Second, visualization techniques, such as ray casting, ray tracing, global illumination, and high-order rendering, often require significant computational resources to generate high-resolution, high-fidelity images.

Various strategies have been proposed to address this performance issue in large-scale volumetric data visualization. For example, distributed volume visualization methods~\cite{10386434, 5219060} harness the parallel machines to divide the workload and then composite the partially rendered results into the final rendering image. Multi-resolution methods~\cite{Heckbert1999MultiresolutionMF, 10.1145/1186822.1073277, GUTHE200451, 1234567} optimize the arrangement of data segments with different levels of detail (LOD) to improve the rendering latency while maintaining a high rendering quality. To improve the smoothness of volumetric data exploration, visualization systems also utilize caching and prefetching~\cite{10.1145/279361.279372, 10.1145/2907071} to reduce out-of-core data movement across the system memory hierarchy, thereby decreasing input latency while rendering continuous frames along a user's exploratory trajectory. The user behavior while exploring volumetric data can be learned~\cite{10549835, 8365978} to better assist the prefetching through the prediction of view parameters.

Recent image synthesis techniques utilizing deep learning-based super-resolution in both spatial and temporal domains~\cite{10.1145/3485132,10.1007/978-3-030-00563-4_11, liu2022video} are being adopted for visualizing large-scale volumetric datasets with responsive input latency. Given a partially rendered image from a view angle, where only a small percentage of the total pixels are rendered, the idea is to predict or interpolate the missing pixels through a trained neural network in constant time by leveraging the parallelization of the GPU instead of going through expensive rendering algorithms to calculate color blending for each pixel of the final image. The network that reconstructs an incomplete image into a full image is called a reconstruction neural network, and we refer to such a network as \emph{RecNN} in the rest of this paper. Those methods rely on training an implicit neural representation~\cite{NEURIPS2020_53c04118, 10.1007/978-3-031-19809-0_5} using the original large-scale dataset and a complex rendering pipeline, and once trained, the representation enables fast inference, making it well-suited for interactive visualization tasks. Volume visualization methods based on RecNNs have achieved state-of-the-art performance in rendering latency~\cite{8237743, 9903564}, particularly when processing large-scale scientific datasets with complex lighting effects.

The bag of pixels to render for each partially rendered image is determined by a binary sampling pattern provided by the specific reconstruction neural network used (e.g., downsampling patterns for super-resolution RecNNs and foveal patterns for foveated rendering RecNNs). However, such a pattern is a predefined hyperparameter before training, and it does not adapt to any of the key visualization parameters like the user's views, transfer functions, or features of the underlying dataset. Recently proposed Learning Adaptive Sampling (LAS)~\cite{9264699} tries to predict an arbitrary selection of important pixels, called the Importance Mask (IM), from a low-res rendering image. The ratio of the important pixels to the total rendering pixels is called the Important Pixel Ratio (IPR). Although LAS reduces the number of pixels to render for the downstream RecNN, it suffers from several key drawbacks that limit its practicality and lead to suboptimal performance gains: 1) Rendering the low-res input image still takes a relatively long time. 2) The IM can only be learned from the 2D space with the same resolution of the final rendering image, which can be very high for scientific visualization (normally $\ge512\times512$), resulting in a large number of samples to render for a specific IPR with a certain quality level. 3) LAS only supports the super-resolution type of RecNNs and needs a long time to retrain its weights. In this work, we provide a unified Importance Mask Learning Network (IML Net) to learn and distill the IM from a low-res intermediate embedding supporting various types of downstream RecNNs through the proposed compaction/decompaction layers. IML Net learns the IM directly from the sampling pattern by jointly considering the dataset, the user's view parameters, and the downstream RecNN. We also proposed an Importance Mask Synthesis Network (IMS Net), which will directly synthesize the IM from novel views during the user's exploration. Our neural rendering network (IML/S Net) combines the two networks with the downstream RecNN, achieving much faster rendering than LAS with similar rendering quality under the same IPR.
We use two state-of-the-art image inpainting RecNNs for volume visualization, EnhanceNet~\cite{8237743} for super-resolution, and FoVolNet\cite{9903564} for foveated rendering, to showcase how our method can further reduce the rendering latency for free without sacrificing noticeable rendering quality. We also perform a comprehensive comparison with LAS for both rendering quality and latency.
The main contributions of this work include:
\setlist{nolistsep}
\begin{itemize}[leftmargin=*]
  \item Importance Mask Learning Network(IML Net), a unified neural network to learn the IM from low-res intermediate embedding derived from diverse sampling patterns by jointly considering the volumetric dataset, the user's view parameters, and the downstream reconstruction neural network.
  \item Importance Mask Synthesis Network(IMS Net), a regressor designed to efficiently predict the importance mask directly from view parameters instead of computationally expensive input.
  \item IML/S Net + RecNN, a visualization neural rendering network to further improve the already fast volume visualization methods using RecNN for free with similar rendering quality.
  \item Our networks can also be efficiently trained independently of the downstream RecNN to optimize the rendering latency for off-the-shelf pre-trained RecNN.
\end{itemize}

\section{Related Work} 
\label{related work}
\subsection{Implicit Neural Representation}
Implicit Neural Representation (INR) refers to a method of representing data, such as images, 3D shapes, or other types of continuous signals, using a neural network. For handling large-scale volume visualization, compression methods utilizing implicit neural representation \cite{lu2021compressive, tang2020deep} have been proposed to reduce network size and optimize I/O-intensive operations. To improve the rendering latency of the visualization system, Wu et al.~\cite{10175377} utilize hash encoding-based INR to accelerate interactive volume visualization with high reconstruction quality. Yariv et al.~\cite{NEURIPS2021_25e2a30f} improve geometry representation and reconstruction in neural volume rendering by modeling the volume density as an implicit function of the geometry. Weiss et al.~\cite{weiss2022fast} introduce fV-SRN as a novel extension of SRN (Scene Representation Networks) to achieve significantly accelerated reconstruction performance of volume rendering. Despite the use of powerful GPUs, the training time for INR remains significantly long when processing complex, large-scale scientific datasets~\cite{TANG2024103874, 10175377}. Sun et al.~\cite{sun2025f} propose F-Hash to significantly speed up the convergence of training INR from large-scale time-varying scientific data through a feature-based multi-resolution Tesseract input encoding. Both the proposed IML and IMS Nets were trained as INRs with learned properties of the dataset, view parameters, and the downstream RecNN.

\subsection{Reconstruction Neural Network}

A Reconstruction Neural Network (RecNN) in visualization refers to a type of neural network used to reconstruct or generate detailed representations of data from incomplete, noisy, or compressed input~\cite{10.1007/s10462-022-10147-y, cao2021video}. The goal of RecNN is to recover or restore a high-quality, detailed version of an object or dataset from a simplified or partial form.
The most commonly used type of RecNN is super-resolution neural networks that reconstruct high-res data from low-res data. The sampling pattern of super-resolution RecNN is a downsampling pattern. The EnhanceNet, proposed by Sajjadi et al.~\cite{8237743}, demonstrates high reconstruction quality in generating volume visualization images~\cite{9264699}. Weiss et al.~\cite{8918030} propose a volumetric isosurface rendering with deep learning-based super-resolution. Tang et al.~\cite{TANG2024103874} propose STSR-INR, which extends the super-resolution to both spatial and temporal domains. Another important RecNN is the foveated rendering RecNN, which does reconstruction from partial rendering near the region where the user is focusing their gaze. Together with the techniques in eye-tracking-related research~\cite{sundstedt2022systematic, 8900970}, foveated rendering provides a solution to balance the user experience and computational cost. The sampling pattern of foveated rendering RecNN is a foveal pattern. Kaplanyan et al. propose DeepFovea~\cite{10.1145/3355089.3356557}, the first foveated reconstruction method utilizing generative adversarial networks (GAN) to speed up the rendering frame rate for gaming. Then Bauer et al. propose FoVolNet~\cite{9903564}  to improve the foveated rendering in the volume visualization domain and achieve state-of-the-art rendering latency. In this work, we select two RecNNs, EnhanceNet and FoVolNet, as representations of super-resolution RecNN and foveated rendering RecNN to demonstrate how much rendering latency can be further reduced using our method.

\subsection{Image Synthesis}

Deep learning-based image synthesis in volume visualization refers to the use of deep learning techniques to generate or enhance visual representations of 3D volumetric data, such as medical scans (MRI, CT), scientific simulations, or geological models~\cite{lochmann2016real}. Volumetric data can be very large, making real-time rendering computationally expensive. Generative models can efficiently synthesize rendering results by learning the underlying structure of the data, thereby reducing the need for expensive pixel-level computation and enabling faster generation of visual results. Berger et al.~\cite{8316963} utilize a GAN framework to directly generate the volume visualization image from view parameters and transfer functions. He et al.~\cite{8805426} provide a regressor to directly generate visualization images from simulation and visualization parameters for fast parameter space exploration. Jun et al. propose VCNet, a new deep-learning approach for volume completion by synthesizing missing subvolumes. Our IMS Net also utilizes image synthesis techniques to generate an importance mask directly from view parameters. Although training generative models, especially for large-scale or high-dimensional data, requires large training datasets and significant computational resources, the importance masks learned by the proposed IML Net are low-res 2D binary matrices that can be efficiently learned by the IMS Net.

\section{Methods}
\subsection{Overview}
\begin{figure}[t]
    \centering 
    \includegraphics[trim=0 0 0 0,clip,width=\linewidth]{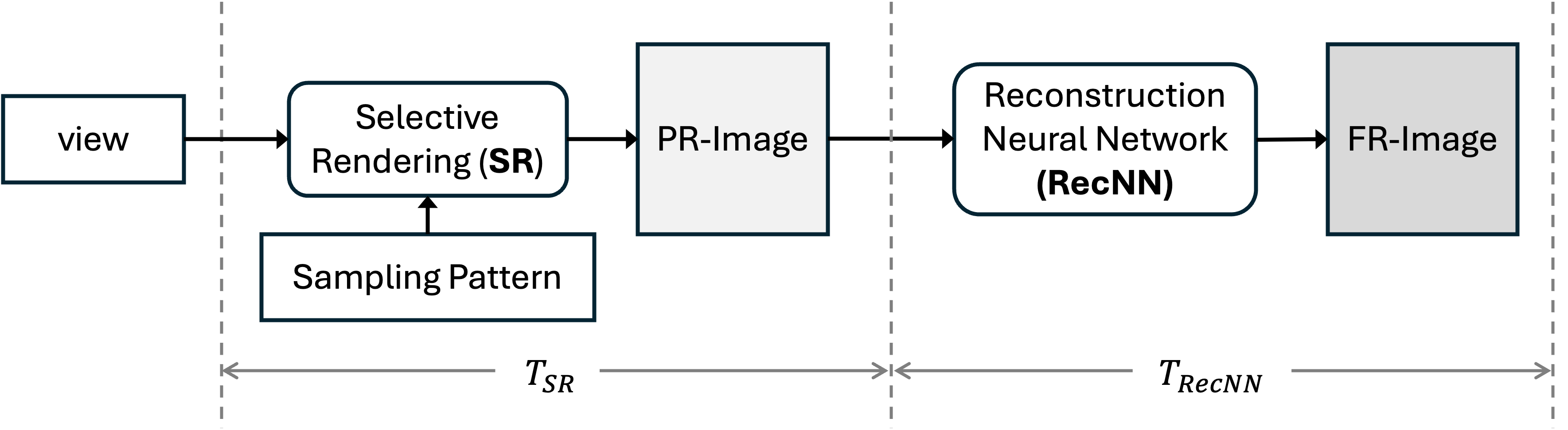}
    \caption{Typical pipeline of interactive volume visualization using RecNN.}
    \label{fig:pipeline}
\end{figure}

For a general volume visualization system, the rendering latency (RL) is the time difference between the moment the user makes a data- or view-dependent operation and the time when the rendering of the visualization image is finished. For interactive volume visualization using RecNN, as shown in \cref{fig:pipeline}, the rendering latency ($T_{RL}$) is the sum of two components, which are the time of Selective Rendering (SR) for generating the partially rendered image (PR-Image), $T_{SR}$, and the inferencing time of the RecNN to generate a fully rendered image (FR-Image) from PR-Image, $T_{RecNN}$:

\begin{equation}
T_{RL}=T_{SR} + T_{RecNN}
\end{equation}

Once the RecNN is defined and trained, its inferencing time ($T_{RecNN}$) can be accelerated by GPU in constant time. We define the pixels covered by the sampling pattern as $S_{sp}$. For a specific sampling pattern, $T_{SR}$ is the sum of time used to render each pixel in the sampling pattern defined by the specific RecNN.

\begin{equation}
T_{SR}= \sum_{k \in S_{sp}}T_{k}
\end{equation}

The time used to render the $k$th pixel, $T_k$, is determined by the complexity of the chosen rendering technique (ray casting~\cite{795213}, ray tracing~\cite{10.1145/1198555.1198754}, global illumination~\cite{10.1145/2448196.2448205} etc.) and the interpolation method (linear or high-order~\cite{sun2024mfa}). The time used to render the PR-Image, $T_{SR}$, is the main overhead of the overall rendering latency. Because of the high computational cost of pixel rendering, decreasing the size of $S_{sp}$ can significantly reduce the rendering latency. However, the size of $S_{sp}$ is predefined by the sampling pattern, which is a fixed hyperparameter for existing RecNNs. In order to further decrease the computational cost, a new set of important pixels $S_{im}$ needs to be selected from $S_{sp}$, and the rendering of $S_{im}$ can losslessly reconstruct the rendering of $S_{sp}$. We refer to the binary mask that filters out $S_{im}$ from $S_{sp}$ as the Importance Mask (IM) for the remainder of the paper. For a chosen rendering function $R()$ and reconstruction function $f()$. The optimal $S_{im}$ can be derived from:

\begin{equation}
S_{im} = \arg \min_{\left|R(S_{sp})-f(R(S_{im}))\right|<\epsilon}\left| S_{im}\right|
\end{equation}
where $\epsilon$ is a small predefined interval of error metric between the rendering of $S_{sp}$ and the reconstruction from the rendering of $S_{im}$. $\left| S_{im}\right|$ is the number of important pixels. Our objective is to select an optimal subset from the sampling pattern with the fewest possible pixels while ensuring that the rendering of this subset can accurately reconstruct the full sampling pattern with minimal reconstruction error. We name the percentage of pixels when using optimal $S_{im}$ as the Optimal Rendering Percentage (ORP).

The volumetric datasets, mostly collected from scientific domains, show strong spatial correlations in their 3D domains. Reflected on the 2D image domain, such spatial correlation can also be observed. This provides opportunities to compress the complex pixel-level rendering into compact representations. Although there are numerous existing works to extract superpixels directly from 2D datasets~\cite{10.1145/3652509}, $S_{im}$ needs to be selected by not only considering the data itself but also jointly considering the users' input and downstream RecNN. Inspired by the saliency map~\cite{6248093} demonstrating which parts of the input data (e.g., pixels in an image) contribute most to the predictions of a neural network, we provide a method to automatically learn a selection of pixels that are the most important to reconstructing the rendering of the sampling pattern. Instead of calculating the gradient of output with respect to the input, as the saliency map does, we utilize an autoencoder architecture to learn the IM as a latent feature through backpropagation from the training dataset. In this work, we propose two neural networks: the Importance Mask Learning Network (IML Net) and the Importance Mask Synthesis Network (IMS Net). IML Net learns the importance mask from training datasets, which are visualization images generated from the respective user's view-dependent operations. Once the importance masks are learned, they will then be used as a training dataset for the IMS Net, which will learn the mapping from the user's view to the IM through novel view synthesis. 

\begin{figure}[t]
    \centering 
    \includegraphics[trim=0 0 0 0,clip,width=\linewidth]{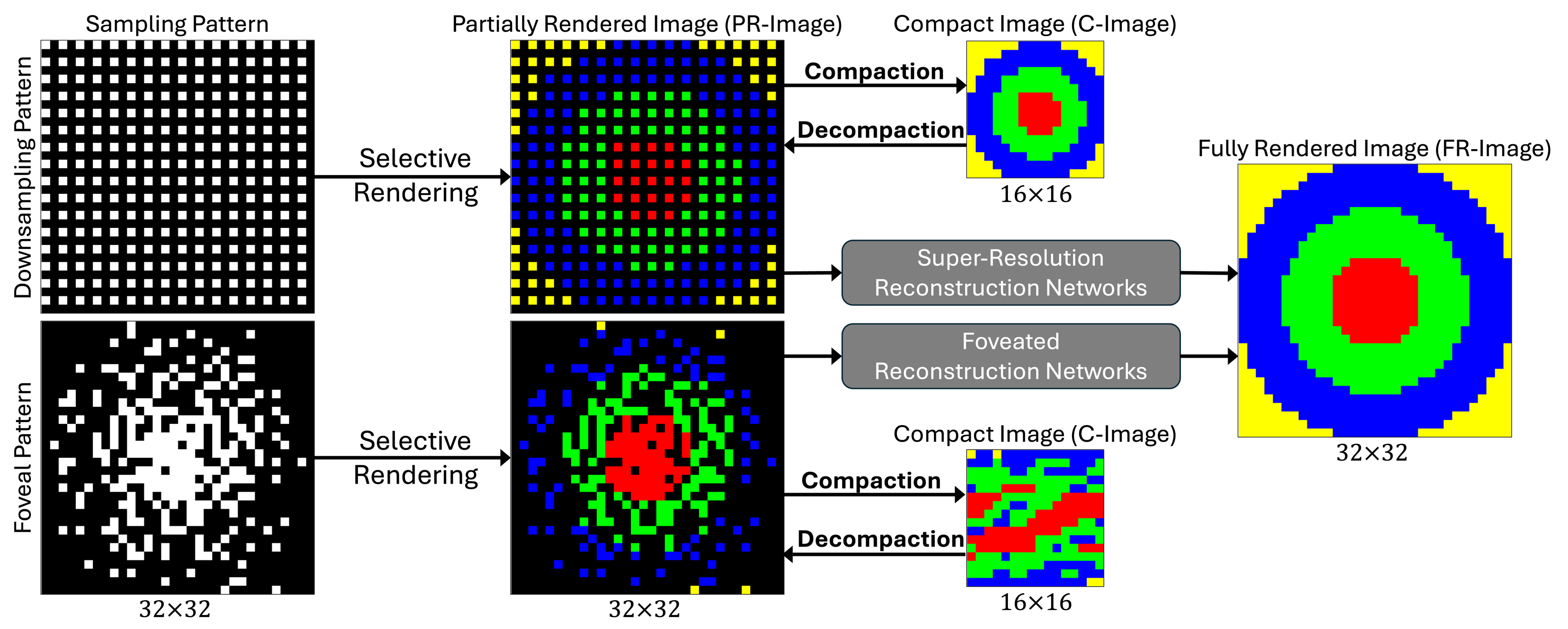}
    \caption{Demonstration of compaction and decompaction for super-resolution RecNN and foveated rendering RecNN.}
    \label{fig:compaction}
\end{figure}

\begin{figure*}[t]
    \centering 
    \includegraphics[trim=0 0 0 0,clip,width=\textwidth]{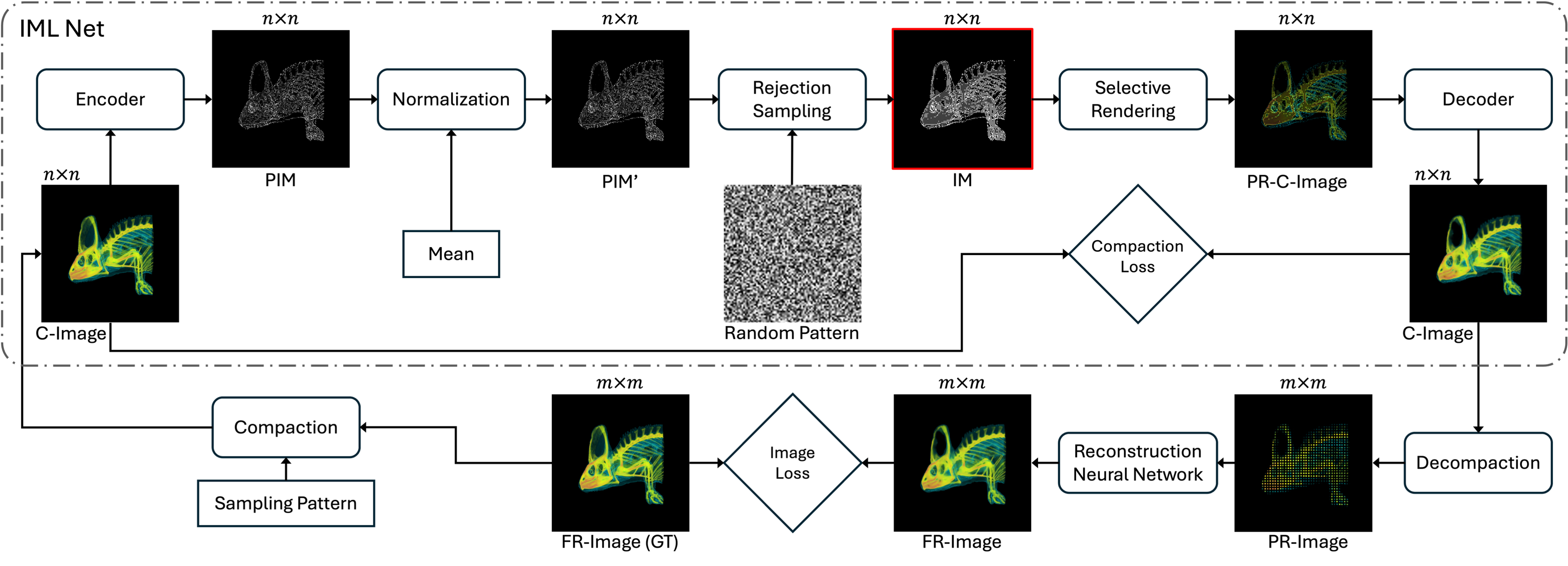}
    \caption{Network architecture of IML Net. IML Net works on a compact 2D embedding domain (C-Image) with resolution of $n\times n$, which is much smaller than the full resolution output with a resolution of $m\times m$ ($m \gg n$).}
    \label{fig:iml-net}
\end{figure*}

\subsection{Differentiable Compaction and Decompaction}
The goal of our work is to derive an optimal $S_{im}$ from $S_{sp}$, however, the shape of the sampling pattern varies for different RecNN. For example, super-resolution RecNN uses a downsampling pattern while the foveated rendering RecNN uses a foveal pattern. The downsampling pattern is evenly distributed across the PR-Image. The foveal pattern generally refers to the region of an image where the eye's fovea (central vision) is focused, which is where human vision is most detailed. In foveated rendering, only the pixels in this high-resolution foveal area are rendered in full detail, while the surrounding areas corresponding to peripheral vision are rendered at a much lower resolution. To uniformly process sampling patterns with different distributions, we introduce a compaction layer to compact the pixels in the sampling pattern into a regular 2D matrix. \cref{fig:compaction} shows how the PR-Image is selectively rendered according to downsampling and foveal patterns, and how the PR-Image is compacted into a regular 2D matrix. We name this 2d matrix a compact image (C-Image). The compaction collects each pixel within the sampling pattern by following a straightforward Raster scan order on the PR-Image. If the C-Image is not filled after compaction, zeros are padded to its empty pixels. We can see that the C-Image of super-resolution RecNN is the corresponding low-res version of the FR-Image, while the C-Image of foveated rendering RecNN is more irregular. The compaction layer is a practice of dataset distillation where only informative pixels are gathered for efficient training. Decompaction is the reverse of compaction. The compaction and decompaction layers are 2D point transformations between PR-Image space and C-Image space. To make the compaction and decompaction layers differentiable, a transformation is implemented as a translation matrix multiplying each pixel in one space to map its value to the transformed location of another space. During compaction, the C-Image is calculated through the linear transformation:
\begin{equation}
C{\text -}Image = PR{\text -}Image\times C_{RecNN}
\end{equation}
Where $C_{RecNN}$ is the Compaction Matrix, which is a 2D constant matrix defined by the specific RecNN. During decompaction, the PR-Image can be retrieved:
\begin{equation}
PR{\text -}Image = C{\text -}Image\times D_{RecNN},\quad D_{RecNN} = C_{RecNN}^{-1}
\end{equation}
Where the Decompaction Matrix $D_{RecNN}$ is the inverse of the $C_{RecNN}$. The benefit of incorporating compaction and decompaction layers includes: 1) Reduce the dimension of the redundant training data into a compact subset for efficient Meta Learning. 2) Decrease the number of parameters of the IML network. 3) Improve the speed of convergence during training. 4) Improve the speed of inferencing for interactive visualization. 5) Allow the proposed IML and IMS Networks to remain agnostic to diverse sampling patterns within a unified network architecture. Compared to the LAS, the proposed compaction/decompaction enables learning from a low-res intermediate embedding (C-Image), therefore, further reducing the expensive pixel calculation for faster rendering.

\begin{figure}[t]
    \centering 
    \includegraphics[trim=0 0 0 0,clip,width=\linewidth]{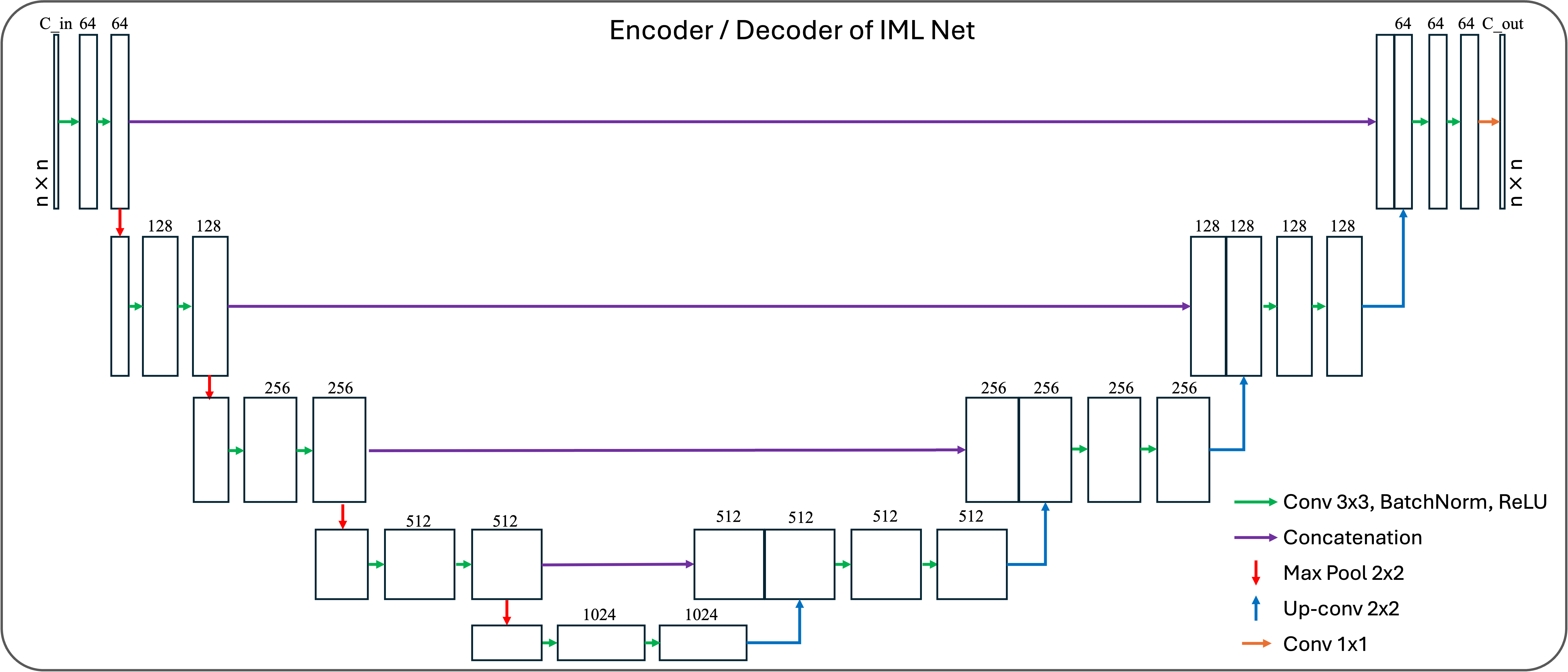}
    \caption{Network architecture of the encoder and decoder in IML Net. The input image resolution is $n\times n$.}
    \label{fig:endecoder}
\end{figure}

\subsection{Importance Mask Learning Network}
The functions of IML Net include: 1) For each rendered sampling pattern, learn an IM that selects only the important pixels from it to render for the downstream reconstructions. 2) Learn a decoder that reconstructs the full rendering of the C-Image from the partially rendered C-Image (PR-C-Image) based on the learned IM.

\subsubsection{Network Architecture}
\cref{fig:iml-net} shows the network architecture of IML Net. IML Net is an autoencoder neural network that takes a C-Image after compaction as input and outputs an image as close as the C-Image. The number of pixels and their distribution on C-Image are defined by the specific sampling pattern selected. The IM is learned on top of it by optimizing the network parameters through backpropagation during training. Key components of the network are:

\textbf{Encoder and Decoder:}
The first and the last subnetworks of the IML Net are encoder and decoder sharing the same network structure as shown in \cref{fig:endecoder}. We utilize a standard U-Net~\cite{10.1007/978-3-319-24574-4_28} to construct the encoder/decoder network. Our IML Net is invariant to the input C-image resolution due to the use of convolutional layers. For the encoder, C\_in is set to 3 for taking the RGB C-Image after compaction as input, C\_out is set to 1 for outputting a grayscale image where the value of each pixel correlates to the probability of being an important pixel. The U-Net structure of the encoder can effectively learn such pixel values to find the IM. After the last layer of the encoder, a softplus layer is used to convert the output value into the positive range for the following normalization layer. For the decoder, both C\_in and C\_out are set to 3 for taking the partially rendered C-Image (PR-C-Image) as input and outputting an image close to the input C-Image of the encoder. The decoder of our IML network performs a low-level in-painting to complete the missing pixels in the PR-C-Image to reconstruct the complete C-Image.

\textbf{Normalization:} The output of the encoder presents preliminary information about the importance of each pixel, and we name this output Preliminary Importance Mask (PIM). We apply a similar practice proposed by Weiss et al.~\cite{9264699} to normalize the PIM with a hyperparameter, called mean value, which will control the percentage of C-Image pixels selected as important pixels.
The normalized PIM, can be computed:
\begin{equation}
PIM'_{ij} = l+PIM_{ij}\frac{u - l}{u_{PIM}+\delta}
\end{equation}
where the $i$ and $j$ are pixel indices, $u$ is the prescribed mean value, and $l$ is the lower bound on the sample distribution, which is set to 0 in this work. $u_{PIM}$ is the mean of PIM. $\delta$ is a small constant set as $10^{-7}$ to avoid zero division. All the above operations ensure that the normalization layer remains differentiable. It is worth noticing that while the normalization step can also be performed using a softmax layer, this approach may result in a loss of control over the proportion of C-Image pixels designated as important through the prescribed mean. Our normalization step enables the hyperparameter tuning to investigate the relationship between the sparsity of important pixels and reconstruction performance.

\textbf{Rejection Sampling:}
The rejection sampling layer is to generate a binary IM from the $PIM'$ with continuous values. The practice is filtering out the important pixel through a comparison between the pixel value in $PIM'$ and a randomly sampled value. The rationale behind this practice is due to the spatial correlation found in the prediction from image generative neural networks. This fact makes the direct learning of the important pixel as a blob of connected pixels, which doesn't distinguish the underlying structure. Comparing the prediction with an uncorrelated random pattern can solve this issue and create isolated pixels that reflect the importance of the underlying structure. The random pattern $P$ is a grayscale image with the same dimensions as C-Image, with its pixel value randomly sampled from a uniform distribution between 0 and 1. We use plastic sampling to generate $P$ for the superior quality of the learned IM compared to other sampling strategies like random sampling, regular sampling, and Halton sampling. The IM can be retrieved through an independent Bernoulli process through rejection sampling.
\begin{equation}\label{eq:rs}
    IM_{ij}= 
\begin{cases}
    1,& \text{if } PIM'_{ij} > P_{ij}\\
    0,              & \text{otherwise}
\end{cases}
\end{equation}
To make the rejection sampling layer differentiable, the IM can be calculated as:
\begin{equation}
IM_{ij} = Sigmoid(\alpha(PIM'_{ij} - P_{ij})),\quad Sigmoid(x)=\frac{1}{1+e^{-x}}
\end{equation}
where $\alpha$ is a positive scalar determining the steepness of the Sigmoid function. The pixel value of each IM is in the range $[0, 1]$. The above equation is only used for training the IML Net. When using IML Net during the inferencing, which will be discussed in \cref{section:inferencing}, only the regular rejection sampling of \cref{eq:rs} is used. In that case, a binary mask using 1 bit per pixel is enough to represent the importance mask.

\textbf{Selective Rendering:}
The selective rendering layer will generate the partially rendered C-Image (PR-C-Image) according to the IM. During training, this step is basically element-wise multiplication between the fully rendered C-Image (FR-C-Image) of the training dataset and IM:
\begin{equation}
PR{\text -}C{\text -}Image = IM\times FR{\text -}C{\text -}Image
\end{equation}
During the inferencing stage, this layer is replaced by actual pixel-level rendering computation defined by the chosen volume rendering algorithm for learned important pixels.

\begin{figure}[t]
    \centering 
    \includegraphics[trim=0 0 0 0,clip,width=\linewidth]{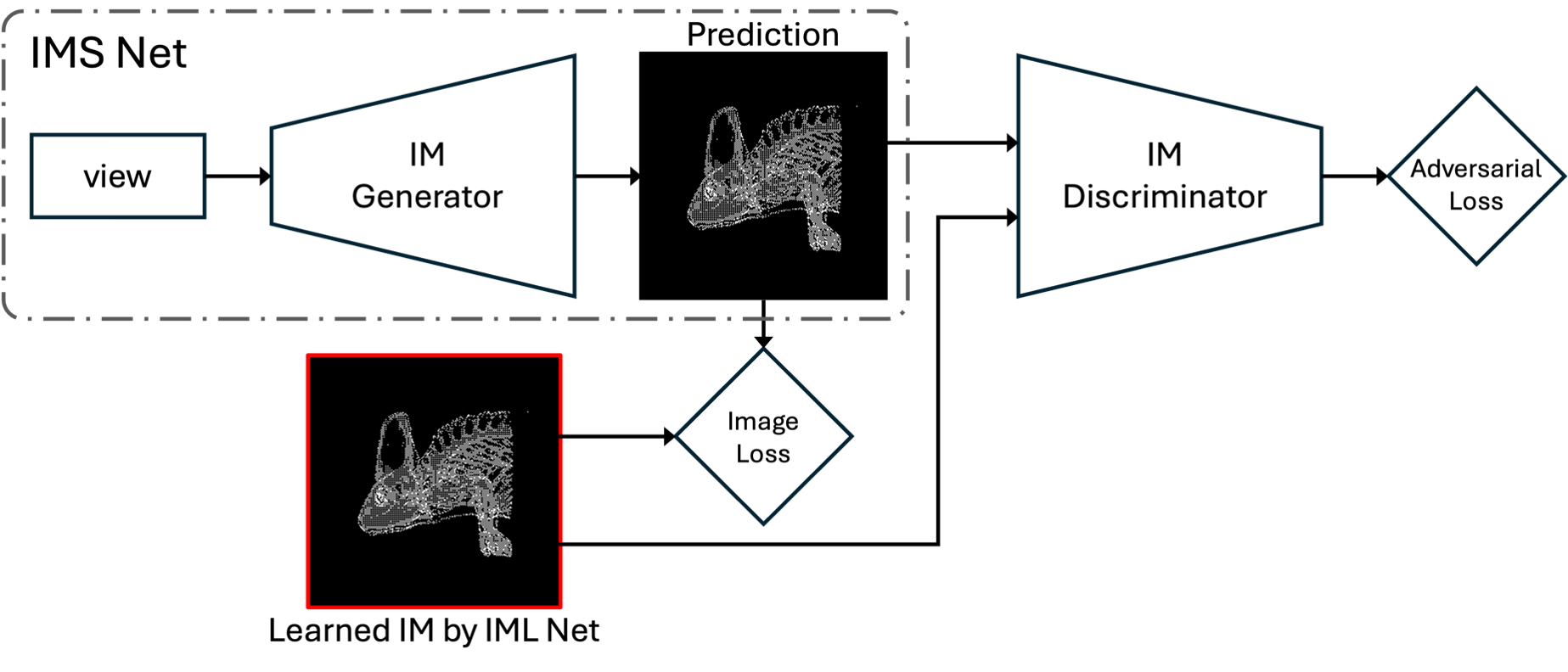}
    \caption{Network architecture of IMS Net.}
    \label{fig:ims-net}
\end{figure}

\subsubsection{Loss Functions}
The IML Net can be trained by itself with a compaction loss function, as shown inside the IML Net block of \cref{fig:iml-net}, which is an image loss function between the input C-Image and reconstructed C-Image. This practice is suitable for optimizing the off-the-shelf pre-trained RecNNs without going through the long end-to-end training process resulting from adding a complex downstream RecNN. The IML Net can also be trained together with downstream RecNNs selected for various sampling patterns. In this case, the loss function is the weighted sum of the previous compaction loss and the image loss between the predicted reconstructed full-resolution image (FR-Image) and the ground truth (GT) image rendered using the baseline rendering algorithm without using RecNN. This end-to-end training will give better reconstruction results by jointly optimizing the parameters from both the IML Net and the downstream RecNN. In our work, we use the Mean Squared Error (MSE) as the image loss function and 0.5 as the weight for both image losses if trained end-to-end.

\subsection{Importance Mask Synthesis Network}
Once the IML Network is learned, we collect the view that generates the C-Image and the learned IM as input-label pairs to train the IMS Net through supervised learning. The function of IMS Net is to quickly and directly predict the IM from a new set of view parameters.

\subsubsection{Network Architecture}
\begin{figure}[t]
    \centering
    \begin{subfigure}[b]{\linewidth}
        \centering
        \includegraphics[trim=0 0 0 0,clip,width=\linewidth]{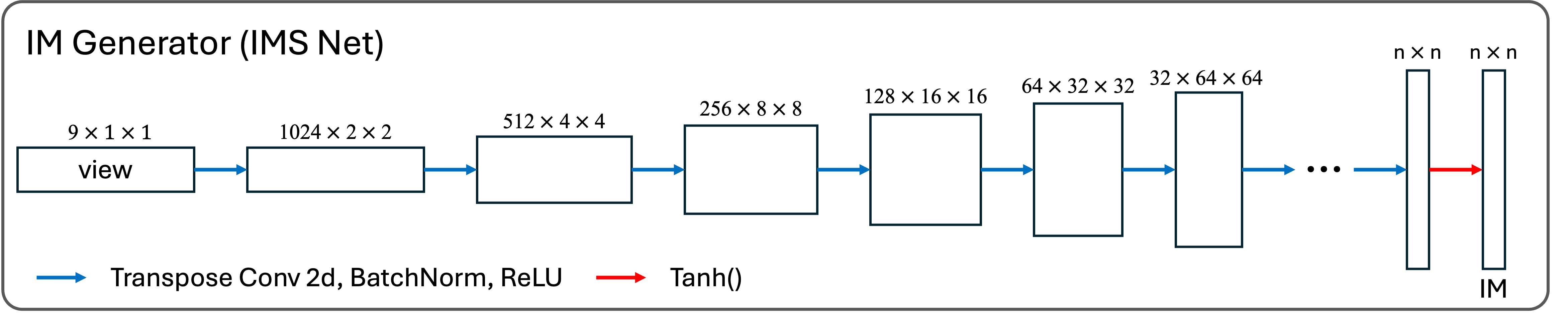}
        \caption{IM generator}
        \label{fig:im_generator}
    \end{subfigure}
    \vskip\baselineskip
    \begin{subfigure}[b]{\linewidth}
        \centering 
        \includegraphics[trim=0 0 0 0,clip,width=\linewidth]{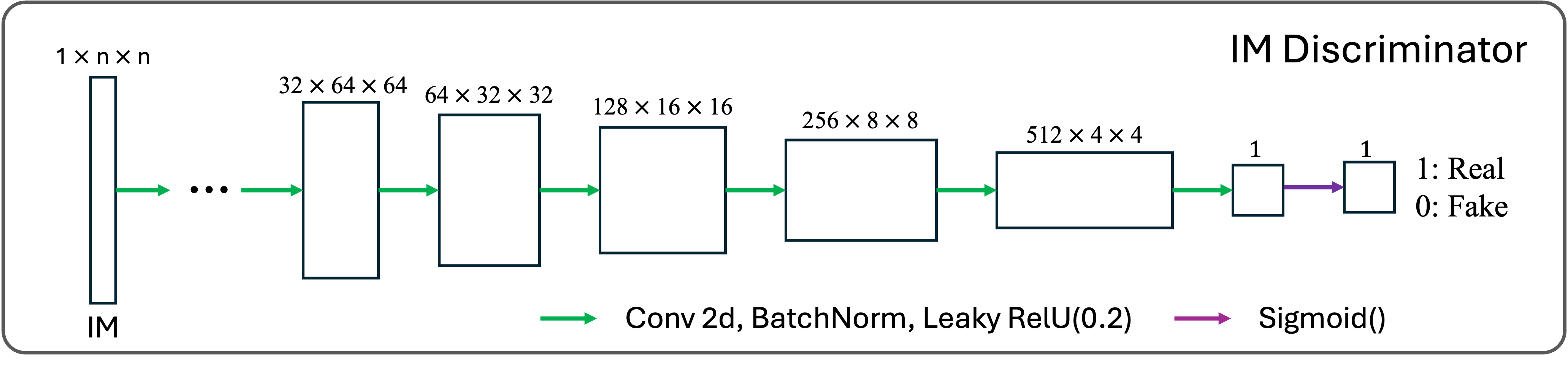}
        \caption{IM discriminator}
        \label{fig:im_discriminator}
    \end{subfigure}
    \caption{Network architecture of the IM generator of IMS Net (a) and the IM discriminator for training the GAN (b).}
    \label{fig:gendis}
\end{figure}

The focus of this paper is on visualizing scientific data, which is a time-sensitive application. Therefore, we need to keep the main steps of the pipeline, including training and inferencing, finished as fast as possible. Our training includes two steps: training the IML Net and training the IMS Net. We provide a method to speed up the training of IML network, as mentioned in \cref{feasibility}, by training a standalone IML Net without connecting to the downstream RecNN. However, we cannot carry out the same optimization on training the IMS Net. As a result, we have to select a GAN with less training/inferencing complexity. DCGAN~\cite{radford2015unsupervised} provides a better trade-off between the complexity and the reconstruction quality than other GAN architectures like Conditional GAN (cGAN)~\cite{Isola_2017_CVPR} and StyleGAN~\cite{8977347} whose training/inferencing times are too long for scientific visualization. \cref{fig:ims-net} shows the network architecture of IMS Net and the discriminator used for training. IMS Net is a generator or regressor that takes a sparse set of view parameters to generate a 2D binary image as the IM. The IMS Net is trained through the generative adversarial network (GAN). Key components of the network are:

\textbf{Generator:}
The input to the generator is a set of view parameters. We adopted the OpenGL convention to define the view parameters as 9 float numbers:
\begin{equation}
view = (eye\_x/y/z, \; lookAt\_x/y/z, \; up\_x/y/z)
\end{equation}
where the $eye$ is the camera position, $lookAt$ is the focal point where the camera is looking, and $up$ is the "up" direction of the camera. IMS Net is a DCGAN with its detailed structure shown in \cref{fig:im_generator}. It consists of several transposed convolutional layers with batch normalization and ReLU activation functions.

\textbf{Discriminator:}
\cref{fig:im_discriminator} shows the detailed structure of the binary classifier as a discriminator. The discriminator consisted of several convolutional layers with batch normalization and Leaky ReLU activation functions.


\subsubsection{Loss Functions}
As shown in \cref{fig:ims-net}, we consider two losses to train the IMS Net: image loss on the IM and adversarial loss. The adversarial loss is composed of two parts: the generator loss and the discriminator loss. The discriminator loss $L_D$ is the cross-entropy of the binary classifier measuring how well the discriminator can distinguish between real and generated images.
 
\begin{equation}
L_D = -\mathbb{E}_{x \sim p_x(x)}[\log D(x)] - \mathbb{E}_{z \sim p_z(z)}[\log (1 - D(G(z)))]
\end{equation}
where  $D(\mathbf{x})$  is the discriminator’s output for a real image. $D(G(\mathbf{z}))$  is the discriminator’s output for a fake image  $G(\mathbf{z})$. The generator loss $L_G$ used is the non-saturating loss measuring how well the generator can fool the discriminator.

\begin{equation}
L_G = -\mathbb{E}_{z \sim p_z(z)}[\log D(G(z))]
\end{equation}
$p_{x}(x)$ is the probability distribution of real training data and $p_{z}(z)$ is the probability distribution of the noise vector $z$. The loss function selection and configuration are crucial in training a GAN. Detailed results using different loss functions for training IMS Net are discussed in \cref{sec:loss}.

\begin{figure}[t]
    \centering 
    \includegraphics[trim=0 0 0 0,clip,width=\linewidth]{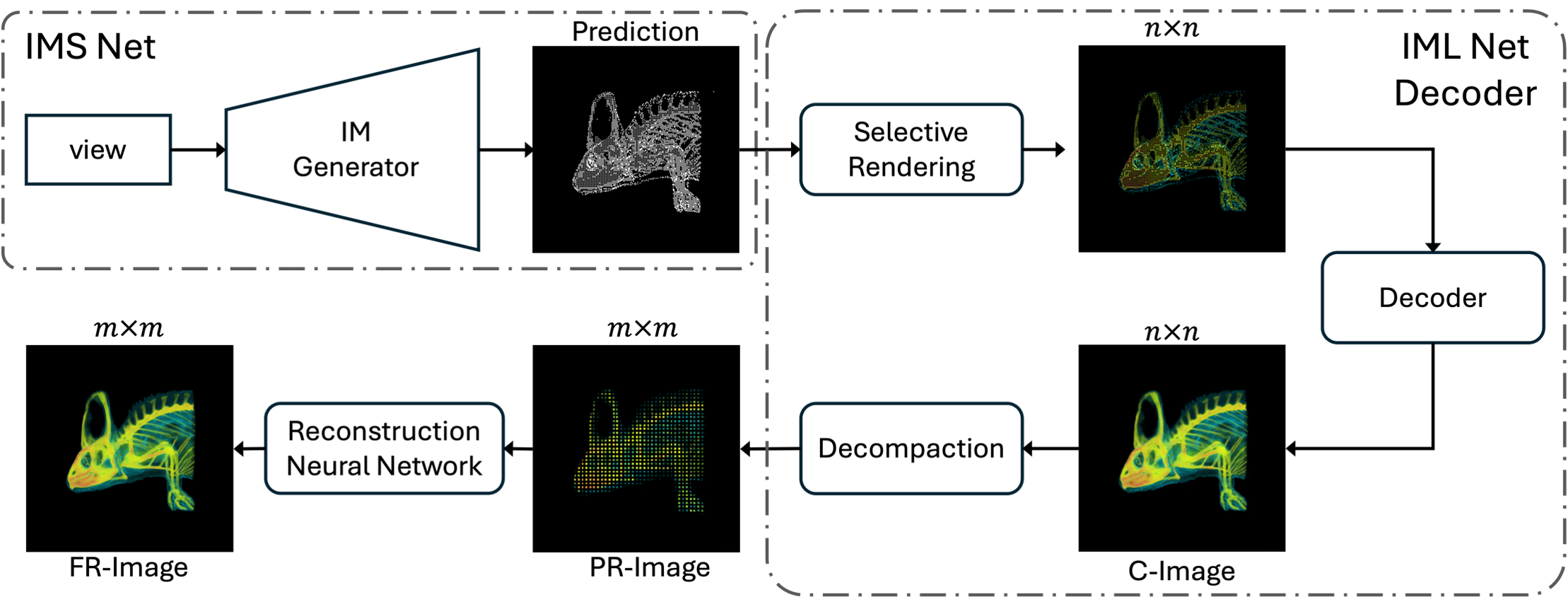}
    \caption{Network architecture of the proposed visualization neural rendering network, IML/S Net + RecNN. IML/S is the trained IMS Net connected with the decoder of the trained IML Net.}
    \label{fig:imls_net}
\end{figure}

\begin{table}[t]
  \caption{Volumetric datasets used in the experiments.}
  \label{tab:datasets}
  \scriptsize%
  \centering%
  \begin{adjustbox}{width=0.4\textwidth}
      \renewcommand{\arraystretch}{0.2} 
      \setlength{\extrarowheight}{-2pt} 
      \begin{tabu}{ c c c c }
          \toprule
          Dataset & Resolution & Size & Data Type \\
          \midrule
          Chameleon       & $1024^2\times1088$ & 2.1 GB & uint16 \\
          \midrule
          Beechnut        & $1024^2\times1546$ & 3.0 GB & uint16 \\
          \midrule
          Rayleigh-Taylor & $1024^3$ & 4.0 GB & float32 \\
          \midrule
          Flame           & $1408\times1080\times1100$ & 6.23 GB & float32 \\
          \bottomrule
      \end{tabu}
  \end{adjustbox}
\end{table}

\begin{figure}[t]
    \centering
    \begin{subfigure}[b]{0.32\linewidth}
        \centering
        \includegraphics[trim=10 0 10 0,clip,width=\linewidth]{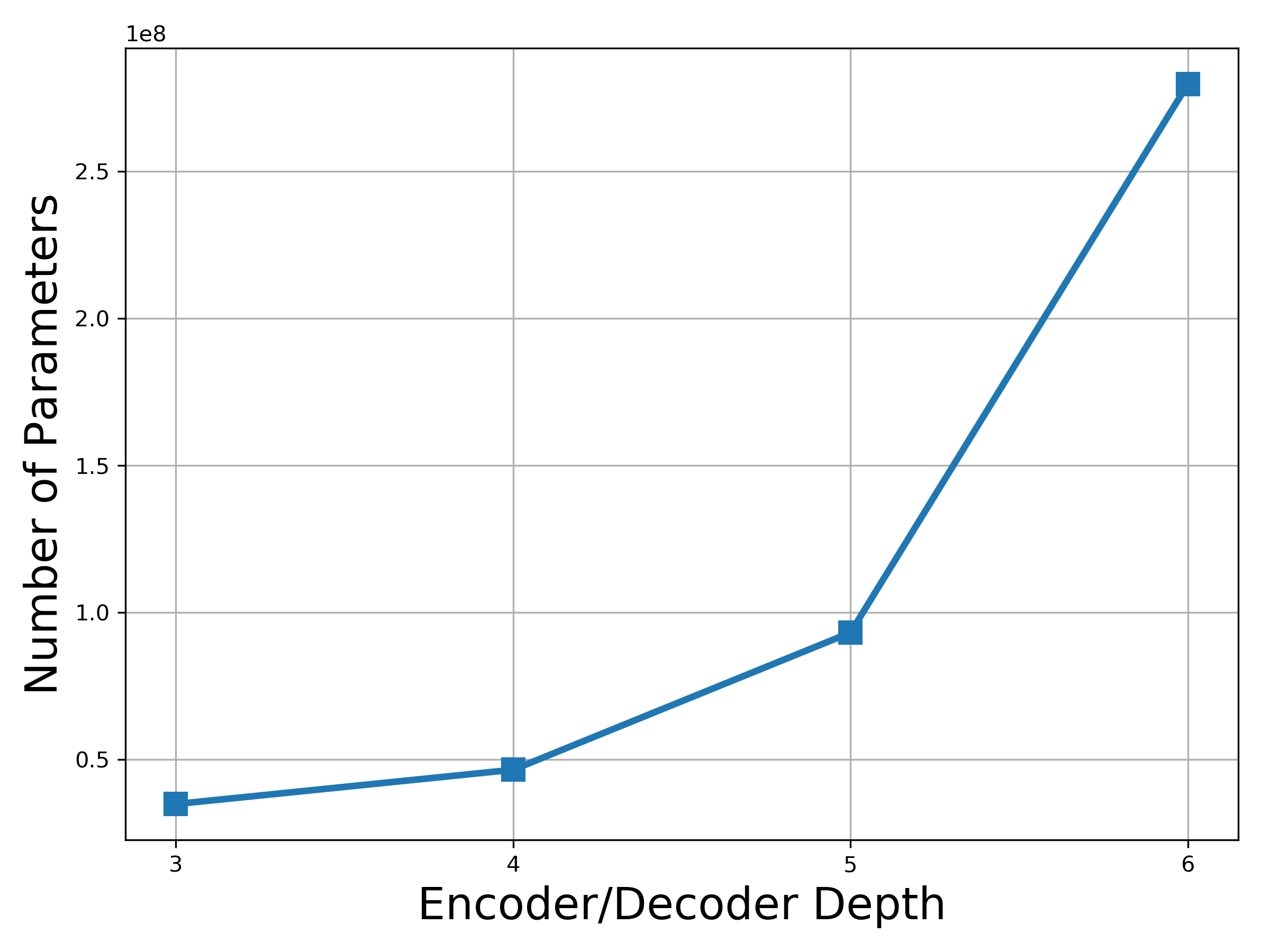}
        \caption{IML network parameter size}
        \label{fig:depth_iml_num_para}
    \end{subfigure}
    \hfill
    \begin{subfigure}[b]{0.32\linewidth}
        \centering 
        \includegraphics[trim=10 0 10 0,clip,width=\linewidth]{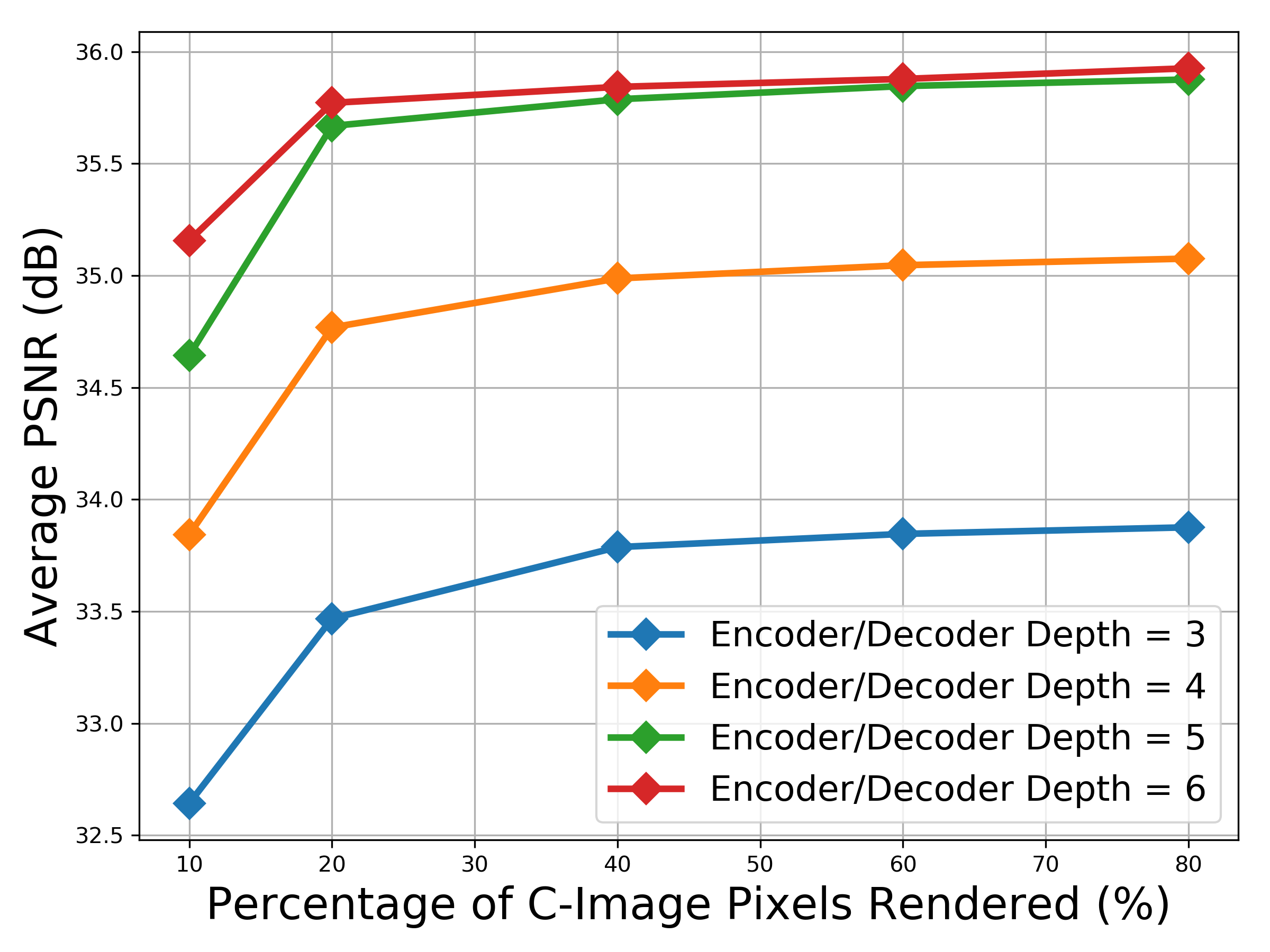}
        \caption{Average PSNR (S-R))}
        \label{fig:depth_iml_psnr_sr}
    \end{subfigure}
    \hfill
    \begin{subfigure}[b]{0.32\linewidth}
        \centering 
        \includegraphics[trim=10 0 10 0,clip,width=\linewidth]{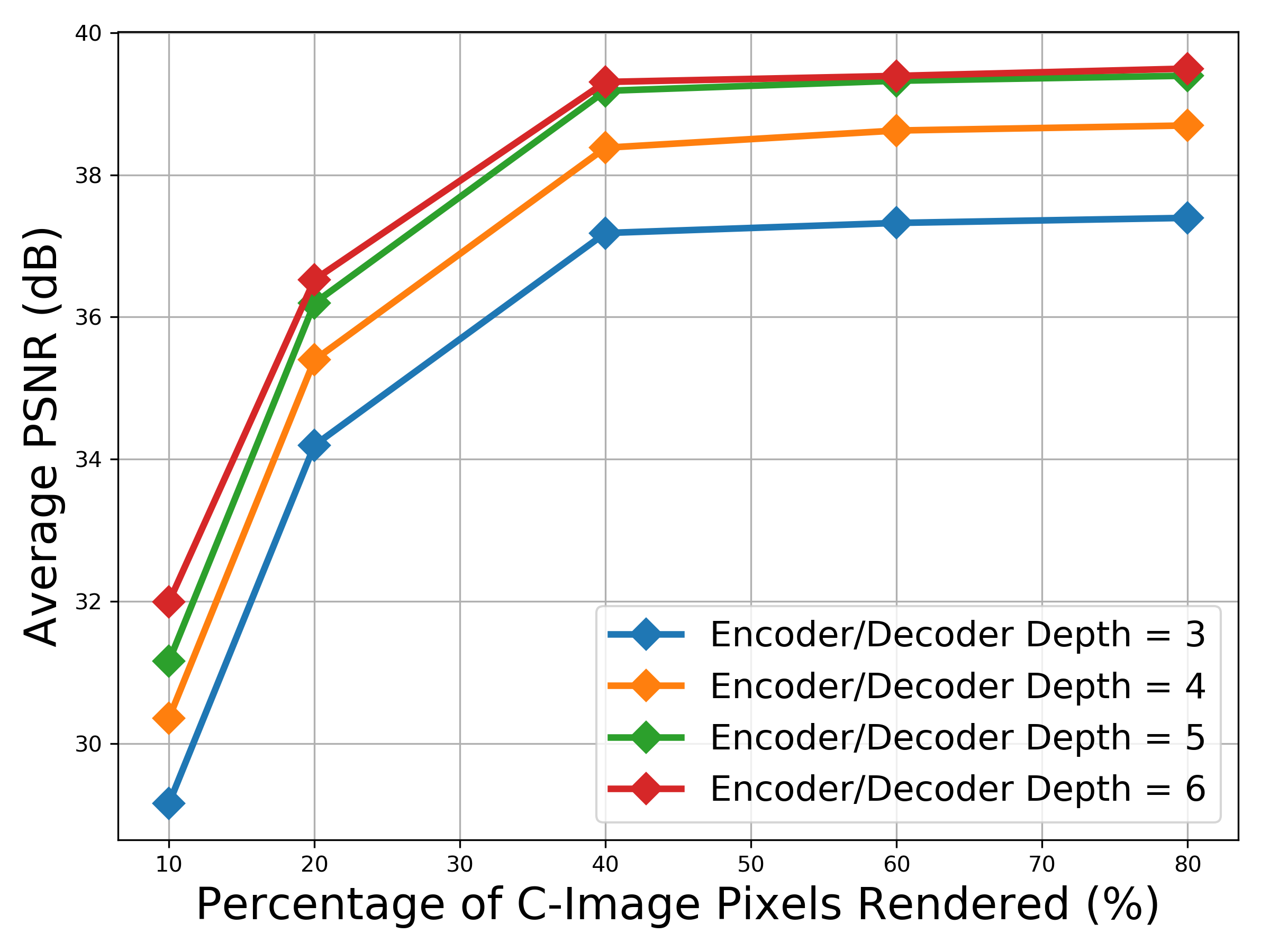}
        \caption{Average PSNR (F-R))}
        \label{fig:depth_iml_psnr_fr}
    \end{subfigure}
    \caption{IML Networks with different depths of the U-Net architecture. Evaluation performed on the Chameleon dataset with a visualization image resolution of $512\times512$. (a) shows the total number of parameters of IML Net. (b) and (c) show the reconstruction quality of super-resolution (S-R) and foveated rendering (F-R) while increasing the percentage of C-Image rendered. }
    \label{fig:depth_iml}
\end{figure}

\begin{figure}[t]
    \centering
    \begin{subfigure}[b]{0.32\linewidth}
        \centering
        \includegraphics[trim=10 0 10 0,clip,width=\linewidth]{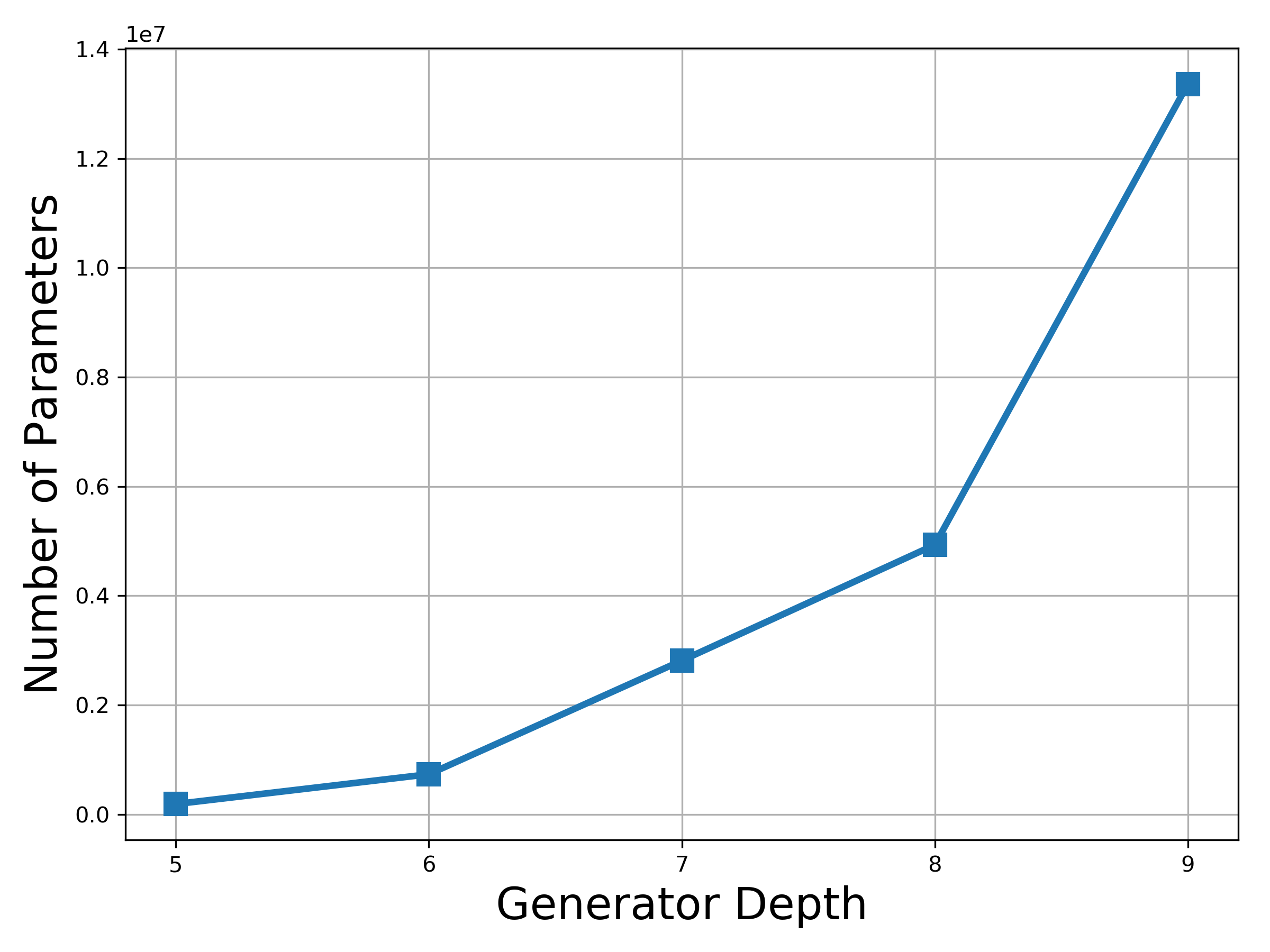}
        \caption{IMS network parameter size}
        \label{fig:depth_ims_num_para}
    \end{subfigure}
    \hfill
    \begin{subfigure}[b]{0.32\linewidth}
        \centering 
        \includegraphics[trim=10 0 10 0,clip,width=\linewidth]{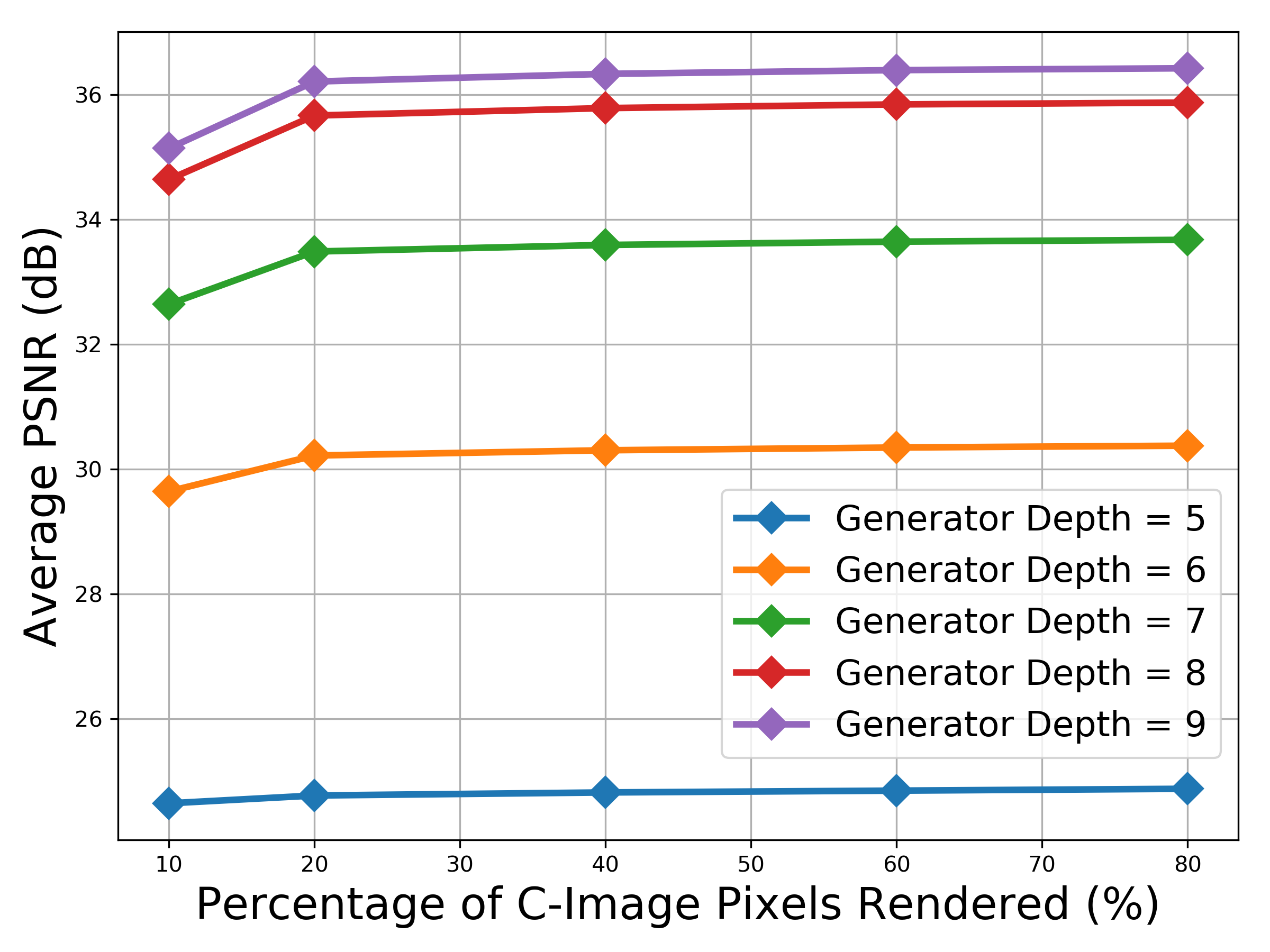}
        \caption{Average PSNR (S-R))}
        \label{fig:depth_ims_psnr_sr}
    \end{subfigure}
    \hfill
    \begin{subfigure}[b]{0.32\linewidth}
        \centering 
        \includegraphics[trim=10 0 10 0,clip,width=\linewidth]{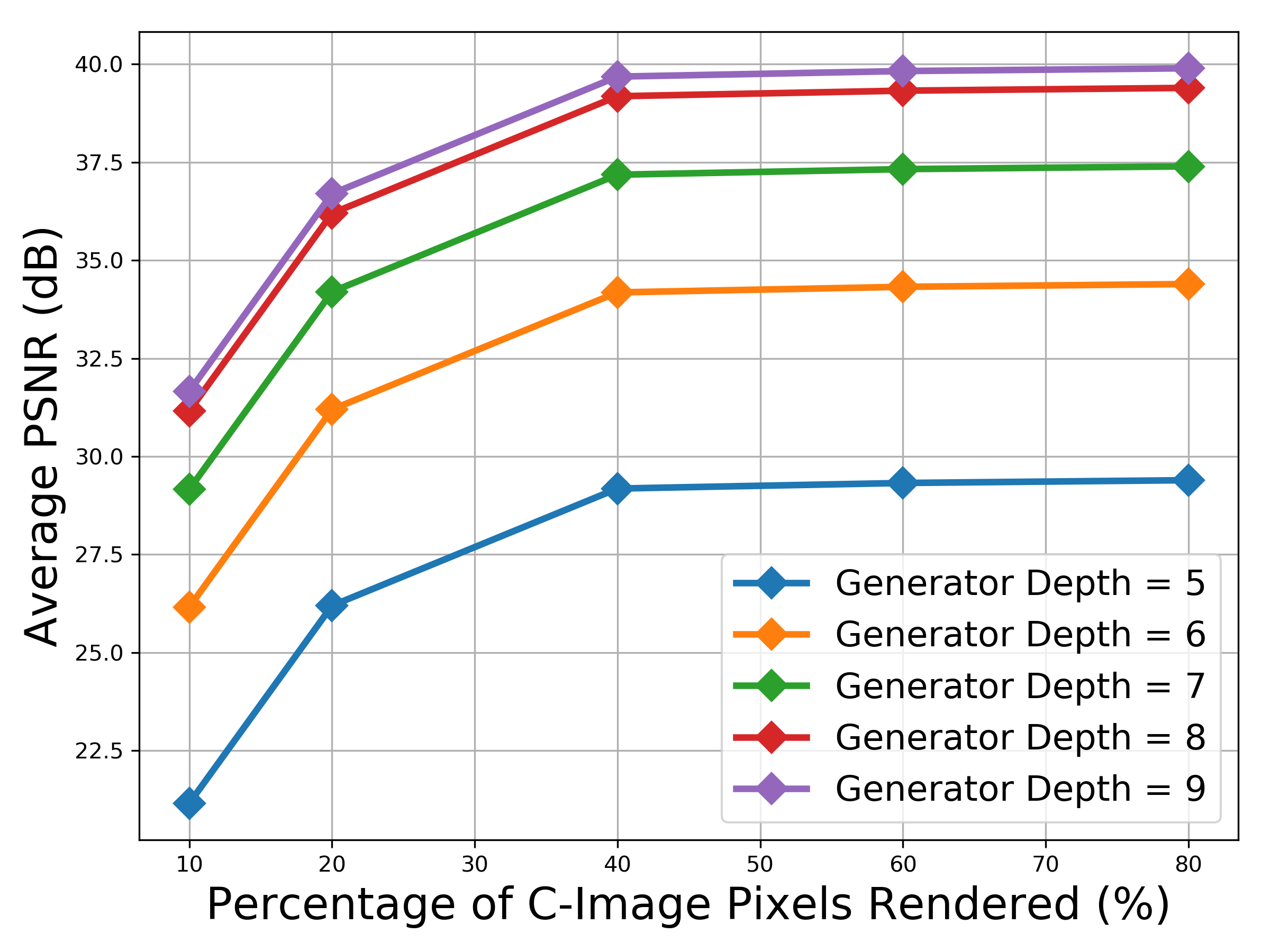}
        \caption{Average PSNR (F-R))}
        \label{fig:depth_ims_psnr_fr}
    \end{subfigure}
    \caption{IMS network with different depth used in its generator. Evaluation performed on the Chameleon dataset with a visualization image resolution of $512\times512$. (a) shows the total number of parameters of the generator. (b) and (c) show the reconstruction quality of super-resolution (S-R) and foveated rendering (F-R) while increasing the percentage of C-Image rendered.}
    \label{fig:depth_ims}
\end{figure}

\subsubsection{Training}
Training a generative model to directly generate volume visualization is challenging~\cite{8316963, 10.5555/3157096.3157346, Zhang_2017_ICCV, brock2018large, pmlr-v70-arjovsky17a} for several reasons: 1) It is difficult for GAN to generate high-res images. 2) Training of GAN tends to be unstable because of the mode collapse, vanishing gradients, imbalanced learning rate, and data overlap issues. 3) While generative models can synthesize convincing volume visualization, they may struggle with very fine details or subtle structures. To prevent the aforementioned issues during the training of our generator, we utilize the following practices:
\begin{itemize}[leftmargin=*]
  \item Avoid learning the IM directly from the sparse view parameters but from the C-image, which presents more structural information, which is the motivation of the proposed IML Net. Our experiments show that the IM directly learned from view parameters is a binary mask whose important pixels are evenly distributed across the region of the volumetric object. In that case, important pixels couldn't properly capture the locations of informative regions of the rendered image.
  \item Train the generator to predict low-res simple patterns. The IM (as highlighted in \cref{fig:iml-net}) that the IMS Net tries to learn is a low-res 2D binary image with only a single channel and finite outputs (-1 or 1).
  \item Utilize batch normalization for both the generator and discriminator to stabilize the training.
  \item Using activation function ReLu for the generator and LeakyRelu(0.2) for the discriminator.
  \item Scale the input binary IM from $\{0, 1\}$ to $\{-1, 1\}$.
  \item Use the hyperbolic tangent function (Tanh) at the end of the generator to constrain the output between -1 and 1.
  \item Use adversarial loss together with advanced perceptual loss~\cite{10.1007/978-3-319-46475-6_43} instead of simple Binary Cross Entropy (BCE) or Mean Squared Error (MSE) loss. Perceptual Loss is a loss function that evaluates the similarity between images based on high-level image features, contents, and patterns, rather than pixel-wise differences. It leverages pre-trained deep neural networks (e.g., VGG~\cite{simonyan2014very}) to compare images at a feature level, focusing on perceptual quality and structural similarity.
\end{itemize}

\subsubsection{Inferencing}\label{section:inferencing}
Once the IMS Net is trained, a visualization neural rendering network is constructed by combining three components: the trained IMS Net, the decoder of the trained IML Net, and the RecNN, as shown in \cref{fig:imls_net}. We name this network IML/S Net + RecNN. The new network directly generates a visualization image from the new view with improved rendering latency than the original RecNN. Detailed quality and latency evaluation of IML/S Net + RecNN is discussed \cref{sec:results}.

\section{Experiments and Evaluation}
\subsection{Datasets}
\subsubsection{Volumetric Datasets}
For quality and performance evaluation, we select 4 large-scale volume datasets with distinct spatial features collected from diverse domains as detailed in \autoref{tab:datasets}. The Chameleon dataset is a CT scan of a chameleon. The Beechnut dataset is a microCT scan of a dried beechnut. The Rayleigh-Taylor dataset is a time step of a density field in a simulation of the mixing transition in Rayleigh-Taylor instability. The Flame dataset is a simulated combustion 3D scalar field. The spacings of all datasets are normalized within the spatial range [-1, 1] with values normalized within the range [0, 1].
\subsubsection{Training Data}
The training dataset used to train the networks is the view and fully rendered image (FR-Image) pairs. First, we randomly select 2000 views around the volume of each volumetric dataset. Second, an FR-Image is rendered using a specific volume rendering algorithm with lighting effect. In this work, we select the commonly used ray casting DVR as the renderer. Third, filter the FR-Image with the sampling pattern to generate the partially rendered image (PR-Image), where only pixels within the sampling pattern are kept. Fourth, apply compaction on PR-Image to generate C-Image. For training the IML Net standalone, the input is C-Image while the output is the predicted C-Image from the decoder. For training the IML Net end-to-end with the RecNN, the input is the C-Image while the output is the predicted FR-image from RecNN. For training the IMS Net, the input is the view parameters while the output is the learned IM derived from the trained IML Net. The training and validation partition ratio is 9:1. Each volumetric dataset also generates a testing dataset consisting of 120 views, forming an exploratory trajectory to simulate real user exploration. The FR-Image is rendered with a resolution of $512\times512$ using the Visualization Toolkit (VTK)~\cite{865875} with a customized timestamp to measure the rendering time for each pixel as detailed in the Appendix. The interpolation is set to trilinear. The sample distance is 0.02. The CPU thread is set as a single thread to faithfully measure the time duration when rendering individual pixels in sequence.

\begin{figure}[t]
    \centering 
    \includegraphics[trim=0 0 0 0,clip,width=\linewidth]{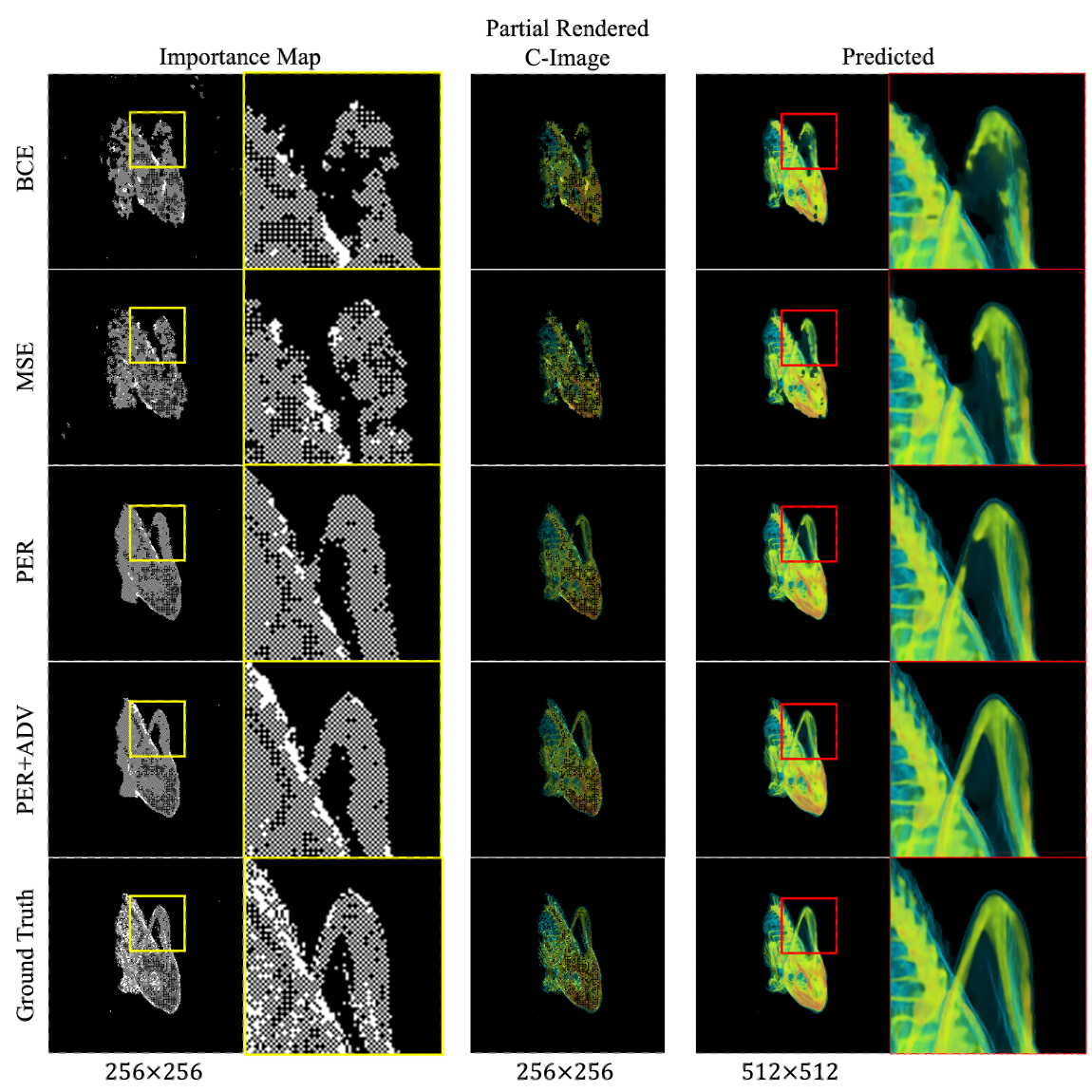}
    \caption{IM prediction quality comparison using different configurations of loss functions on the Chameleon dataset.}
    \label{fig:diff_loss_image}
\end{figure}

\subsection{Experimental Setup}
The experiments are designed to investigate the interactive volume visualization using RecNN. We are going to evaluate both the quality and performance using recent volume visualization leveraging RecNN and the proposed neural rendering pipeline, IML/S Net + RecNN. The input is a sequence of unseen views, and the output is the FR-image. We use PSNR as the main metric to evaluate rendering quality. The computing platform is a desktop featuring an Intel(R) Core(TM) i7-7700K CPU with 8 threads running at 4.20GHz, paired with 16GB of DDR4 DRAM clocked at 3200MHz, and operating on Ubuntu 20.04.4 LTS. The final visualization image has a resolution of $512\times512$ (262144 total pixels). The C-Image resolution is set as $256\times256$ (65535 total pixels).

\subsection{RecNN Selection and Configuration}
We select two state-of-the-art RecNNs used for interactive volume visualization, EnhanceNet and FoVolNet, to showcase how our method can help to further improve their rendering latencies. For the super-resolution RecNN using EnhanceNet, we also use $4\times$ super-resolution setting, the same as the experiment in its paper. Since our evaluation only focuses on one-shot image generation, for a fair comparison, we remove the recurrent connections of the FoVolNet and only keep its W-Net structure for static image prediction. Fixed foveated rendering, where the high-resolution region is static and fixed in the center of the screen, is used in the experiment rather than dynamic foveated rendering. The foveal pattern $FP$ is generated by the following methods:
\begin{equation}
    FP_{ij}= 
\begin{cases}
    1,& \text{if } P_{ij} > E_{ij}\\
    0,              & \text{otherwise}
\end{cases}
\end{equation}
where $P_{ij}$ is a random pattern generated by blue noise, which doesn't generate unevenly distributed samples from low-frequency energy spikes in the spatial domain\cite{9903564}. The $E_{ij}$ is a modulation surface to control the size of the foveated area through $\sigma$:
\begin{equation}
E_{ij} = e^{-0.5(f_{x}^{2}+f_{y}^{2})\sigma}
\end{equation}
where $f_x$ and $f_y$ are the center coordinates of the foveal pattern.

\begin{figure}[t]
    \centering
    \begin{subfigure}[b]{0.485\linewidth}
        \centering
        \includegraphics[trim=10 0 20 0,clip,width=\linewidth]{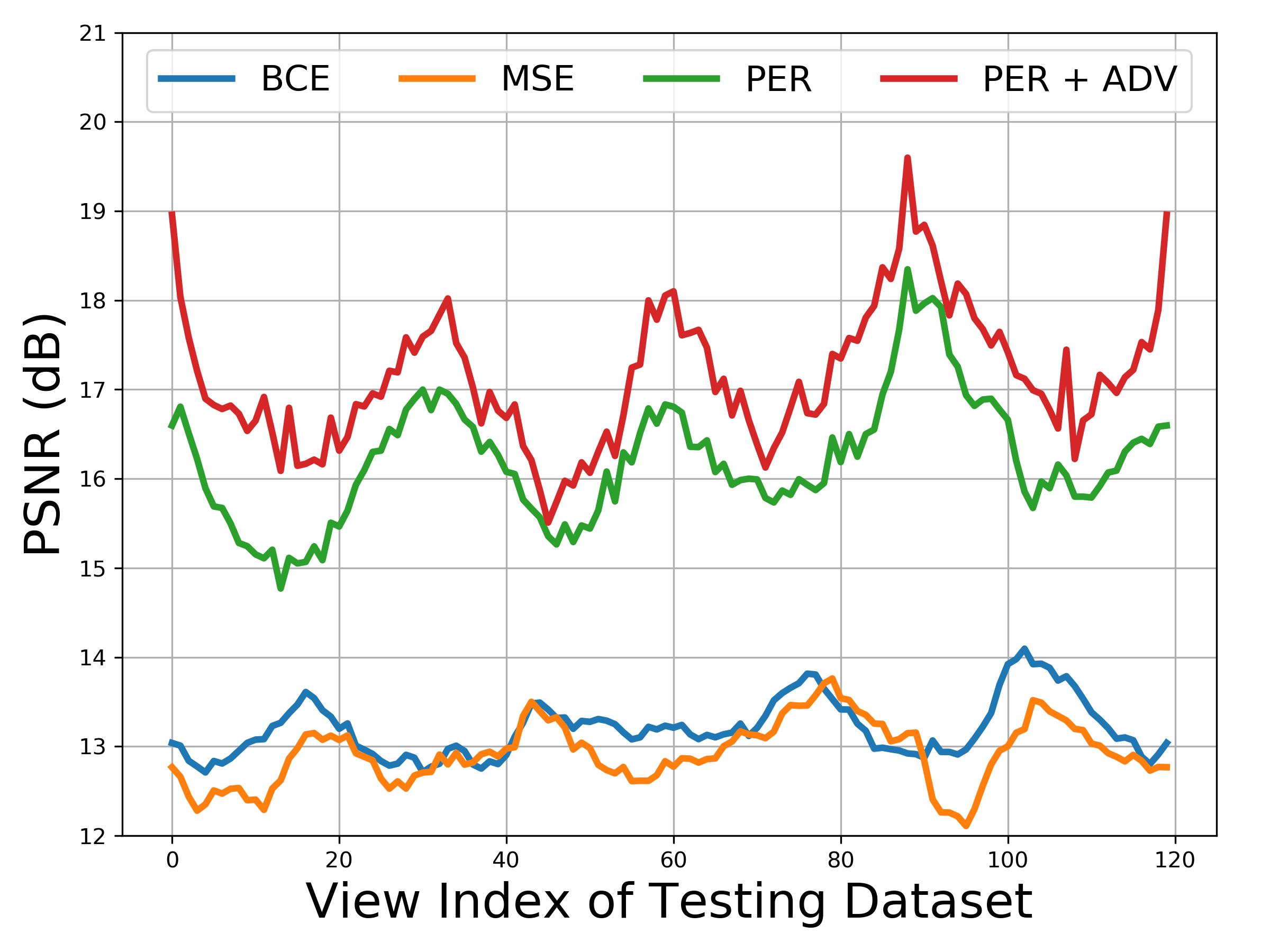}
        \caption{IMS Net}
        \label{fig:iml_loss}
    \end{subfigure}
    \hfill
    \begin{subfigure}[b]{0.485\linewidth}
        \centering 
        \includegraphics[trim=10 0 20 0,clip,width=\linewidth]{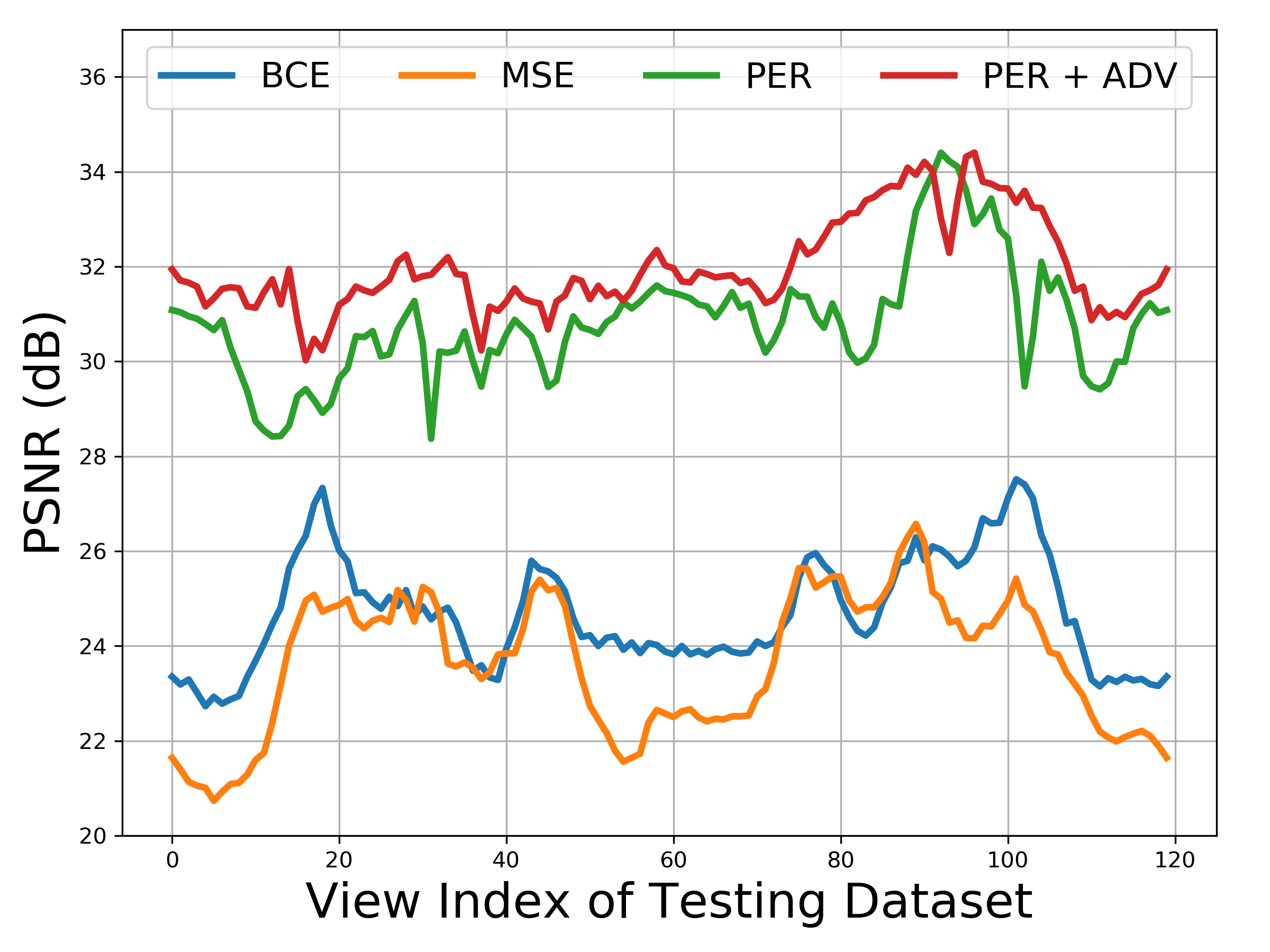}
        \caption{IML/S Net + EnhaneNet}
        \label{fig:ims_loss}
    \end{subfigure}
    \caption{Qualitative comparison of IM prediction from IMS Net and final reconstruction from IML/S Net + EnhanceNet using different configurations of loss functions on the Chameleon testing dataset.}
    \label{fig:diff_loss_curve}
\end{figure}

\subsection{Training}
The proposed neural networks are trained using the PyTorch software stack to accelerate the training and inferencing performance on a single NVIDIA RTX A6000 GPU. The Adaptive Moment Estimation (Adam) optimizer is used with an initial learning rate of 0.001 for training both IML and IMS Nets. A validation dataset is selected to detect overfitting on the training dataset and terminate the network parameters from updating through the early stop mechanism. The patience of the early stop is set as 20 epochs for IML Net and 1000 for IMS Net. The IML Net can be trained standalone or trained end-to-end with RecNN. 

\subsection{Ablation Study}
In this section, we investigate how the configuration of the networks and hyperparameters would influence the model's performance.

\subsubsection{Network Configuration}
\hspace{\parindent}\textbf{Encoder and Decoder of IML:} The encoder of the IML Net is critical to learning an accurate IM for it provides a comprehensive guess of the IM. Its U-Net architecture is capable of distinguishing fine-grained details through its skip connections and recommending candidates of important pixels. Compared with the encoder that serves as an information filter, the decoder serves as an information reconstruction that reconstructs the PR-C-Image to a full C-Image. We investigate how the network configuration of the IML encoder and decoder affects the performance of the proposed rendering pipeline. Since the decoder of the IML Net mirrors the encoder's configuration, we modify both together by adjusting the depth of their U-Nets. We fixed the architecture of IMS Net and only adjusted the IML Net. As the depth of the encoder and decoder increases, its reconstruction quality increases sublinearly (as shown in \cref{fig:depth_iml_psnr_sr} and \cref{fig:depth_iml_psnr_fr}) while the number of parameters of the network increases superlinearly (as shown in \cref{fig:depth_iml_num_para}). While a deeper IML Net can provide better reconstruction quality, its larger size slows down inference, leading to higher rendering latency. We select the depth of both the encoder and decoder as 5 for the balanced performance between rendering quality and inferencing latency.

\begin{figure*}[t]
    \centering
    \begin{subfigure}[b]{0.49\linewidth}
        \centering
        \includegraphics[trim=0 0 0 0,clip,width=\linewidth]{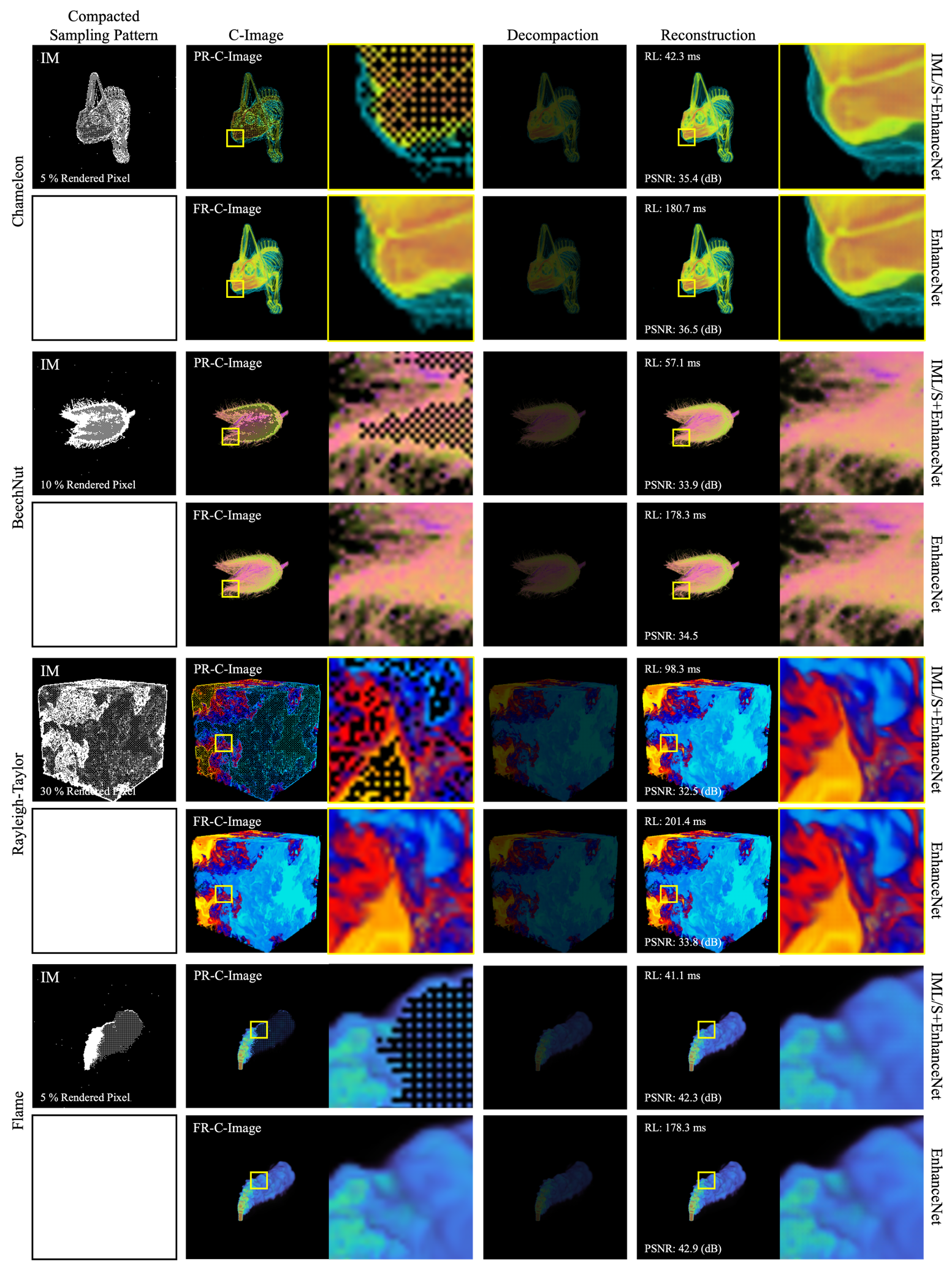}
        \caption{EnhanceNet and IML/S Net + EnhanceNet}
        \label{fig:super_images_comp}
    \end{subfigure}
    \hfill
    \begin{subfigure}[b]{0.49\linewidth}
        \centering 
        \includegraphics[trim=0 0 0 0,clip,width=\linewidth]{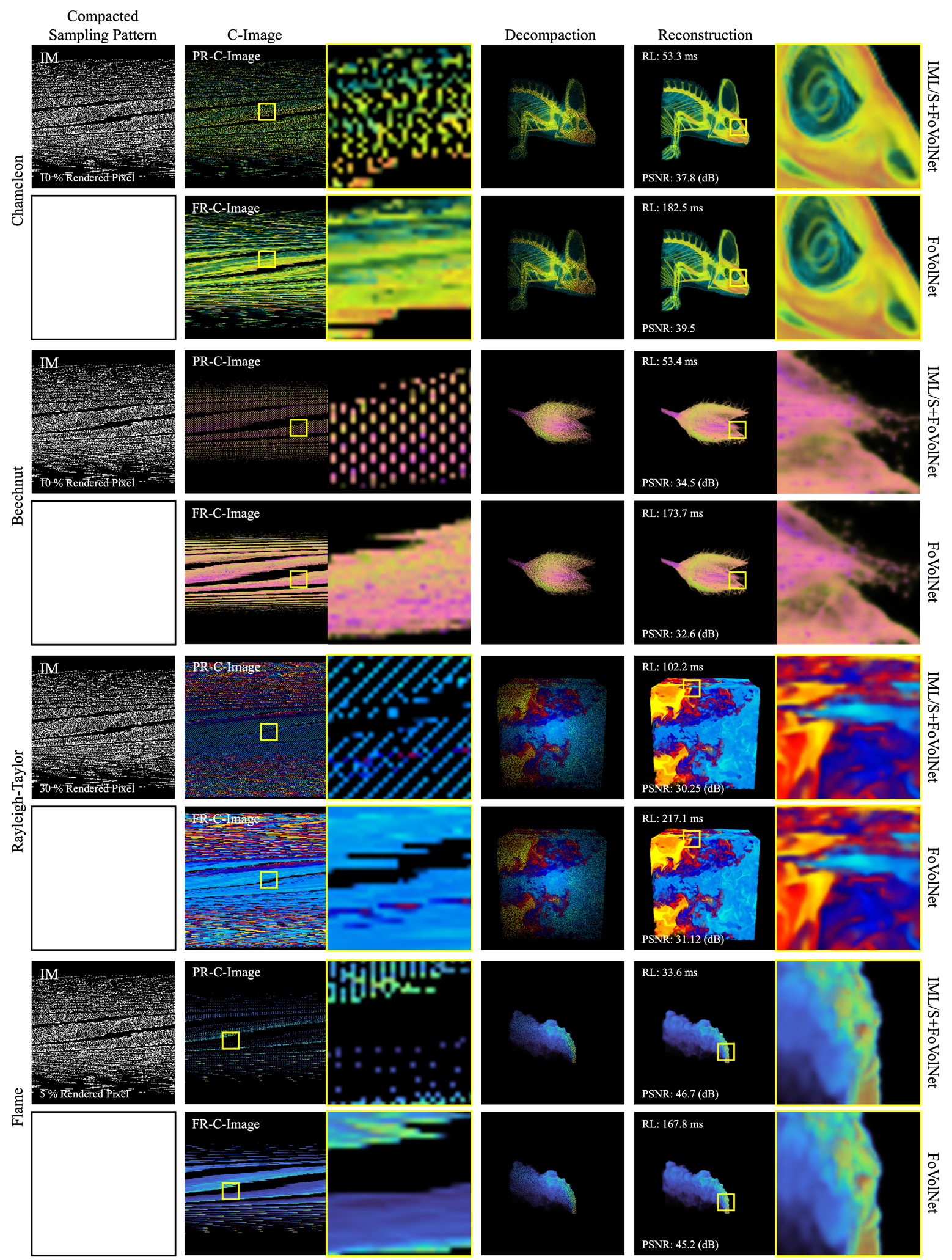}
        \caption{FoVolNet and IML/S Net + FoVolNet}
        \label{fig:foveat_images_comp}
    \end{subfigure}
    \caption{Visual comparison of reconstruction quality.}
    \label{fig:images_comp}
\end{figure*}

\textbf{Generator of IMS:} The generator of IMS Net takes a novel view and generates the corresponding IM. The accuracy of the generated IM determines the quality of the input to the downstream IML decoder for reconstruction. We evaluate the reconstruction capability of IMS Net by using different numbers of transposed convolutional layers, which is also the depth of the generator. We fixed the architecture of IML Net and only adjusted the IMS Net. As the number of transposed convolutional layers increases, similarly to adjusting the depth of IML Net, its reconstruction quality increases sublinearly (as shown in \cref{fig:depth_ims_psnr_sr} and \cref{fig:depth_ims_psnr_fr}) while the number of parameters of the network increases superlinearly (as shown in \cref{fig:depth_ims_num_para}). We also observed that adjusting the number of transposed convolutional layers in IMS Net has a greater impact on the reconstruction quality than the depth of the U-Net in IML Net. We select the layer size as 8 for IMS Net to balance rendering quality and inferencing latency.

\subsubsection{Hyperparameter Tuning}
\hspace{\parindent}\textbf{Mean Value for Rejection Sampling: }
The hyperparameter of the mean value plays an important role in determining the percentage of pixels in the C-Image as important pixels. The IML Net trained on a specific mean value, after the rejection sampling, will converge to have the percentage of important pixels of C-Image approximately equal to the $mean\times 10$. For a given reconstruction quality tolerance $\epsilon$, we can find the optimal rendering percentage (ORP) through a binary search on the mean value within the range $[0.01, 1]$. The $\epsilon$ is set as 1 dB in our experiment to search for the optimal mean value and ORP. Using the ORP for IML/S Net will give the RecNN a free improvement on the rendering latency without noticeable quality degradation.


\textbf{Loss functions: }\label{sec:loss}
For training the IMS Net for the best results, we tried multiple image-based loss functions, including Binary Cross Entropy (BCE) loss, Mean Squared Error (MSE) loss, Perceptual Loss (PER) loss, and Adversarial (ADV) loss. We found that combining PER and ADV loss gives the best result, as shown in \cref{fig:diff_loss_image}. BCE and MSE are not capable of learning an accurate IM. The PER loss is much better at accurately predicting the correct shape of the IM. However, PER loss struggles to predict the correct details of the IM. Using a loss that combines both the PER (weight = 1) and ADV (weight = 0.01) will give the closest result to the ground truth. The quantitative evaluation on the accuracy of the IM prediction from IMS Net and reconstruction from IML/S Net + EnhanceNet using various loss configurations is listed in \cref{fig:diff_loss_curve}.


\subsection{Results}\label{sec:results}
\subsubsection{Quality Evaluation}
We compare the reconstruction quality between the visualization pipelines using RecNN alone and our proposed IML/S Net + RecNN.

\begin{figure}[t]
    \centering
    \begin{subfigure}[b]{0.48\linewidth}
        \centering
        \includegraphics[trim=10 0 10 0,clip,width=\linewidth]{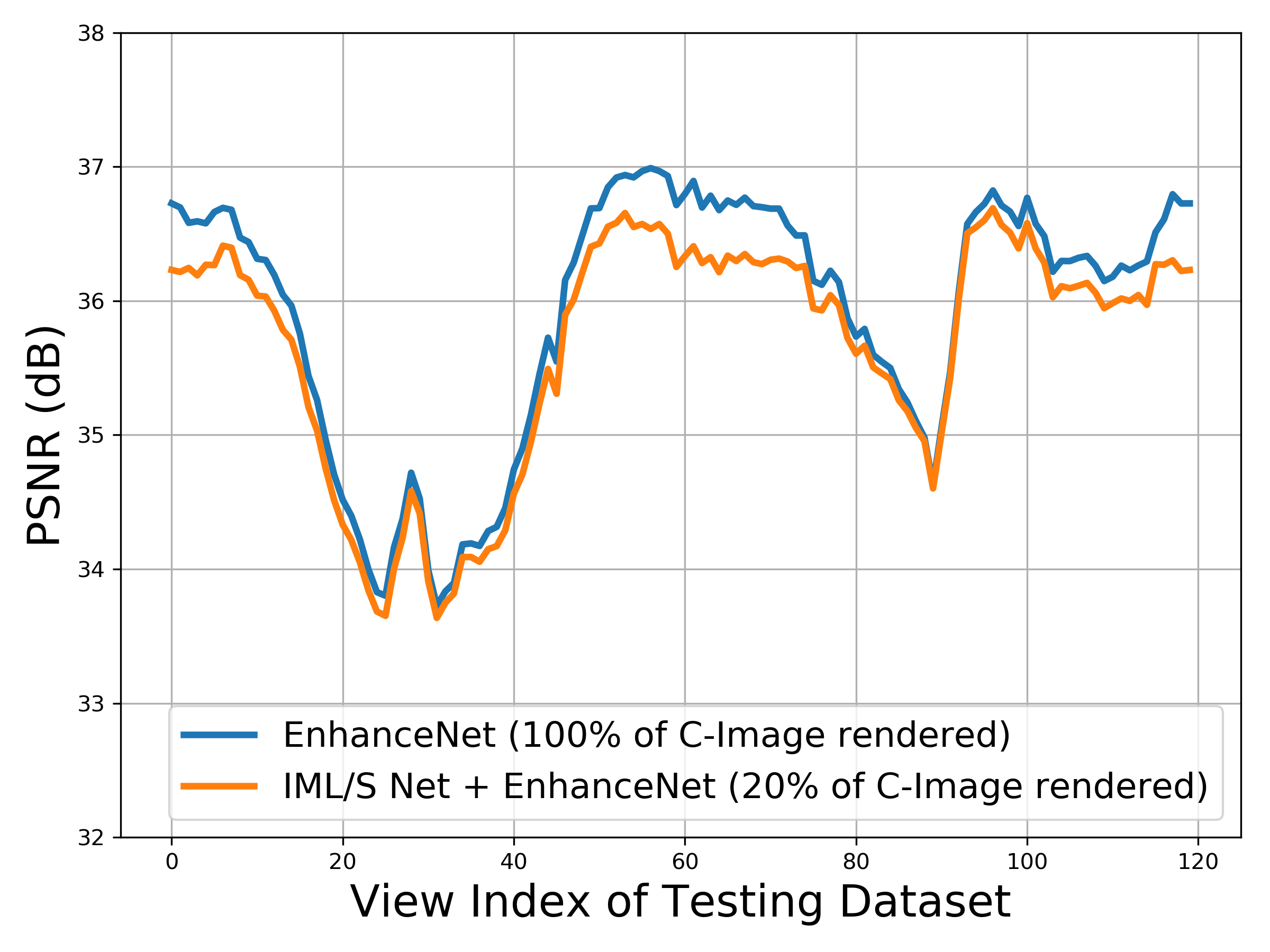}
        \caption{PSNR of each view}
        \label{fig:psnr_20p_super}
    \end{subfigure}
    \hfill
    \begin{subfigure}[b]{0.48\linewidth}
        \centering 
        \includegraphics[trim=10 0 10 0,clip,width=\linewidth]{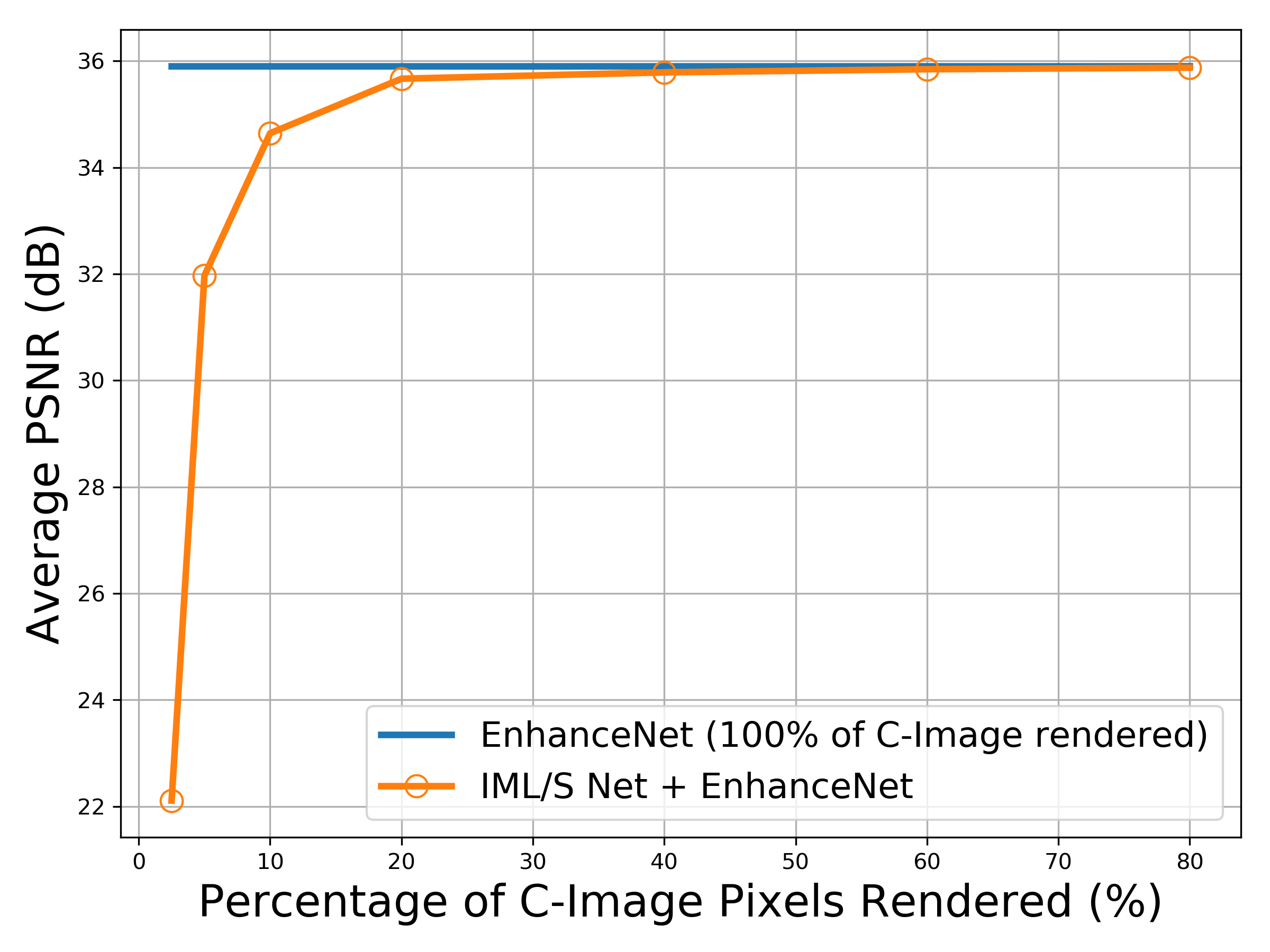}
        \caption{Average PSNR}
        \label{fig:psnr_all_super}
    \end{subfigure}
    \caption{Quantitative comparison of reconstruction quality between EnhanceNet and IML/S Net + EnhanceNet on the Chameleon dataset. (a) demonstrates the reconstruction quality from each view in the testing dataset using $20\%$ of C-Image pixels rendered. (b) shows the trend of average reconstruction quality while increasing the percentage of C-Image rendered.}
    \label{fig:psnr_qutitative_super}
\end{figure}

\textbf{Super-resolution:}
\cref{fig:super_images_comp} shows the visual comparison between the EnhanceNet and our IML/S Net + EnhanceNet of all 4 volumetric datasets. For each dataset, integrating our networks with EnhanceNet delivers comparable rendering quality to using EnhanceNet alone, but with significantly reduced rendering latency (2 to 4 times faster). In the PR-C-Images, the importance mask is successfully learned to extract a subset of the C-Image as important pixels. From the IM, we can observe that image areas with more dynamic content, like the edges of the object and the boundaries between distinct regions, were assigned more important pixels by our IML Net. This observation demonstrates that our network is capable of distinguishing complex regions from simpler ones and automatically sampling more important pixels in the complex region to achieve optimal reconstruction quality. The zoomed-in view with a smaller background area requires a higher percentage of important pixels to be rendered in order to achieve a quality similar to the zoomed-out view. Complex datasets like Rayleigh-Taylor also needs more important pixels compared to simple datasets like the Flame. \cref{fig:psnr_qutitative_super} shows the quantitative PSNR metric when rendering all the views in the Chameleon testing dataset. \cref{fig:psnr_20p_super} shows our method gives similar PSNR when only using 20\% of the C-Image pixels rendered. We observe from \cref{fig:psnr_all_super} that as the percentage of important pixels increases, our method infinitely approaches the quality of the standalone RecNN results, and this approximation converges quickly as the percentage of important pixels in the C-Image increases.

\begin{figure}[t]
    \centering
    \begin{subfigure}[b]{0.48\linewidth}
        \centering
        \includegraphics[trim=10 0 10 0,clip,width=\linewidth]{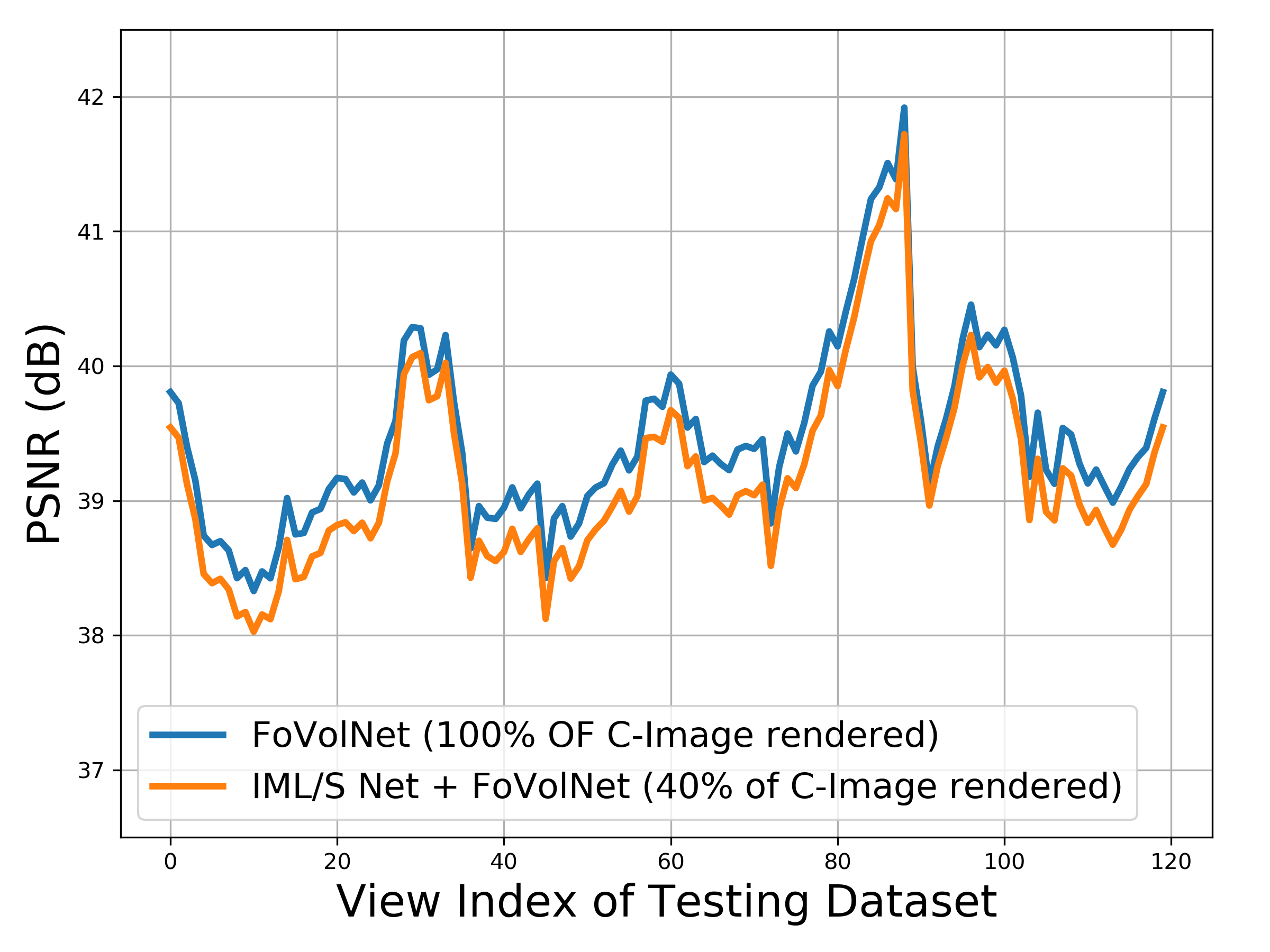}
        \caption{PSNR of each view}
        \label{fig:psnr_40p_foveated}
    \end{subfigure}
    \hfill
    \begin{subfigure}[b]{0.48\linewidth}
        \centering 
        \includegraphics[trim=10 0 10 0,clip,width=\linewidth]{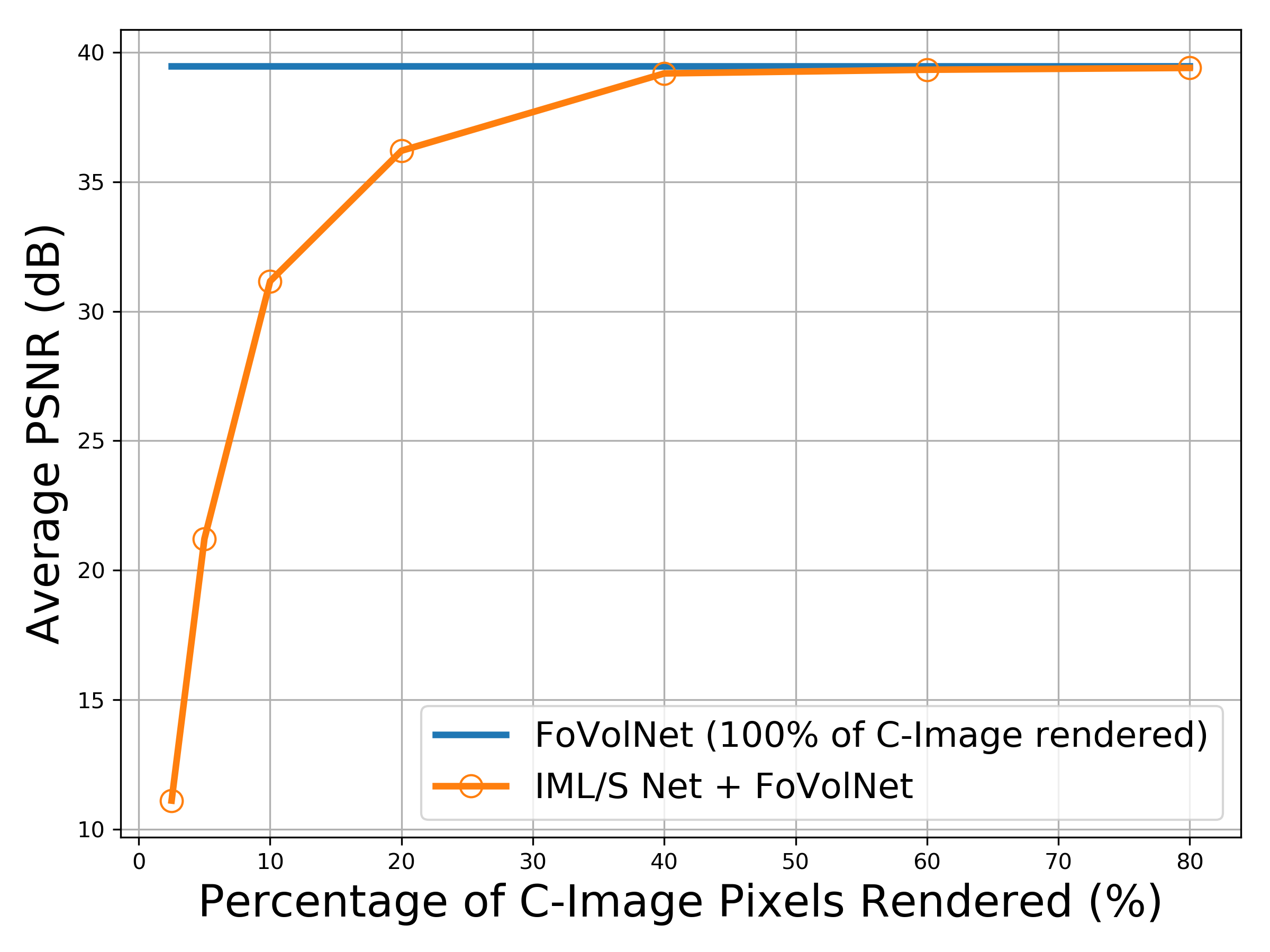}
        \caption{Average PSNR}
        \label{fig:psnr_all_foveated}
    \end{subfigure}
    \caption{Quantitative comparison of reconstruction quality between FoVolNet and IML/S Net + FoVolNet on the Chameleon dataset. (a) demonstrates the reconstruction quality from each view in the testing dataset using $40\%$ of C-Image pixels rendered. (b) shows the trend of average reconstruction quality while increasing the percentage of C-Image rendered.}
    \label{fig:psnr_qutitative_foveated}
\end{figure}

\textbf{Foveated Rendering:} 
\cref{fig:foveat_images_comp} shows the visual comparison between the FoVolNet and our IML/S Net + FoVolNet of all 4 volumetric datasets. We can observe similar results from the super-resolution using RecNN. Although the C-Images compacted from the foveal pattern does not preserve human-readable features like the ones compacted from the downsampling pattern, the IML Net can still find the important pixel from it, and the IMS Net can still predict the correct IM successfully. \cref{fig:psnr_40p_foveated} shows our method gives similar PSNR when only using 40\% of the C-Image pixels rendered. A similar convergence in quality can be observed as the percentage of important pixels in the C-Image increases from \cref{fig:psnr_all_foveated}. We can also observe that the convergence in the foveated rendering of the same Chameleon testing dataset is slower compared to super-resolution rendering. This is due to the foveal pattern, which limits the sampling to the focal area, restricting the possible locations the IML Net can choose from. 


\begin{figure}[t]
    \centering
    \begin{subfigure}[b]{0.485\linewidth}
        \centering
        \includegraphics[trim=10 0 10 0,clip,width=\linewidth]{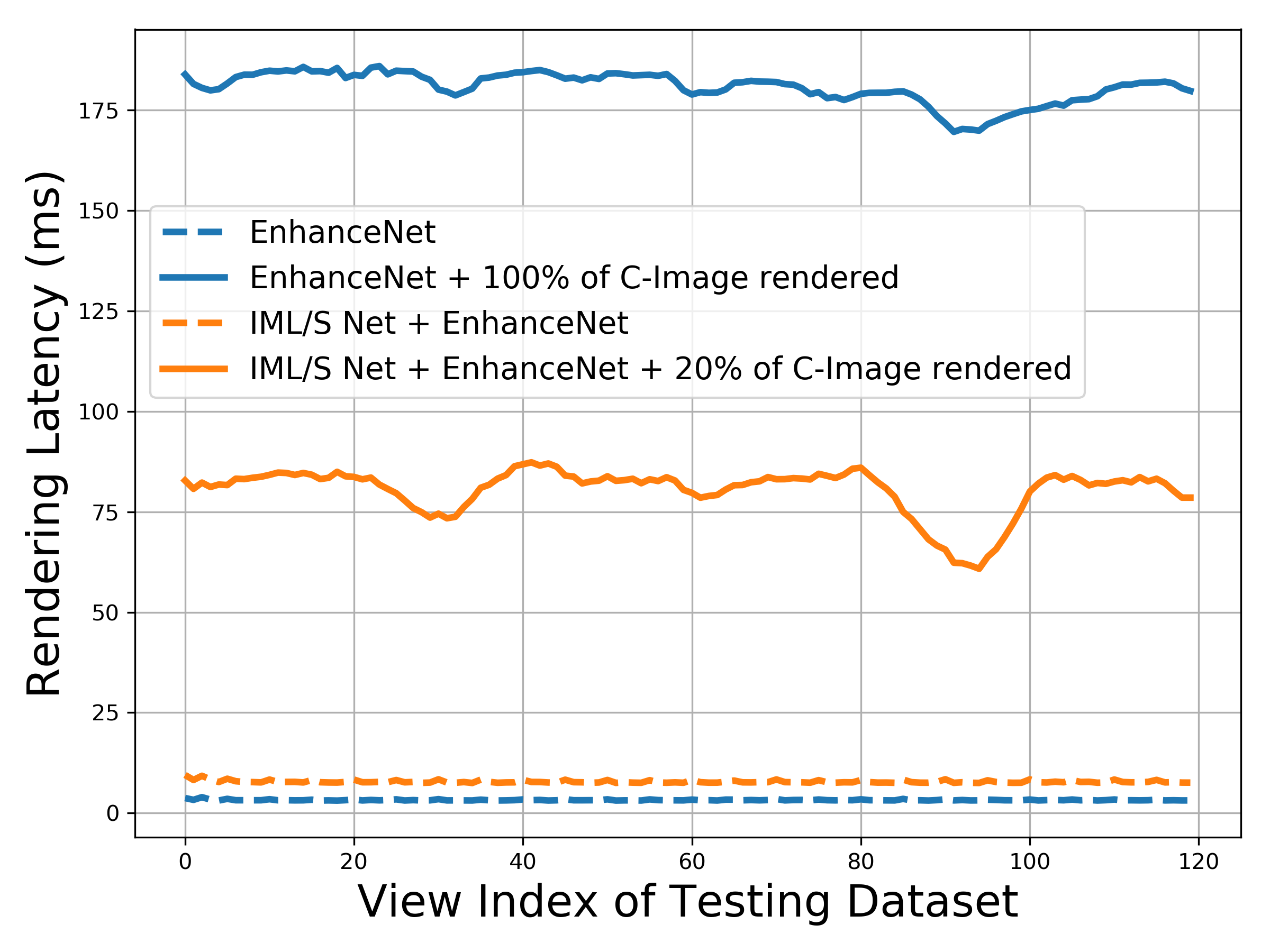}
        \caption{Rendering latency for each view}
        \label{fig:time_super_a}
    \end{subfigure}
    \hfill
    \begin{subfigure}[b]{0.485\linewidth}
        \centering 
        \includegraphics[trim=10 0 10 0,clip,width=\linewidth]{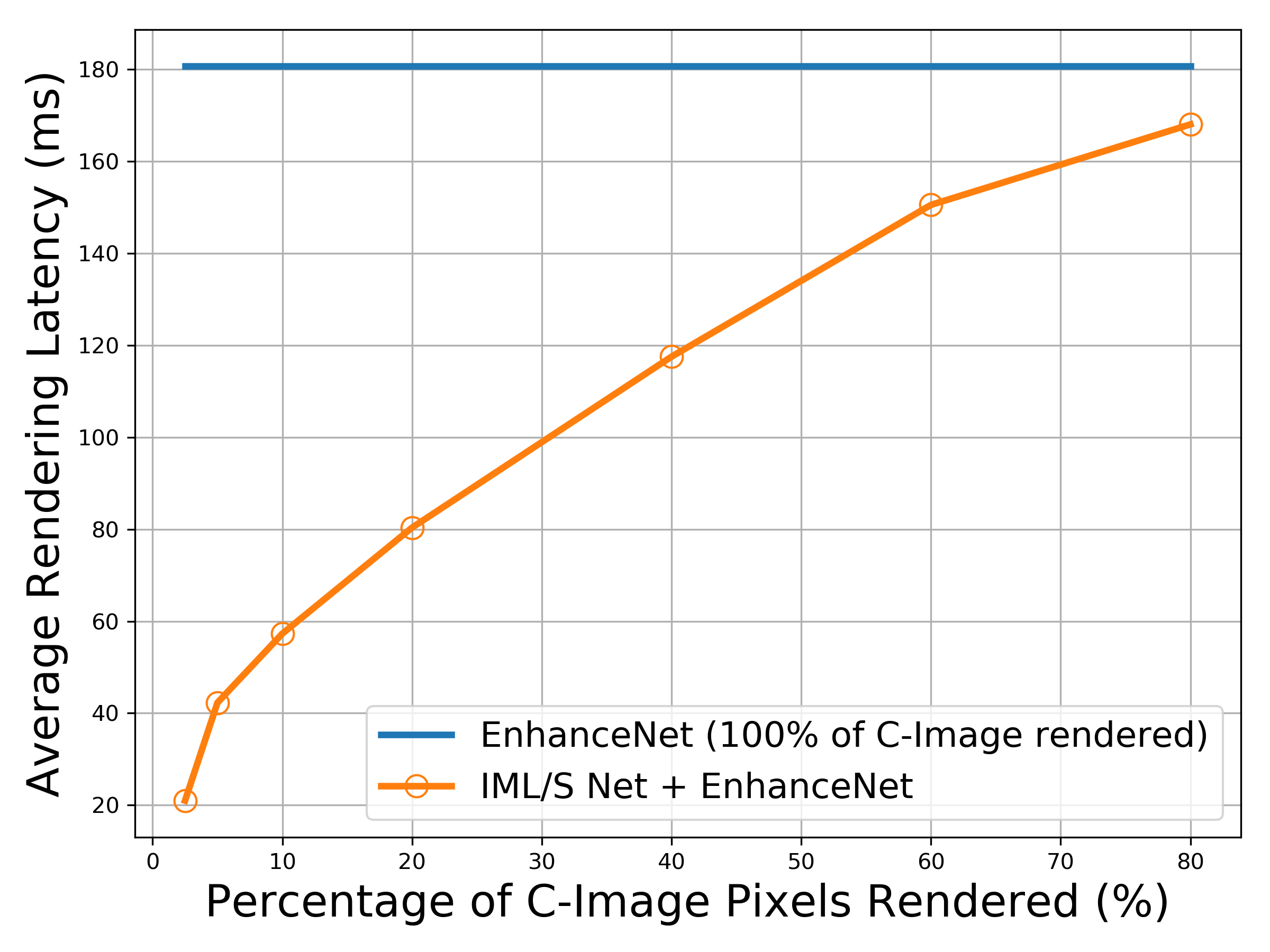}
        \caption{Average rendering latency}
        \label{fig:time_super_b}
    \end{subfigure}
    \caption{Quantitative comparison of rendering latency between EnhanceNet and IML/S Net + EnhanceNet on the Chameleon dataset. (a) demonstrates the rendering latency from each view in the testing dataset using $20\%$ of C-Image pixels rendered. (b) shows the trend of average rendering latency while increasing the percentage of C-Image rendered.}
    \label{fig:time_super}
\end{figure}

\begin{figure}[t]
    \centering
    \begin{subfigure}[b]{0.485\linewidth}
        \centering
        \includegraphics[trim=10 0 10 0,clip,width=\linewidth]{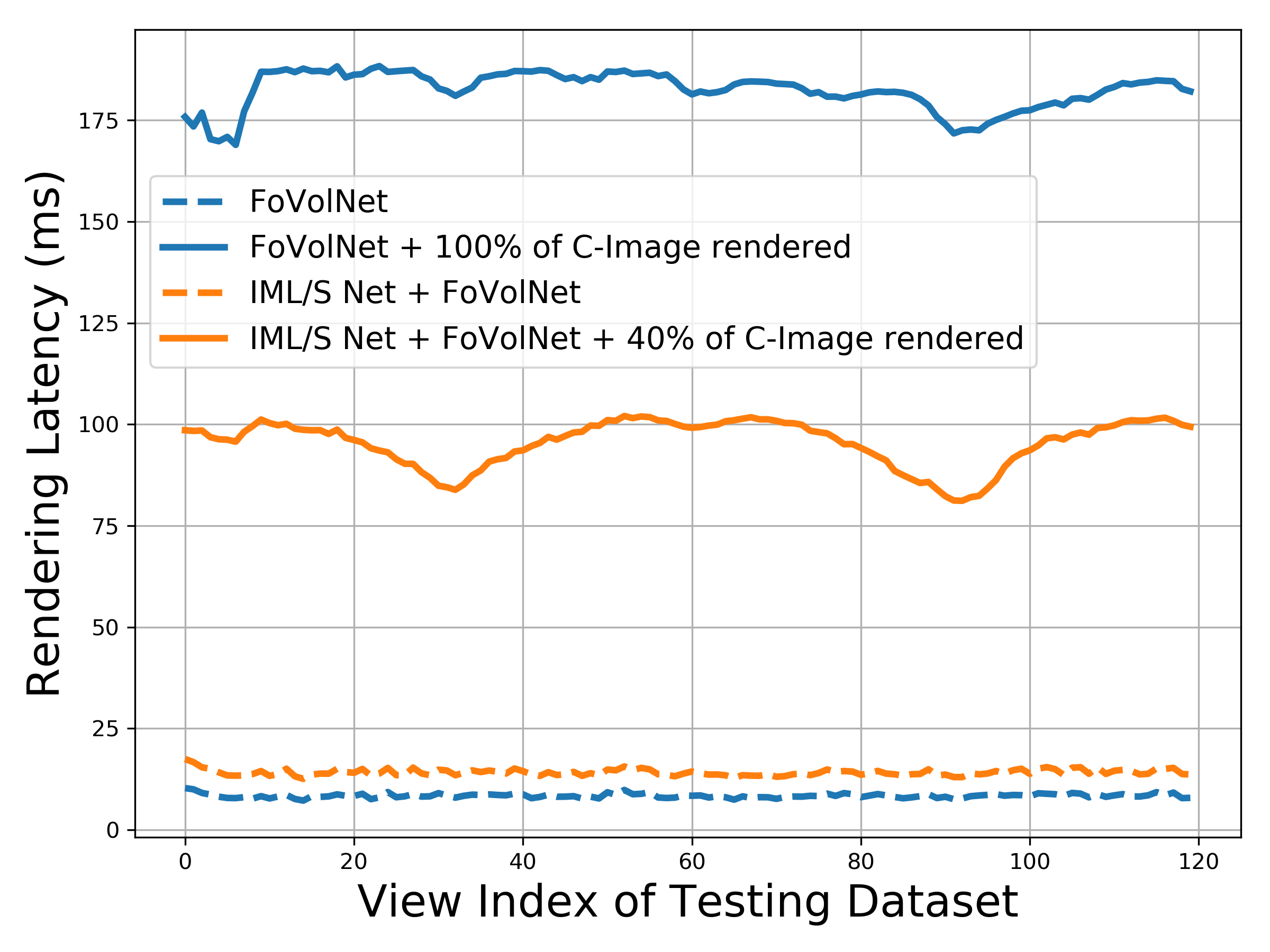}
        \caption{Rendering latency for each view}
        \label{fig:time_foveated_a}
    \end{subfigure}
    \hfill
    \begin{subfigure}[b]{0.485\linewidth}
        \centering 
        \includegraphics[trim=10 0 10 0,clip,width=\linewidth]{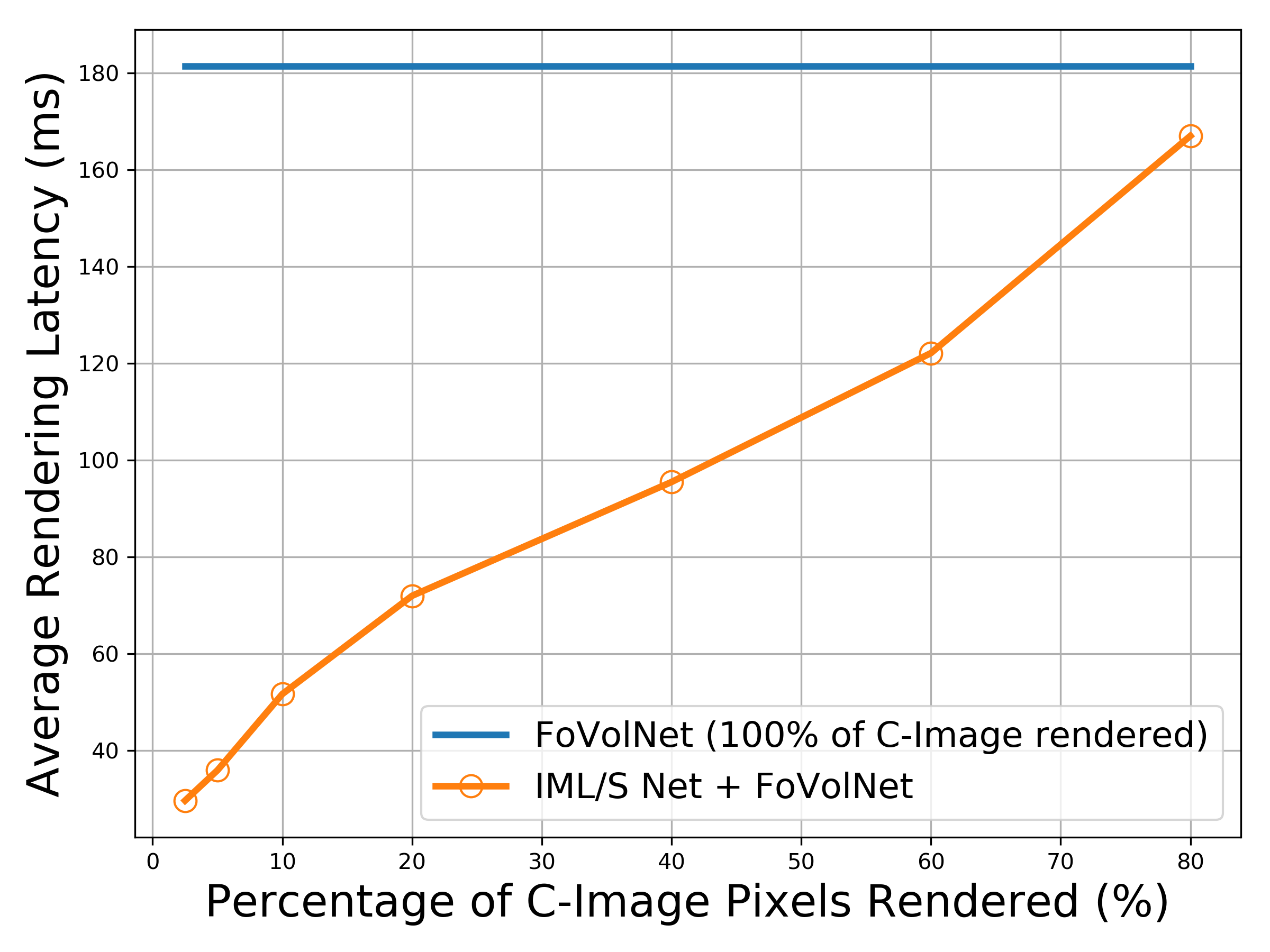}
        \caption{Average rendering latency}
        \label{fig:time_foveated_b}
    \end{subfigure}
    \caption{Quantitative comparison of rendering latency between FoVolNet and IML/S Net + FoVolNet on the Chameleon dataset. (a) demonstrates the rendering latency from each view in the testing dataset using $40\%$ of C-Image pixels rendered. (b) shows the trend of average rendering latency while increasing the percentage of C-Image rendered.}
    \label{fig:time_foveated}
\end{figure}

\begin{figure}[t]
    \centering
    \begin{subfigure}[b]{\linewidth}
        \centering
        \includegraphics[trim=0 0 0 0,clip,width=\linewidth]{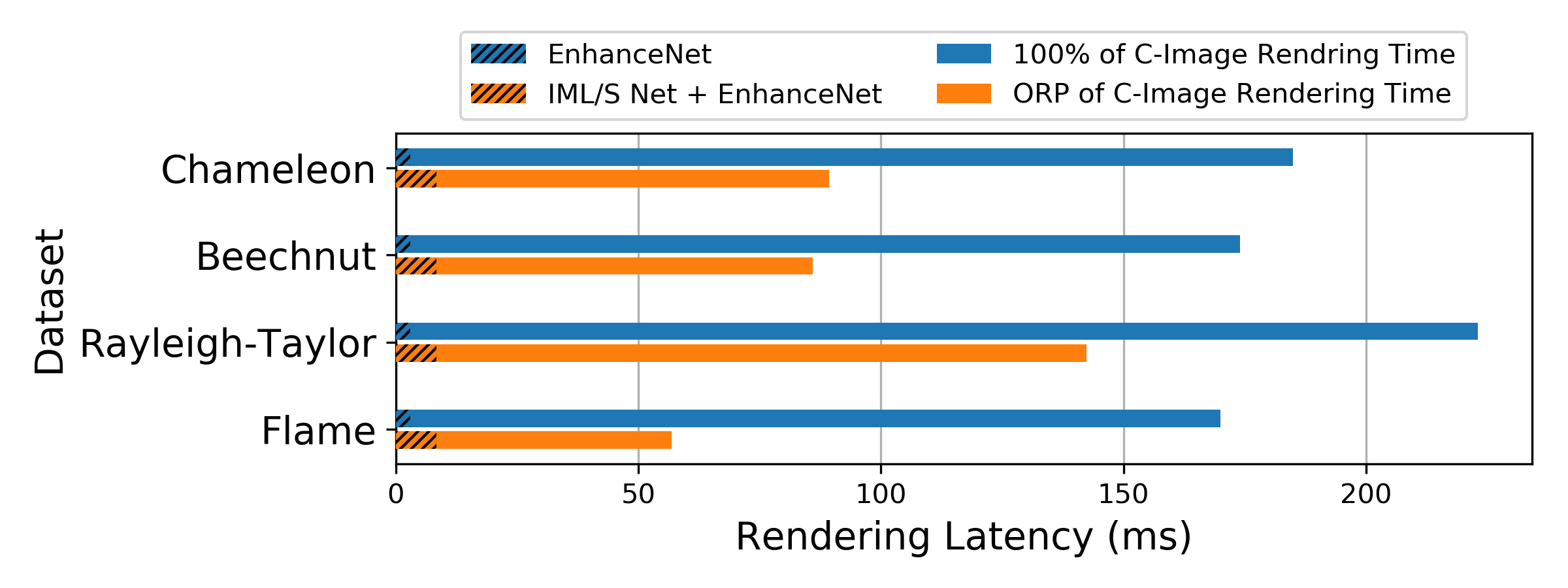}
        \caption{Super-resolution RecNN}
        \label{fig:orp_a}
    \end{subfigure}
    \begin{subfigure}[b]{\linewidth}
        \centering 
        \includegraphics[trim=0 0 0 0,clip,width=\linewidth]{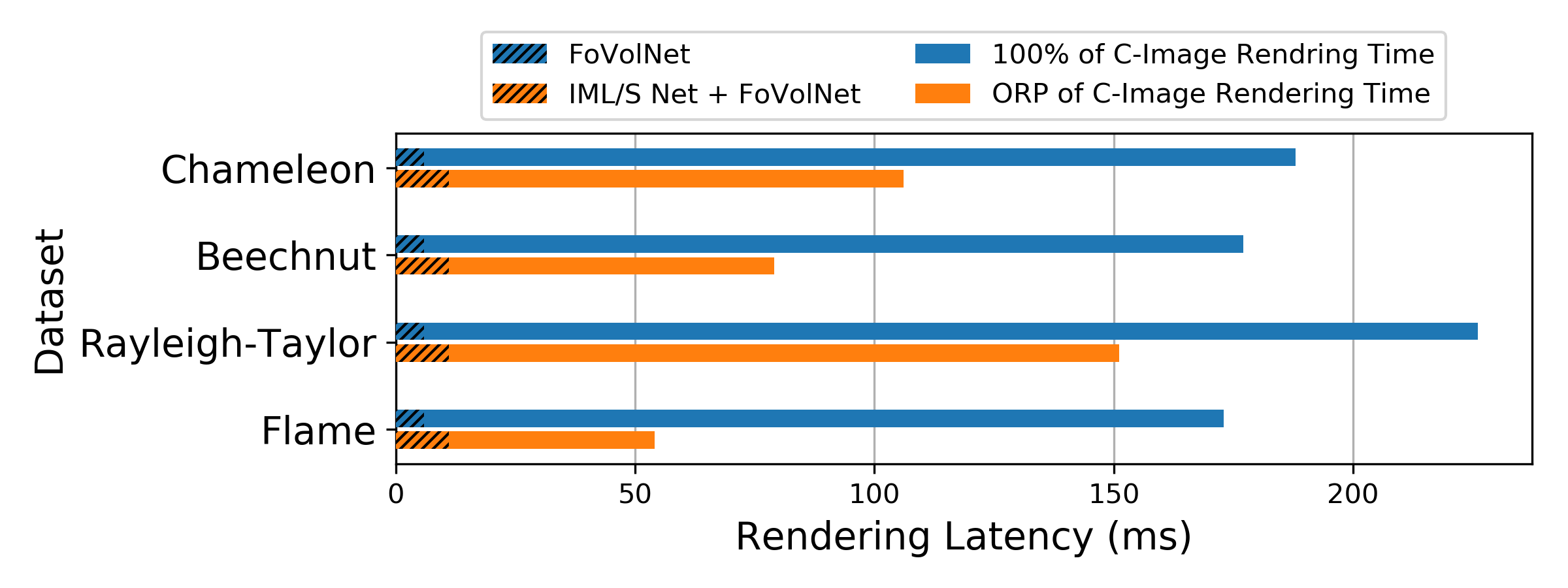}
        \caption{Foveated rendering RecNN}
        \label{fig:orp_b}
    \end{subfigure}
    \caption{Average rendering latency using ORP. The shaded regions are the time used to inference the network. The solid regions are the time used by the rendering algorithm to compute the pixel values in C-Image.}
    \label{fig:orp}
\end{figure}

\begin{table}[t]
  \caption{Detailed ORP for free rendering latency improvement.}
  \label{tab:orp_table}
  \scriptsize%
	\centering%
  \begin{adjustbox}{width=0.49\textwidth}
      \begin{tabu}{ c | c c | c c | c c  }
      \toprule
      Dataset & \multicolumn{2}{|c|}{ORP} & \multicolumn{2}{|c|}{Avg Rendering Latency (ms)}  & \multicolumn{2}{|c}{Avg Speed Up}  \\
        &
      \multicolumn{2}{|c|}{Sup-Res $\downarrow$  \ \ Foveated $\downarrow$} &
      \multicolumn{2}{|c|}{Sup-Res $\downarrow$  \ \ \ \ \ \ \ \ Foveated $\downarrow$} &
      \multicolumn{2}{|c}{Sup-Res $\uparrow$  \ \ Foveated $\uparrow$}\\
      \midrule
      Chameleon & 20\%  & 40\% & 89.4  & 106.1 & 2.07$\times$ & 1.77$\times$\\
      \midrule
      Beechnut & 18\% & 17\% & 86.0 & 79.1 & 2.02$\times$ & 2.24$\times$\\
      \midrule
      Rayleigh-Taylor & 40\% & 42\% & 142.4 & 151.1 & 1.75$\times$ & 1.49$\times$\\
      \midrule
      Flame & 11\% & 10\% & 56.9 & 54.1 & 2.98$\times$ & 3.20$\times$\\
      \bottomrule
      \end{tabu}
 \end{adjustbox}
\end{table}

\begin{table}[t]
  \caption{Training time and the number of epochs for training the IML and IMS Net. End-to-end trains the IML Net together with the downstream RecNN. Standalone only trains the IML Net.}
  \label{tab:training_time}
  \scriptsize%
	\centering%
  \begin{adjustbox}{width=0.49\textwidth}
      \begin{tabu}{ c | c c c c | c c }
      \toprule
      Dataset & \multicolumn{4}{|c|}{IML Net (epoches/hours $\downarrow$)} & \multicolumn{2}{|c}{IMS Net (epoches/hours $\downarrow$)} \\
        &
      \multicolumn{4}{|c|}{EnhanceNet   \ \ \ \ \ \ \ \ \ \ \ \ \ \ \ \  FoVolNet } &
      \multicolumn{2}{|c}{EnhanceNet   \ \ FoVolNet } \\
        &
      \multicolumn{2}{|c}{End-to-end   \ \ Standalone } &
      \multicolumn{2}{c|}{End-to-end   \ \ Standalone } &
      \multicolumn{2}{|c}{    \ \   } \\
      \midrule
      Chameleon & 83 / 5.1 &  92 / 2.0 & 107 / 7.2 & 117 / 2.5 & 1744 / 2.9  &  1687 / 3.1 \\
      \midrule
      Beechnut & 95 / 6.3 &  102 / 2.0& 43 / 3.3 & 73 / 1.5& 1632 / 3.2 & 1722 / 3.5 \\
      \midrule
      Rayleigh-Taylor & 136 / 8.2 & 141 / 2.9 & 161 / 10.2 & 180 / 3.5& 2443 / 3.1 & 2901 / 3.8 \\
      \midrule
      Flame & 152 / 9.1 & 120 / 2.5 & 255 / 15.0 & 163 / 2.8& 1711 / 3.0 & 1804 / 3.2 \\
      \bottomrule
      \end{tabu}
 \end{adjustbox}
\end{table}

\begin{figure}[t]
    \centering
    \begin{subfigure}[b]{0.24\linewidth}
        \centering
        \includegraphics[trim=10 0 11 0,clip,width=\linewidth]{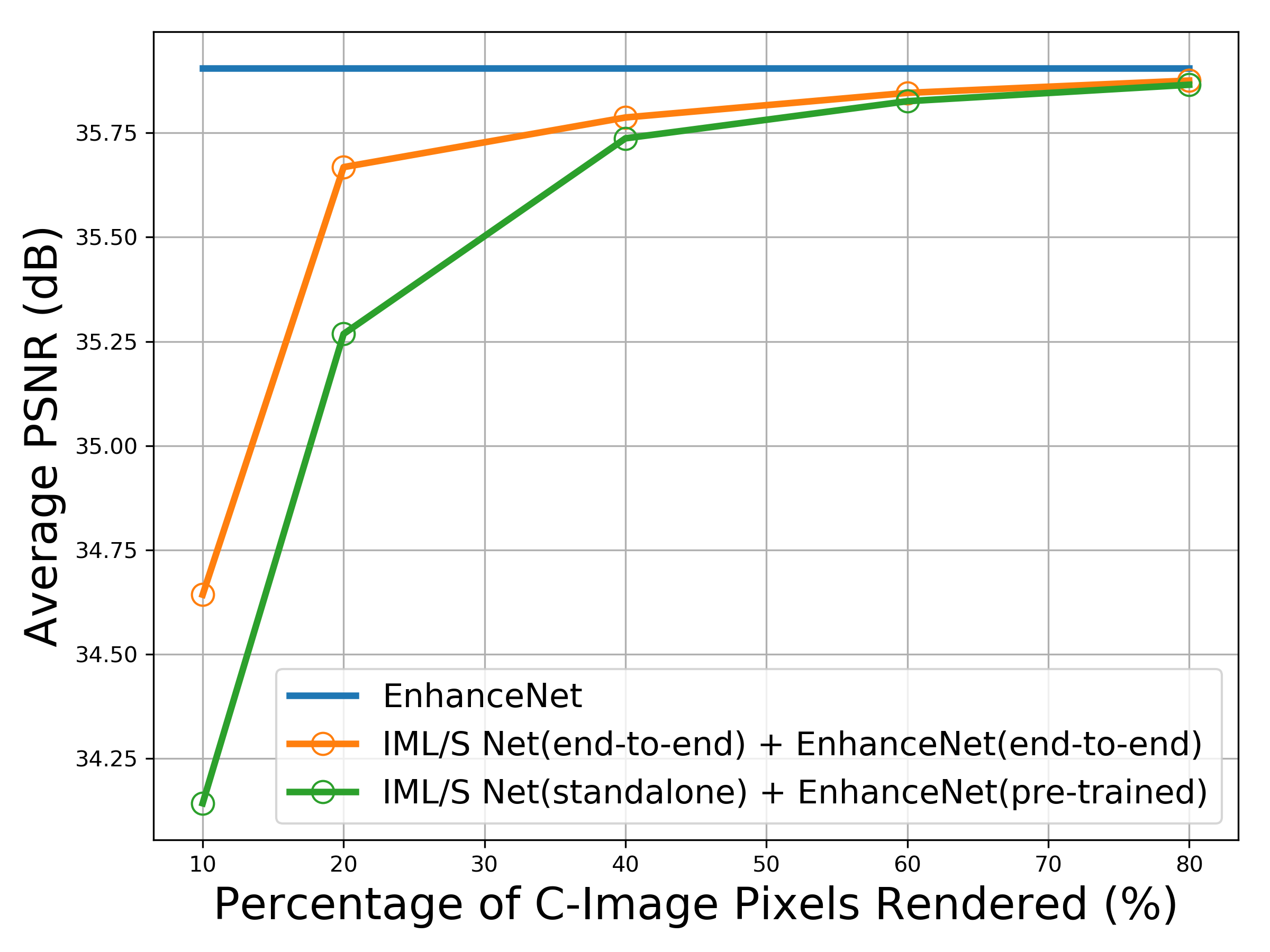}
        \caption{Chameleon (S-R)}
        \label{fig:test_1_sequences}
    \end{subfigure}
    \hfill
    \begin{subfigure}[b]{0.24\linewidth}  
        \centering 
        \includegraphics[trim=10 0 11 0,clip,width=\linewidth]{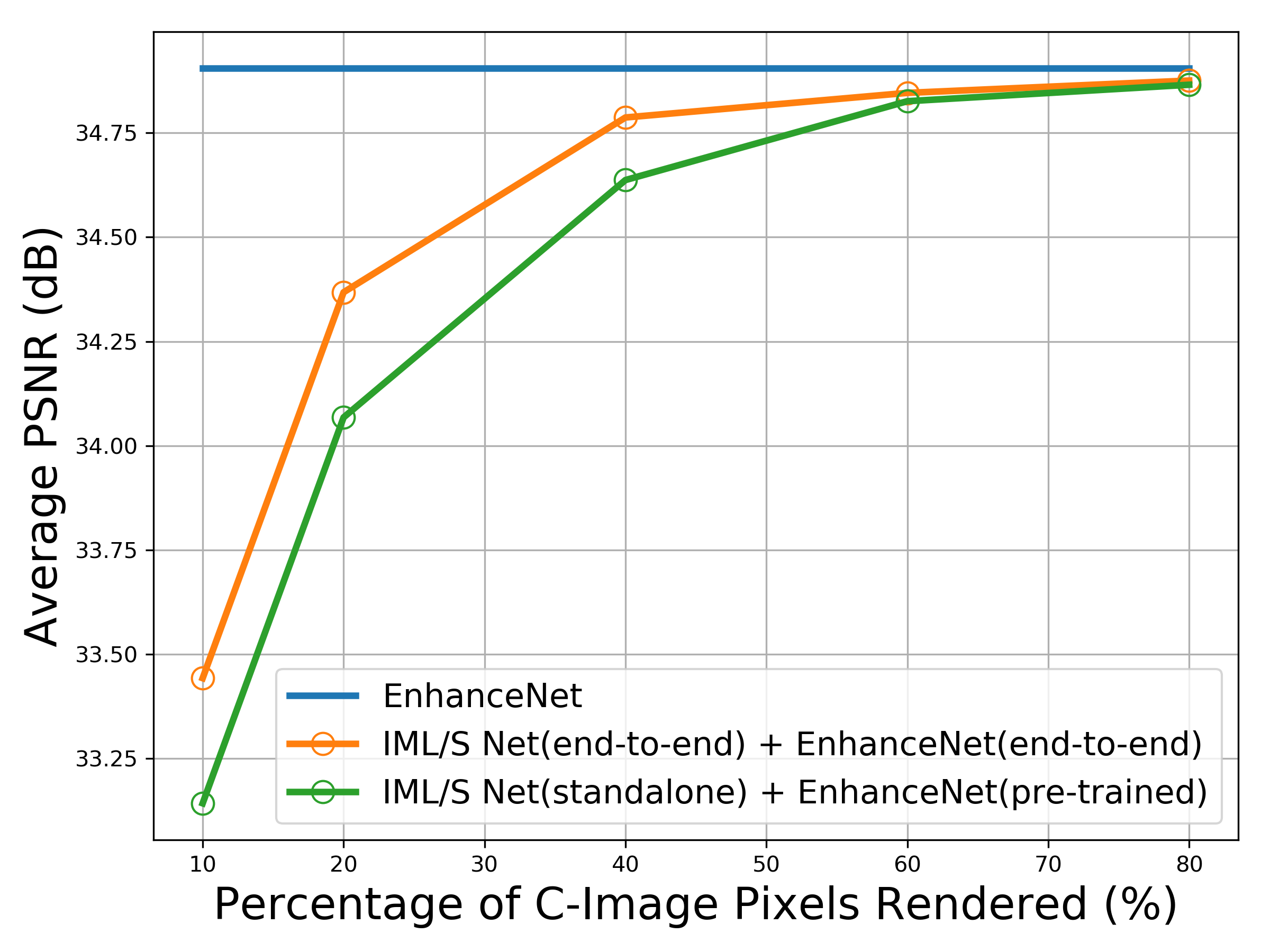}
        \caption{Beechnut (S-R))}
        \label{fig:test_2_sequences}
    \end{subfigure}
    \hfill
    \begin{subfigure}[b]{0.24\linewidth}   
        \centering 
        \includegraphics[trim=10 0 11 0,clip,width=\linewidth]{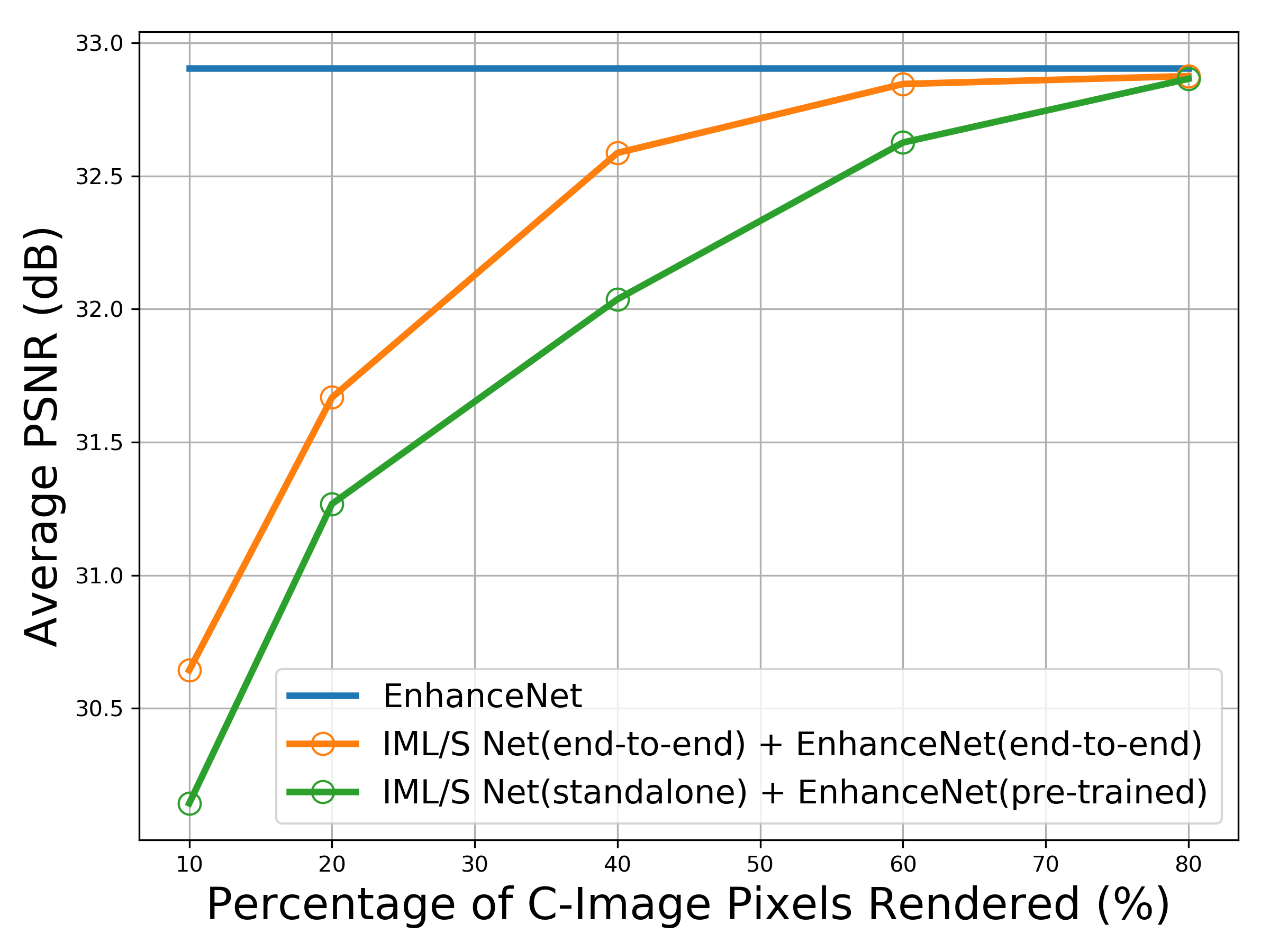}
        \caption{Rayleigh-Taylor (S-R))}
        \label{fig:test_3_sequences}
    \end{subfigure}
    \hfill
    \begin{subfigure}[b]{0.24\linewidth}   
        \centering 
        \includegraphics[trim=10 0 11 0,clip,width=\linewidth]{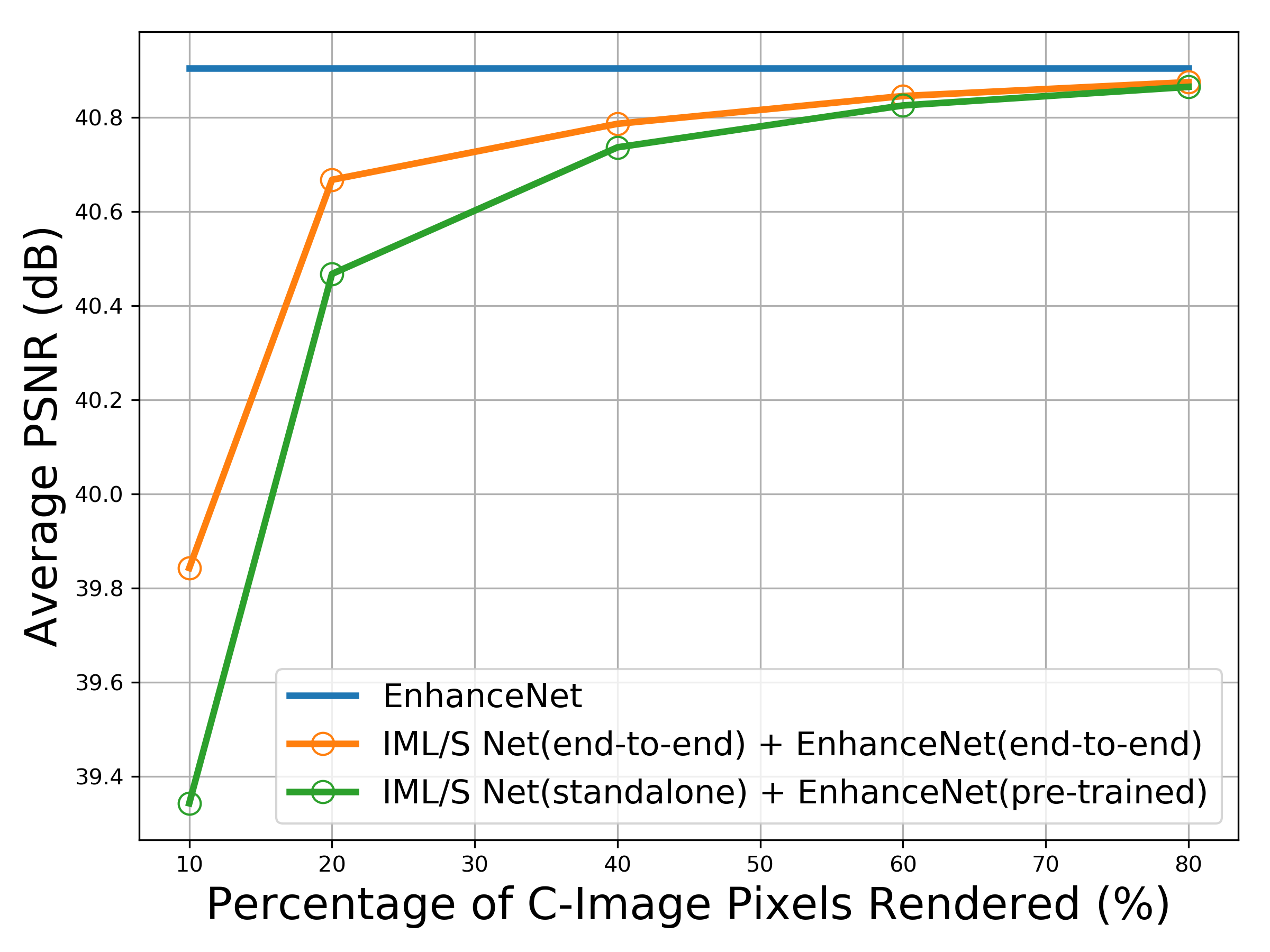}
        \caption{Flame (S-R))}
        \label{fig:test_4_sequences}
    \end{subfigure}
    \vspace{-\baselineskip}
    \begin{subfigure}[b]{0.24\linewidth}   
        \centering 
        \includegraphics[trim=10 0 11 0,clip,width=\linewidth]{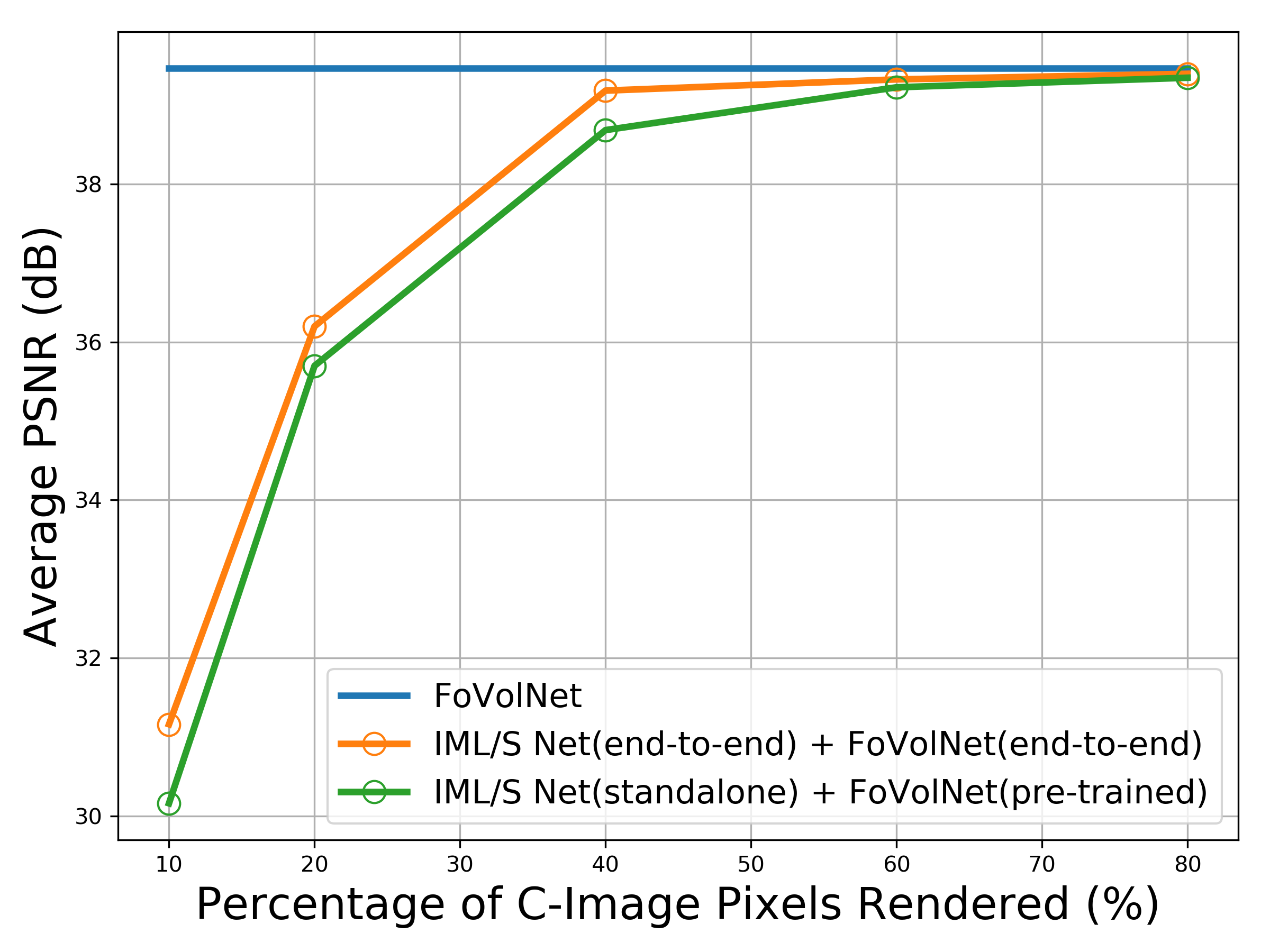}
        \caption{Chameleon (F-R))}
        \label{fig:test_3_sequences}
    \end{subfigure}
    \hfill
    \begin{subfigure}[b]{0.24\linewidth}   
        \centering 
        \includegraphics[trim=10 0 11 0,clip,width=\linewidth]{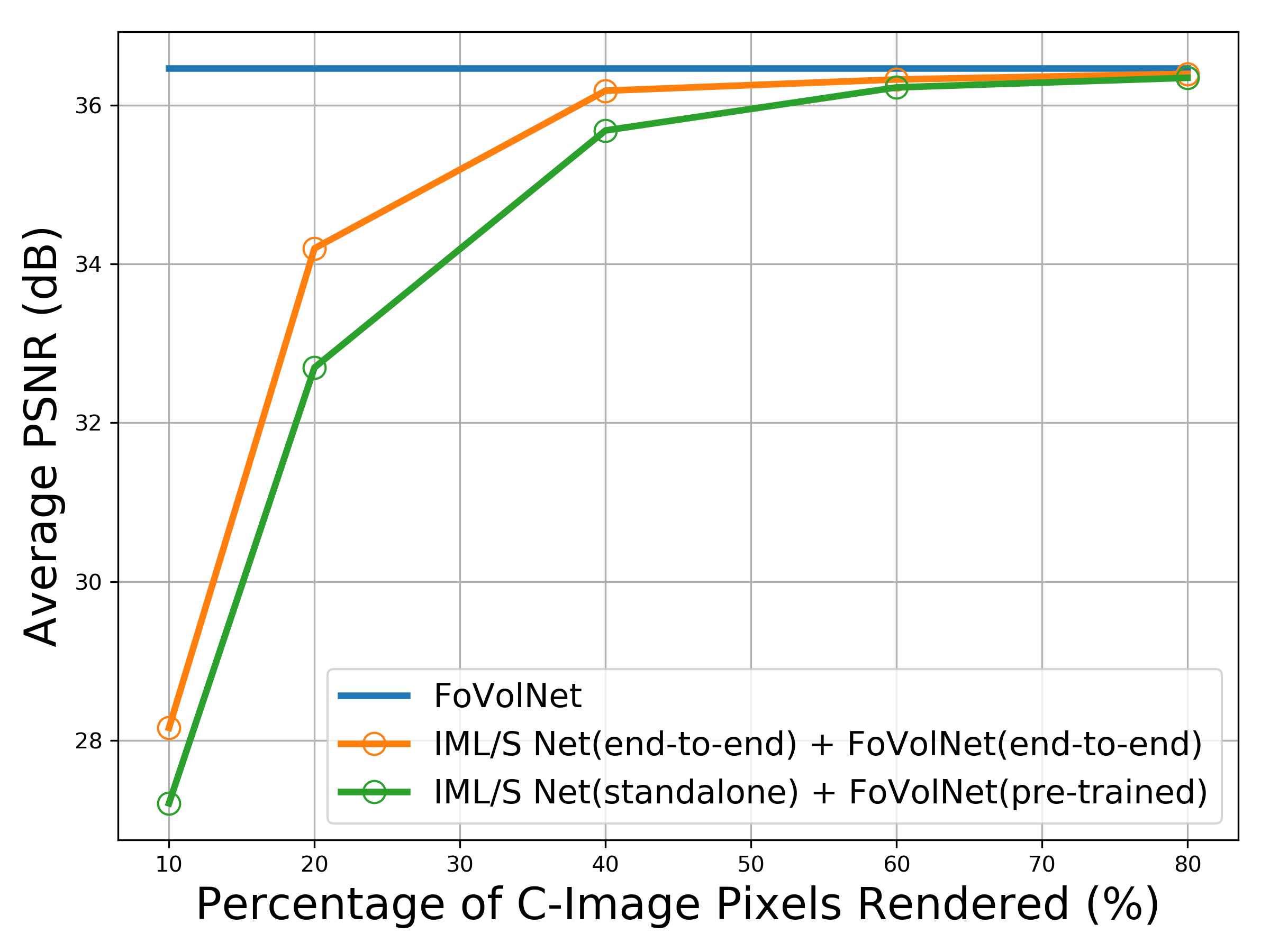}
        \caption{Beechnut (F-R))}
        \label{fig:test_4_sequences}
    \end{subfigure}
    \hfill
    \begin{subfigure}[b]{0.24\linewidth}   
        \centering 
        \includegraphics[trim=10 0 11 0,clip,width=\linewidth]{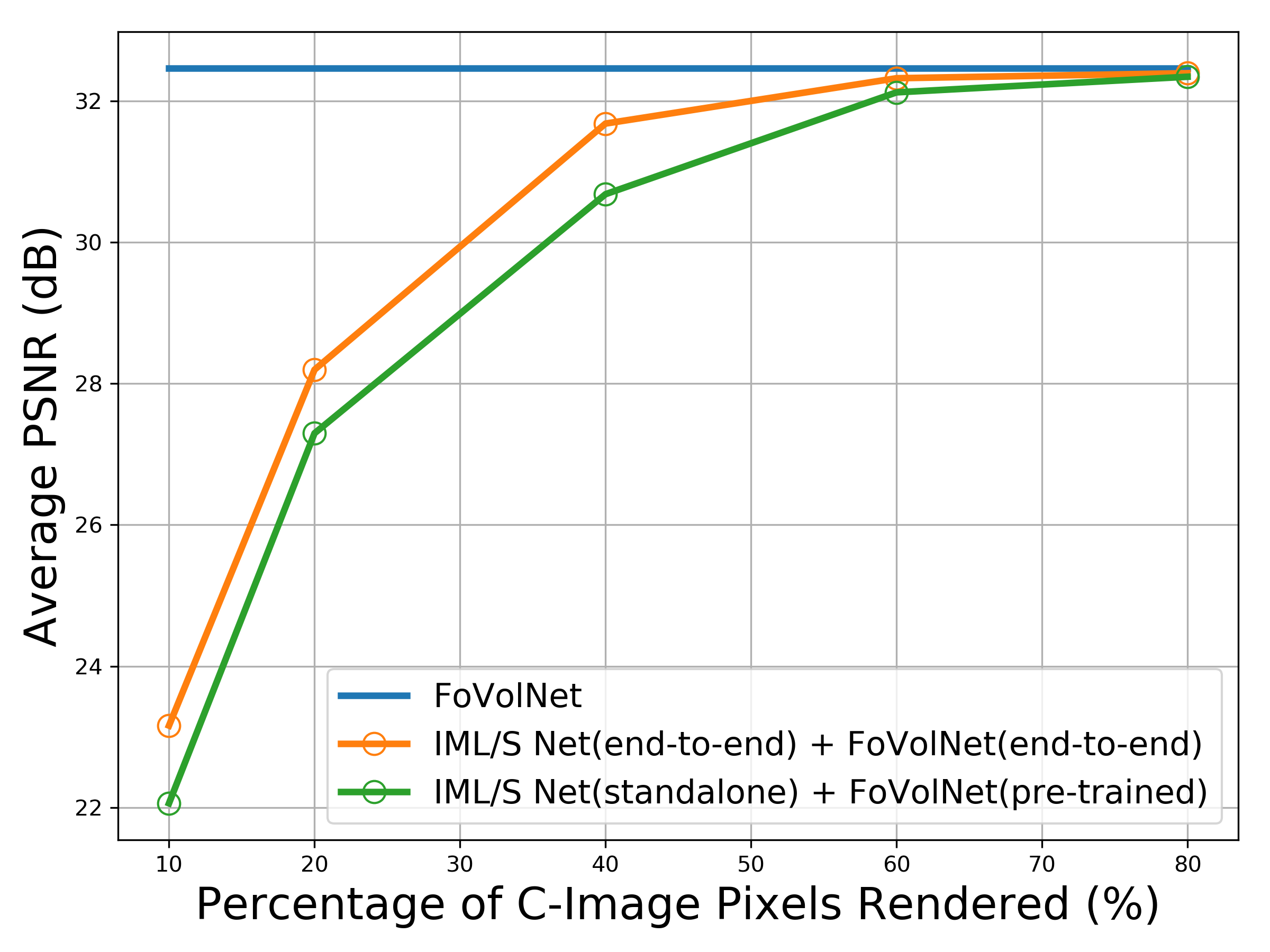}
        \caption{Rayleigh-Taylor (F-R))}
        \label{fig:test_3_sequences}
    \end{subfigure}
    \hfill
    \begin{subfigure}[b]{0.24\linewidth}   
        \centering 
        \includegraphics[trim=10 0 11 0,clip,width=\linewidth]{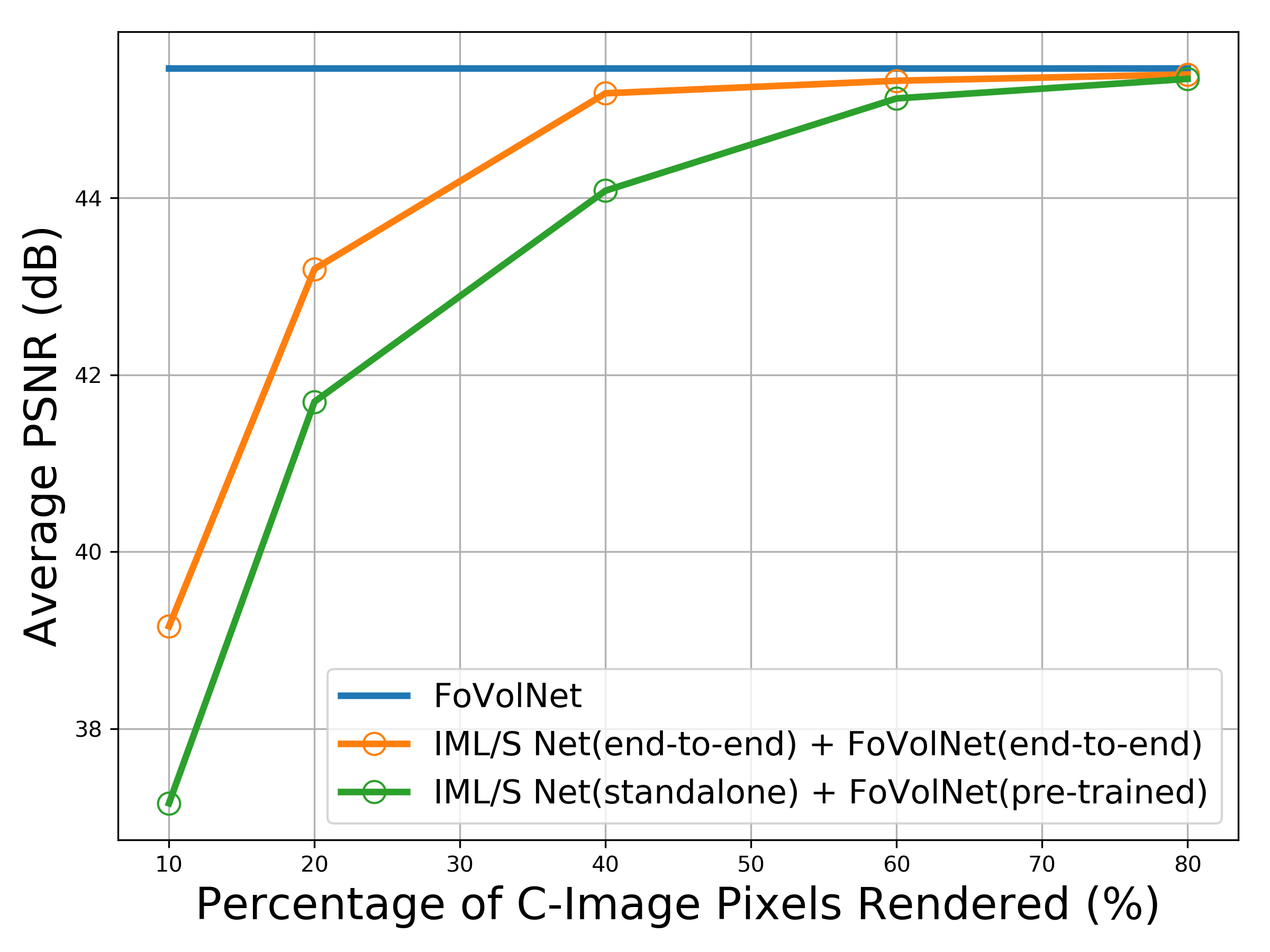}
        \caption{Flame (F-R))}
        \label{fig:test_4_sequences}
    \end{subfigure}
    \caption{Reconstruction quality comparison using RecNN between end-to-end trained IML Net + RecNN and standalone trained IML Net for both super-resolution RNN (S-R) and foveated Rendering RNN (F-R)}
    \label{fig:end2end_vs_standalone}
\end{figure}

\subsubsection{Performance Evaluation}
We compare the overall rendering latency between traditional RecNN rendering 100\% of the C-Image and our proposed IML/S Net together with the RecNN rendering a portion of the C-Image. In this section, we present a detailed analysis of rendering performance using the Chameleon dataset as an example. A joint evaluation of both rendering quality and performance across all testing datasets will be provided in the next section.

\textbf{Super-resolution:}
As we discussed before, the overall rendering latency is the sum of the time to partially render the input image to RecNN and the RecNN inferencing to reconstruct the FR-Image. \cref{fig:time_super_a} presents the quantitative measurement of time usage at each stage. We observe that the inferencing time for both methods is minimal, though the overhead introduced by the IML/S Net slightly increases the total inferencing time compared to using the standalone RecNN. Using the IML/S Net significantly reduces the time required to generate the partially rendered image, as fewer pixels undergo the costly rendering computations. \cref{fig:time_super_b} demonstrates that the growth of average input latency is close to linear with respect to the percentage of important pixels in C-Image.

\textbf{Foveated Rendering:} 
\cref{fig:time_foveated_a} presents the quantitative measurement of time usage at each stage for foveated rendering. Our IML/S Net demonstrates similar performance in foveated rendering RecNN as it does in super-resolution RecNN. Since FoVolNet is more complex than EnhanceNet, the inference time for both methods is higher compared to their counterparts in the super-resolution framework. \cref{fig:time_foveated_b} also demonstrates a linear relationship between the percentage of important pixels in C-Image and the average input latency. Considering both quality and performance results on the Chameleon dataset, we can conclude that, as the percentage of important pixels increases, the quality of reconstruction is more sensitive in super-resolution compared to foveated rendering. This suggests that, for the same reconstruction quality, super-resolution benefits more from our IML/S Net in terms of reducing rendering latency compared to foveated rendering. This observation is related to the nature of the distribution of the Chameleon dataset in the visualization image space. The contents of interest across views are more evenly distributed across the whole image domain rather than only distributed in the center. As a result, the downsampling pattern of super-resolution RecNN covers such a distribution better than the foveal pattern of foveated rendering RecNN.



\subsubsection{Free Rendering Latency Improvement}\label{orp_improvement}
In this section, we further investigate the data-dependent characteristics of various testing datasets on rendering quality and performance using our method. We will also qualitatively evaluate how much rendering latency improvement can be achieved by our method without sacrificing perceivable rendering quality. We find the optimal rendering percentage (ORP) for our IML Net to make IML/S Net reconstruct perspective rendering within 1 dB PSNR difference ($\epsilon$) to the rendering from the standalone RecNN. \cref{fig:orp} shows the average rendering latency using ORP for both EnhanceNet and FoVolNet. On average, our method can improve the rendering latency for each volumetric dataset with similar rendering quality. The detailed results are listed in \cref{tab:orp_table}. We can see that for different datasets, ORP varies to achieve close reconstruction quality to the original RecNN. Rayleigh-Taylor requires the highest ORP to maintain quality due to its dynamic nature across the whole 3D space, which leads to the largest content area in the final rendering compared to other datasets. Flame is simpler and more compact in 3D space than others, resulting in the lowest ORP. Compared to super-resolution, foveated rendering favors datasets where the rendered object is primarily centered in the image, such as Beechnut and Flame, leading to a lower ORP.

\subsubsection{Feasibility Study}\label{feasibility}
In this section, we assess the practicality and viability of the proposed method. To take advantage of the rendering latency improvement provided by the proposed method, the overhead is the extra training step for the IML and IMS networks. The IMS Net is a simple network and its training dataset (view and IM) is also small, so its training process is relatively quick. Due to the complex structure of the RecNNs and their large training data (full-res images), a significant amount of time is dedicated to the end-to-end training of the IML Net and the downstream RecNN. To enhance the feasibility of our method, the proposed IML Net can be trained independently without connecting to the complicated downstream RecNN. Compared to RecNN, the network structure of IML Net is simpler, and its training data is smaller (low-res C-Image). The loss function for training this standalone IML Net is solely based on the compaction image loss. Once the IML Net is trained, its decoder can directly connect to an off-the-shelf pre-trained RecNN to improve the rendering latency. In this way, hours of training time can be saved. The total epochs and training times for training the IML and IMS Net are listed in \cref{tab:training_time}. We can observe that training IML standalone will save a significant amount of training time compared to training IML end-to-end with RecNN. Due to the absence of joint optimization across the entire pipeline, using the independently trained IML Net will result in lower reconstruction quality compared to end-to-end training. However, as shown in \cref{fig:end2end_vs_standalone}, the quality degradation is still acceptable, especially when the IML Net is trained using a larger percentage of important pixels on C-Image.

\begin{table*}[t]
  \caption{Rendering and quality evaluation between Learning Adaptive Sampling (LAS) method and the proposed IML/S with RecNN for super-resolution and foveated rendering. Foveated Rendering is not supported (NS) by LAS. The speed up metric under average rendering latency is measured using IML/S compared to LAS.}
  \label{tab:las_compare}
  \scriptsize%
  \centering%
  \begin{adjustbox}{width=0.8\linewidth}
      \begin{tabu}{ c | c | c | c c c | c c | c c | c c }
      \toprule
      \multirow{3}{*}{Dataset} &  Average Rendering & \multirow{3}{*}{IPR} & \multicolumn{5}{|c|}{Average Rendering Latency (ms) $\downarrow$} & \multicolumn{4}{|c}{Average Rendering Quality (Avg PSNR (dB) $\uparrow$ / Avg SSIM $\uparrow$)} \\
        & Latency of GPU & &
      \multicolumn{3}{|c|}{Super-Resolution} &
      \multicolumn{2}{|c|}{Foveated Rendering} &
      \multicolumn{2}{|c|}{Super-Resolution} &
      \multicolumn{2}{|c}{Foveated Rendering}  \\
        & Ray-caster (ms) & &
      \multicolumn{3}{|c|}{LAS  \ \ \ \ \ IML/S  \ \ \ Speed Up} &
      \multicolumn{2}{|c|}{LAS  \ \ \ \ \ \ \ \ \ IML/S} &
      \multicolumn{2}{|c|}{LAS  \ \ \ \ \ \ \ \ \ IML/S} &
      \multicolumn{2}{|c}{LAS  \ \ \ \ \ \ \ \ \ IML/S} \\
      \midrule
      \multirow{3}{*}{Chameleon} & \multirow{3}{*}{326.5} & 5\% & 
      111.2 & \textbf{42.0} & 2.65$\times$ & NS & 39.6 &
      32.63 / 0.932 & \textbf{32.86} / \textbf{0.940} & NS & 29.11 / 0.883\\
       & & 10\% &
       191.4 & \textbf{60.4} & 3.17$\times$ & NS & 59.3 &
       \textbf{34.65} / \textbf{0.953} & 34.64 / 0.951 & NS & 31.16 / 0.920 \\
       & & 15\% &
       259.9 & \textbf{70.9} & 3.67$\times$ & NS & 66.3 &
       34.80 / 0.955 & \textbf{35.25} / \textbf{0.976} & NS & 34.21 / 0.951\\
       \midrule
       \multirow{3}{*}[-1.5ex]{Beechnut} & \multirow{3}{*}{517.7} & 5\% & 
      118.2 & \textbf{50.3} & 2.35$\times$ & NS & 48.2 &
      31.70 / 0.915 & \textbf{32.98} / \textbf{0.940} & NS & 26.03 / 0.859 \\
       & & 10\% &
       213.6 & \textbf{60.3} & 3.54$\times$ & NS & 57.5 &
       32.86 / 0.937   & \textbf{33.44} / \textbf{0.944}   & NS & 28.16 / 0.891\\
       & & 15\% &
       290.9 &  \textbf{75.0} & 3.88$\times$ & NS & 69.8 &
       33.03 / 0.946   &  \textbf{34.01} / \textbf{0.949}  & NS & 31.05 / 0.908\\
       \midrule
       \multirow{3}{*}[-1.5ex]{Rayleigh-Taylor} & \multirow{3}{*}{354.0} & 10\% & 
      152.7 & \textbf{44.9} & 3.40$\times$ & NS & 39.0 &
      24.19 / 0.816 & \textbf{30.64} / \textbf{0.910} & NS & 23.16 / 0.803 \\
       & & 15\% &
       220.5  & \textbf{62.2} & 3.55$\times$ & NS & 58.9  &
       29.32 / 0.894  & \textbf{31.25} / \textbf{0.917}   & NS & 26.51 / 0.866 \\
       & & 20\% &
       290.8 & \textbf{77.2} & 3.77$\times$ & NS & 71.9 &
       \textbf{32.45} / \textbf{0.942} & 31.67 / 0.923 & NS & 28.20 / 0.889 \\
       \midrule
       \multirow{3}{*}[-1.5ex]{Flame} & \multirow{3}{*}{438.7} & 2.5\% & 
      77.6 & \textbf{22.3} & 3.48$\times$ & NS & 19.5 &
      38.24 / \textbf{0.966} & \textbf{39.03} / 0.965 & NS & 35.33 / 0.980 \\
       & & 5\% &
       124.3  & \textbf{31.3} & 3.97$\times$ & NS & 27.1  &
       38.31 / 0.957  & \textbf{39.31} / \textbf{0.976}   & NS &  36.54 / 0.974\\
       & & 7.5\% &
       175.4 & \textbf{43.4} & 4.04$\times$ & NS & 40.6 &
       39.01 / 0.962 & \textbf{39.57} / \textbf{0.985} & NS & 38.02 / 0.975 \\
      \bottomrule
      \end{tabu}
 \end{adjustbox}
\end{table*}

\begin{figure*}[t]
    \centering
    \includegraphics[trim=0 0 0 0,clip,width=\textwidth]{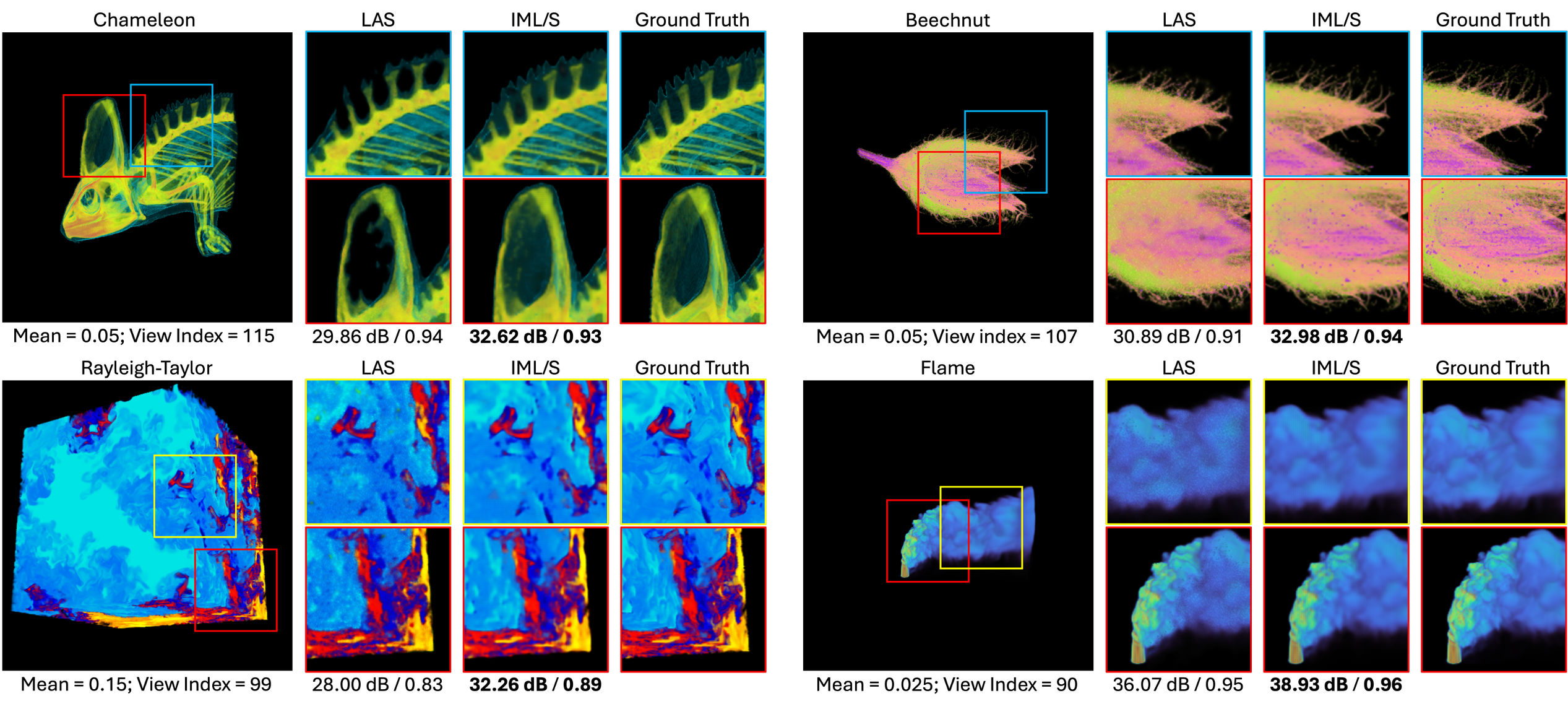}
    \vspace{-0.5\baselineskip}
    \caption{Quantitative evaluation of rendering quality between LAS and IML/S + EnhanceNet from specific views of testing datasets.}
    \label{fig:las_compare_quality}
\end{figure*}

\begin{figure*}[t]
    \centering
    \begin{subfigure}[b]{0.24\linewidth}
        \centering
        \includegraphics[trim=0 12 0 12,clip,width=\linewidth]{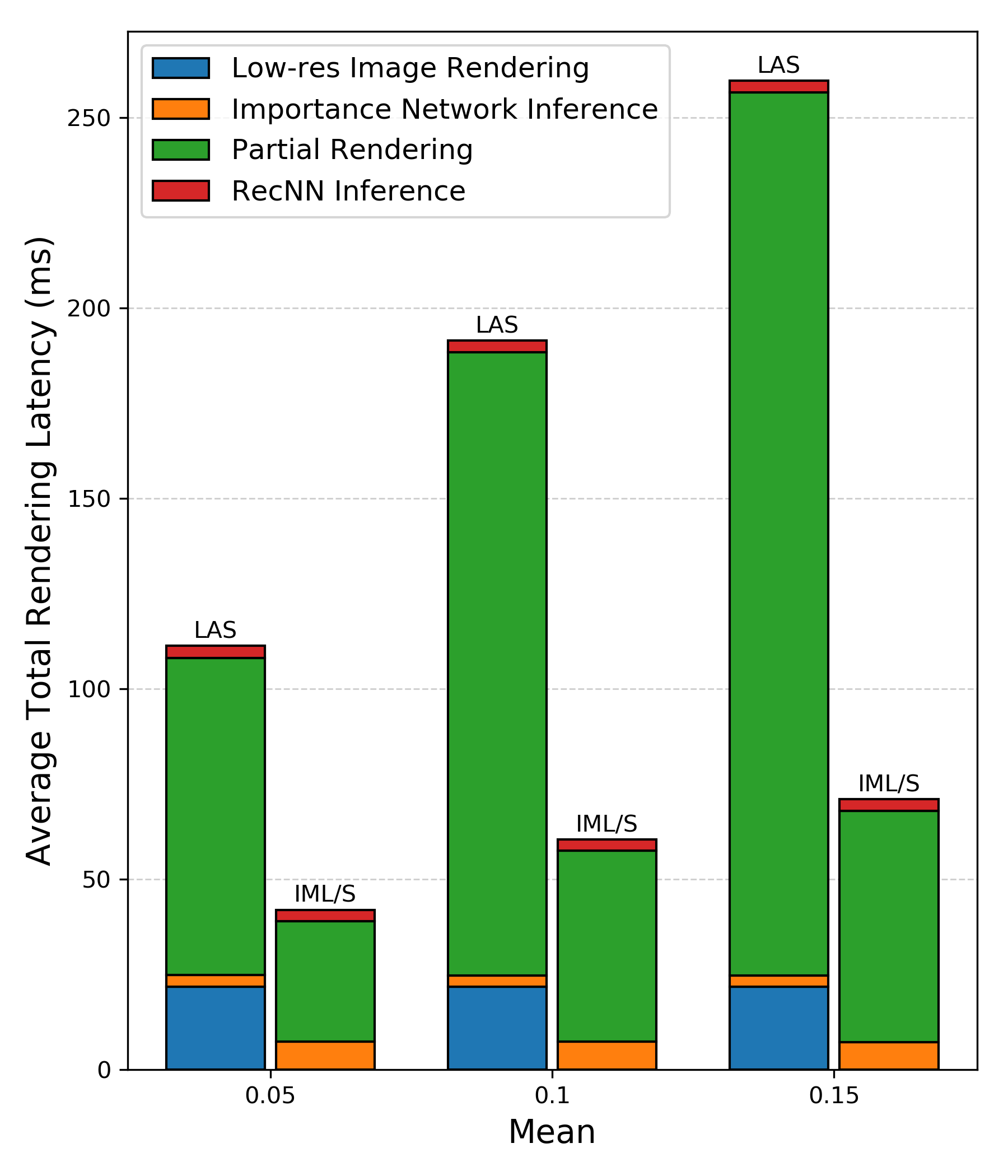}
        \caption{Chameleon}
        \label{fig:bar_chameleon}
    \end{subfigure}
    \hfill
    \begin{subfigure}[b]{0.24\linewidth}  
        \centering 
        \includegraphics[trim=0 12 0 12,clip,width=\linewidth]{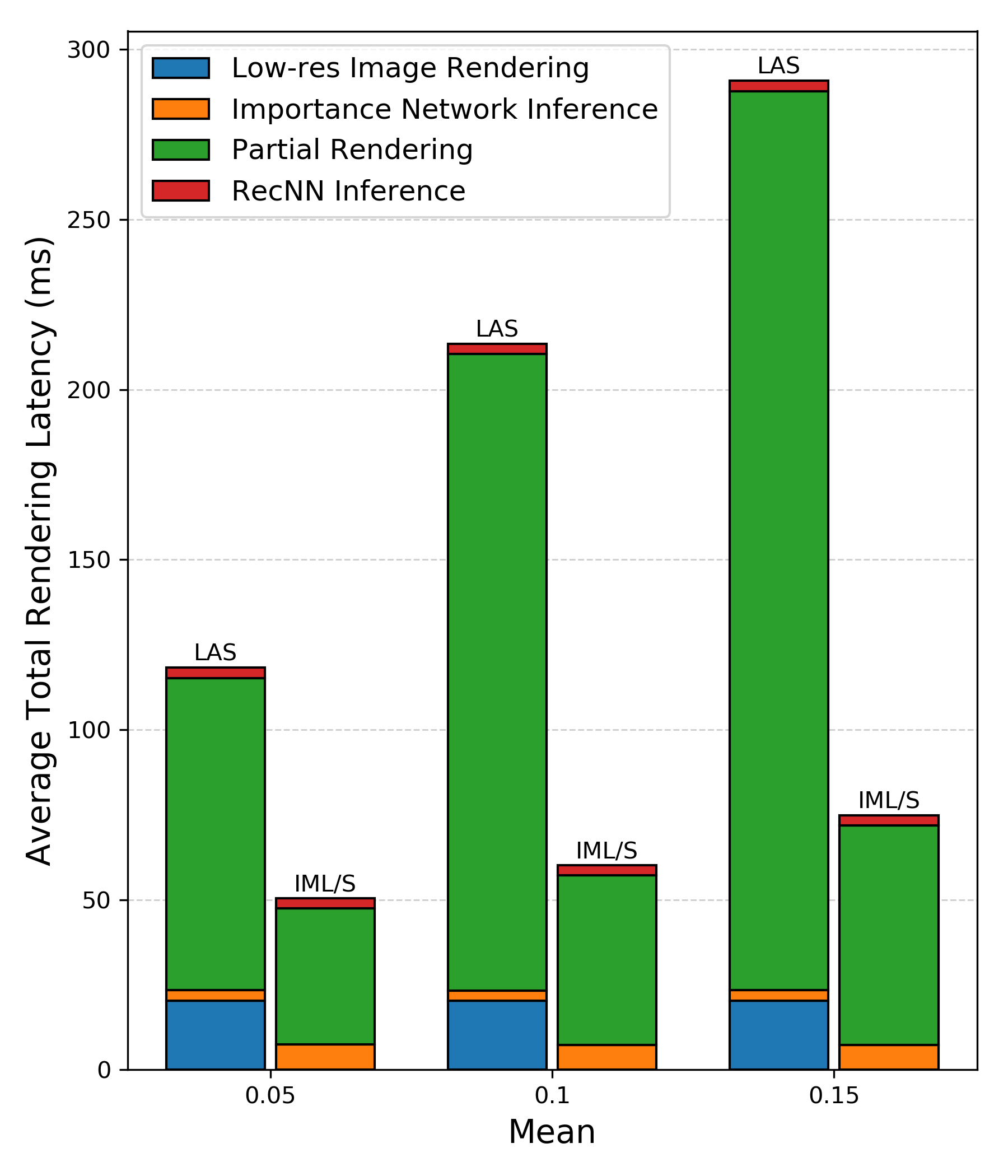}
        \caption{Beechnut}
        \label{fig:bar_beechnut}
    \end{subfigure}
    \hfill
    \begin{subfigure}[b]{0.24\linewidth}   
        \centering 
        \includegraphics[trim=0 12 0 12,clip,width=\linewidth]{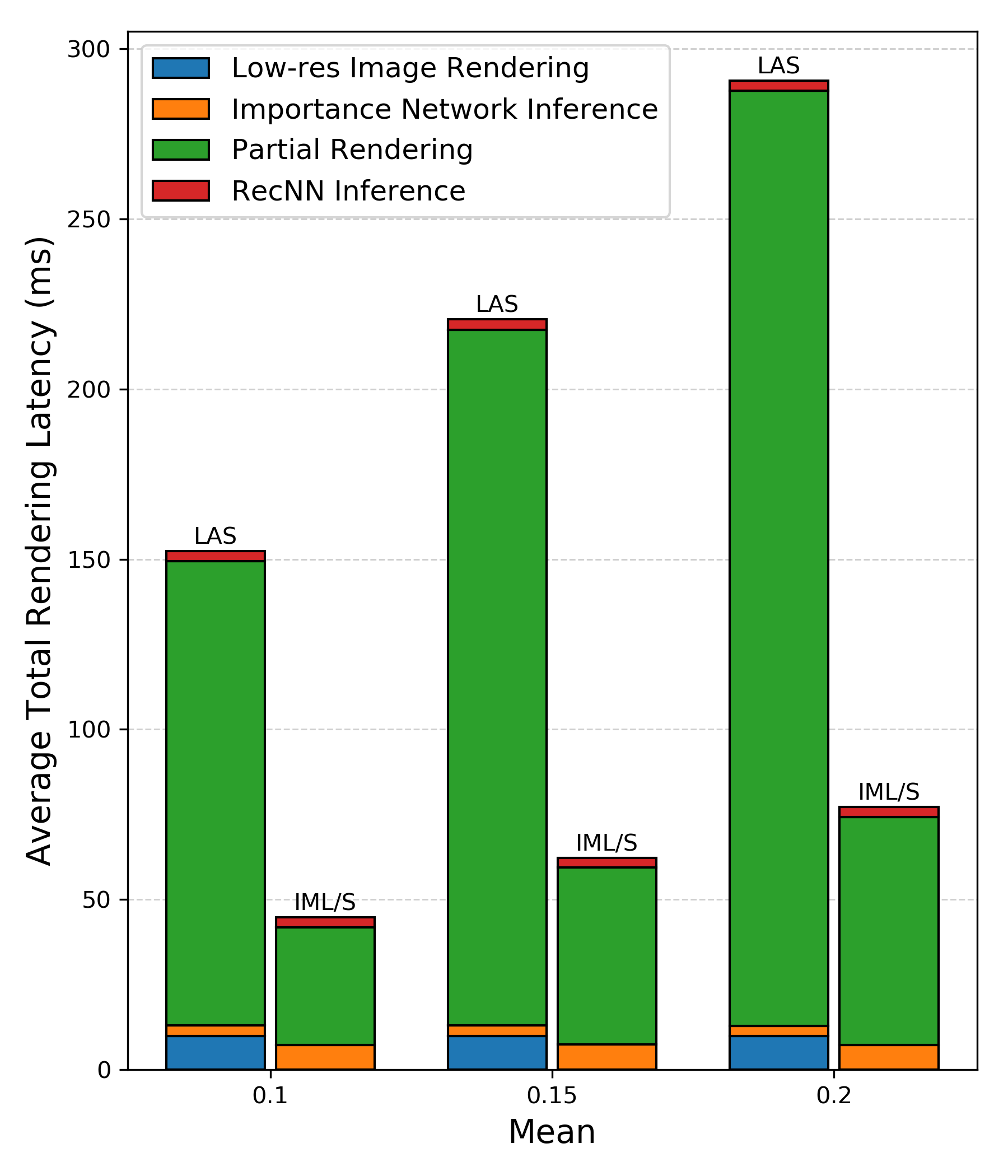}
        \caption{Rayleigh-Taylor}
        \label{fig:bar_miranda}
    \end{subfigure}
    \hfill
    \begin{subfigure}[b]{0.24\linewidth}   
        \centering 
        \includegraphics[trim=0 12 0 12,clip,width=\linewidth]{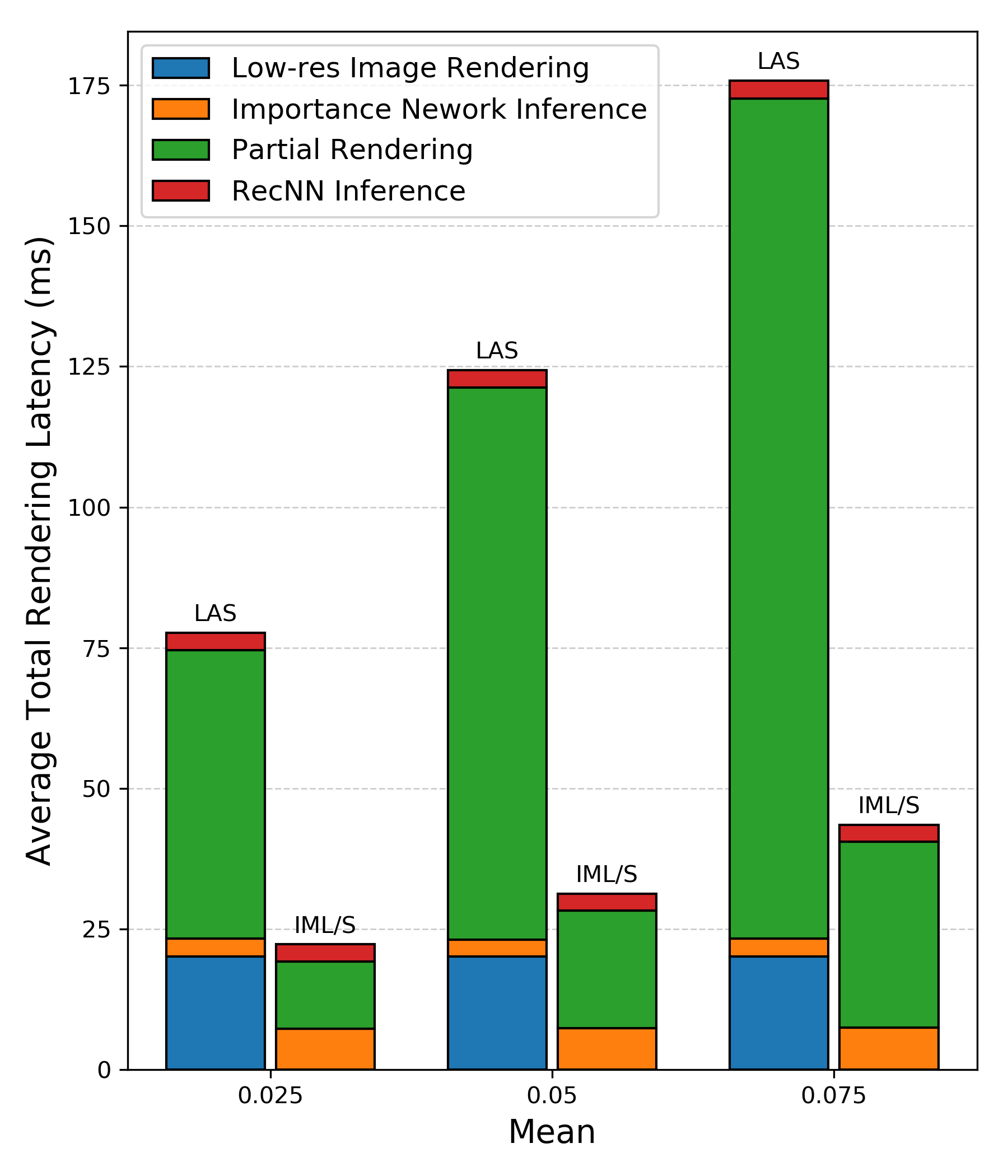}
        \caption{Flame}
        \label{fig:bar_flame}
    \end{subfigure}
    \vspace{-0.5\baselineskip}
    \caption{Time breakdown of average total rendering latency for testing datasets.}
    \label{fig:bar_compare}
\end{figure*}

\subsubsection{Comparison with LAS}
We also compare the proposed IML/S rendering pipeline with the Learning Adaptive Sampling (LAS) method~\cite{9264699} on both rendering quality and latency. Since the LAS doesn't support foveated rendering, only the super-resolution RecNN is used for the comparison. In the experiment, we adopt the same EnhanceNet configuration detailed in the LAS paper, consisting of 10 residual blocks, as the downstream RecNN for both importance mask learning networks. The LAS baseline is the paper's open-source implementation\footnote{https://github.com/shamanDevel/AdaptiveSampling}. The resolution of the rendering is $512\times 512$. Both methods are tested at three levels of IPR for each testing dataset to ensure a thorough evaluation. All the time measurements are the average of 10 trials. Detailed measurements on rendering quality and latency across all the views of the testing datasets can be found in the Appendix.

\textbf{Rendering Quality:} For rendering quality with super-resolution RecNN, as shown in \cref{tab:las_compare}, LAS and IML/S have similar performance for each IPR. This is because, for both methods, the quality of the input to the RecNN is mainly determined by the IPR, although the important pixels are learned through different importance networks. A higher IPR provides both methods with a broader selection of important pixels, resulting in a more informative partially rendered image for the RecNN to reconstruct from. Consistent with \cref{orp_improvement}, we observe that datasets featuring large 3D spaces and complex dynamics, such as Rayleigh-Taylor, require higher IPR to achieve high rendering quality. For spatially compact and simpler datasets such as Flame, relatively high rendering quality can be achieved with just a small set of important pixels from a lower IPR. A qualitative evaluation of rendering quality is shown in \cref{fig:las_compare_quality}.

\textbf{Rendering Latency:} \cref{tab:las_compare} also shows the rendering latency between the proposed method and LAS. We also measure the average rendering latency of a GPU Ray-caster\footnote{https://github.com/NVIDIA/cuda-samples/tree/master/Samples/5\_Domain\_ Specific/volumeRender} accelerated through the CUDA API as a baseline. Detailed time breakdowns of each rendering stage for all the testing datasets are shown in \cref{fig:bar_compare}. IML/S is much faster than LAS for the following reasons: 1) The input to LAS is a low-res image ($64\times 64$), which needs expensive pixel calculation (blue portion of \cref{fig:bar_compare}), while the input to IML/S is just a set of the view parameters, which is known from the user's exploration. 2) Both methods spend time on the three following stages: importance network inference, partial rendering, and RecNet inference. Both methods spend similar time on RecNet inference (red portion of \cref{fig:bar_compare}) because they use the same network. Although IML/S spends more time on the important network inference (orange portion of \cref{fig:bar_compare}) due to its more complicated network architecture, including both the IMS Net and the decoder of IML Net, it saves much more time during the partial rendering stage (green portion of \cref{fig:bar_compare}). This is because IML/S extracts the important pixels from the C-Image, which has a lower resolution ($256\times 256$) than the full-res image ($512\times 512$), where LAS extracts important pixels from. Therefore, for the same IPR, IML/S results in a much smaller number of important pixels to render. Additionally, the C-Image's resolution can be scaled down as needed for the downstream RecNN.

\section{Limitations and Future Work}
In this paper, we provide a way to further optimize the already fast volume visualization method leveraging RecNN. We concentrate on the rationale and validation of the proposed method, as well as the evaluation of how much rendering latency can be reduced while maintaining similar rendering quality. We demonstrate that the network can successfully learn the important pixels from the C-Image by jointly considering various inputs and configurations of a volume visualization system. 

One limitation of the current design is that the mean value used to determine the percentage of C-Image as important pixels is a predefined hyperparameter. While users can adjust the mean value during inference, our experiments indicate that the network trained with a specific mean value $m$ achieves the best reconstruction quality when the inference mean value is also close to $m$. Using a mean value significantly higher than $m$, resulting in a large number of pixels being rendered as important pixels, does not improve reconstruction quality and may even lead to worse results. It would be highly beneficial to develop a method that enables the network to also learn an adaptive mean value, resulting in a dynamic percentage of important pixels in the C-Image that adjusts based on the underlying rendering results. This approach can also assist in quickly identifying the ORP for a given reconstruction quality error tolerance. More comprehensive methods of identifying ORP, rather than setting a 1 dB threshold, are also worth exploring.

Various aspects of the pipeline need detailed investigation in the future. In this work, we only investigate the static fixed foveated rendering, it is worth investigating whether dynamic foveal rendering could impact reconstruction quality, rendering latency, and the pattern of the IM learned. IML Net is trained only using pixel-wise loss (MSE) in our work, other loss configurations, like structural similarity-based loss (SSIM), can be introduced to evaluate how the loss function affects the pattern of the learned IM. It is worth investigating which factors influence the distribution of important pixels and how many pixels are needed to render a specific frame. Factors such as image complexity, region size, and color dynamics may all contribute to determining which pixels are selected as important pixels. We also see the potential of utilizing the proposed IML Net to generate IMs with various levels of detail (LOD), enabling the construction of a multi-resolution representation to further optimize rendering performance.

\section{Conclusion}
In this work, we present IML Net and IMS Net to learn and synthesize an important mask that will help to further improve the already fast volume visualization using a reconstruction neural network. Our work presents the first attempt to directly synthesize an importance mask from a given view for predefined sampling patterns used in various reconstruction neural networks, via a unified interface that leverages the proposed compaction and decompaction layers. We showcase the quantitative improvement of rendering latency using our method on two state-of-the-art reconstruction neural networks, EnhanceNet and FoVolNet, on four distinct volumetric datasets, without noticeable degradation of reconstruction quality. Our proposed IML Net is also flexible to be trained independently from the downstream reconstruction neural network, allowing quick and practical application of our method on top of various off-the-shelf pre-trained reconstruction neural networks.
\section*{Acknowledgement}
This research has been sponsored in part by the National Science Foundation grants IIS-1423487 and IIS-1652846, and Advanced Scientific Computing Research, Office of Science, U.S. Department of Energy, under contracts DE-AC02-06CH11357, program manager Hal Finkel. The authors would like to thank the anonymous reviewers for their insightful comments.

\bibliographystyle{abbrv-doi}

\bibliography{template}


\vspace{2mm}

\begin{wrapfigure}[10]{l}{1in}
        \raisebox{11pt}[\dimexpr\height-0.0\baselineskip\relax]{\includegraphics[width=1in,height=1.25in,clip,keepaspectratio]{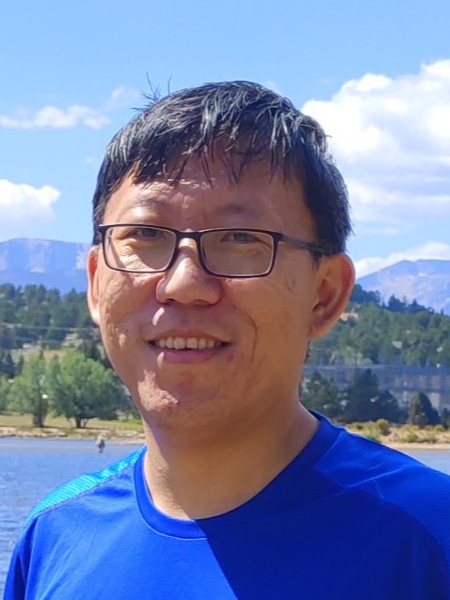}}%
\end{wrapfigure}
\noindent\textbf{Jianxin Sun} is a research assistant professor in the School of Computing at the University of Nebraska-Lincoln. He received a Ph.D. degree in Computer Science from the University of Nebraska-Lincoln and an MS degree in Electrical and Computer Engineering from Purdue University. His research concentrates on AI-driven scientific data modeling, analysis, and visualization through high-performance computing. 

\vspace{5mm}

\begin{wrapfigure}[10]{l}{1in}
        \raisebox{1pt}[\dimexpr\height-0.0\baselineskip\relax]{\includegraphics[width=1in,height=1.25in,clip,keepaspectratio]{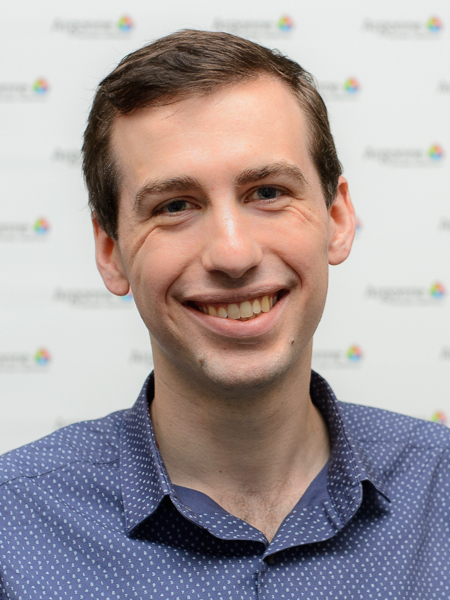}}%
\end{wrapfigure}
\noindent\textbf{David Lenz} is an assistant computer scientist at Argonne National Laboratory. He received his Ph.D. in Mathematics from University of California San Diego. His research focuses on scientific data visualization and analysis, with an emphasis on functional approximation, implicit neural representations, and lossy compression. \; \; \; \; \; \; \; \; \; \; \; \; \; \; \; \; \; \; \; \; \; \; \; \; \; \; \; \; \; \; \; \; \; \; \; \; \; \; \; \; \; \; \; \; \; \; \; \; \; \; \; \; \; \; \; \; \; \; \; \; \; \; \; \; \; \; \; \; \; \; \; \; \; \; \; \; \; \; \; \; \; \; \; \; \; \; \; 

\vspace{5mm}

\begin{wrapfigure}[10]{l}{1in}
        \raisebox{14pt}[\dimexpr\height-0.0\baselineskip\relax]{\includegraphics[width=1in,height=1.25in,clip,keepaspectratio]{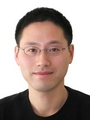}}%
\end{wrapfigure}
\noindent\textbf{Hongfeng Yu} is a Professor in the School of Computing and Director of Holland Computing Center at the University of Nebraska–Lincoln. He earned his Ph.D. in Computer Science from the University of California, Davis. His research focuses on developing theories and technologies for scalable data analysis and visualization, advancing discoveries across scientific and engineering fields through interdisciplinary collaborations. \; \; \; \; \; \; \; \; \; \; \; \; \; \; \; \; \; \; \; \; \; \; \; \; \; \; \; \; \; \; \; \; \; \; \; \; \; \; \; \; \; \; \; \; \; \; \; \; \; \; \; \; \; \; \; \; \; \; \; \; \; \; \; \; \; \; \; \; \; \; \; \; \; \; \; \; \; \; \; \; \; \; \; \; \; \;

\vspace{5mm}

\begin{wrapfigure}[10]{l}{1in}
        \raisebox{14pt}[\dimexpr\height-0.0\baselineskip\relax]{\includegraphics[width=1in,height=1.25in,clip,keepaspectratio]{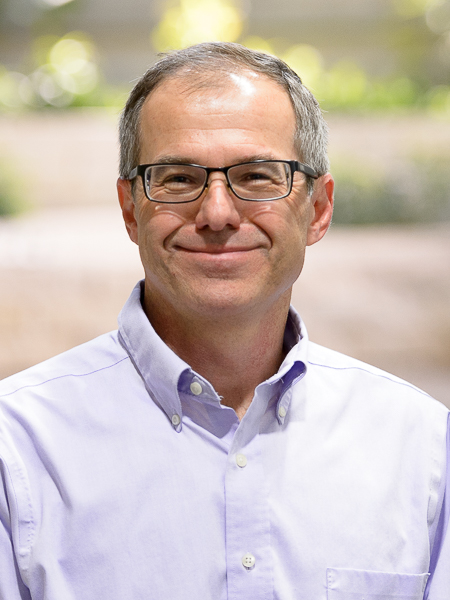}}%
\end{wrapfigure}
\noindent\textbf{Tom Peterka} is a computer scientist at Argonne National Laboratory, scientist at the University of Chicago Consortium for Advanced Science and Engineering (CASE), and fellow of the Northwestern Argonne Institute for Science and Engineering (NAISE). His research interests are large-scale parallel in situ analysis of scientific data. Recipient of the 2017 DOE early career award and five best paper awards, Peterka has published over 150 peer-reviewed articles and papers since earning his Ph.D. in computer science from the University of Illinois at Chicago in 2007.

\end{document}